\documentclass[12pt,epsfig]{article}
\usepackage{amssymb,amsmath, wasysym}
\usepackage{graphicx, float,psfrag}
\usepackage[numbers]{natbib}
\usepackage[]{graphicx,bm}
\usepackage{epsfig}
\newcommand{\lsim}{\lesssim}
\renewcommand{\Re}{\ensuremath{\mathop{\rm Re}}}

\begin{document}
\renewcommand{\theequation}{\arabic{section}.\arabic{equation}}

\title{\vspace{-50mm}
       \vspace{2.5cm}
       \vspace{8mm}
The Higgs Potential in the Type II Seesaw Model}
\vspace{2mm}
\author{A.~Arhrib${}^{1,2}$, R. Benbrik${}^{2,3, 4}$,  M.~Chabab${}^2$, 
        \\ G.~Moultaka${}^*$ ${}^{5,6}$, M.~C. Peyran\`ere${}^{7,8}$, L.~Rahili${}^2$, J.~Ramadan${}^2$ \\[3mm]
{\normalsize \em ${}^1$  D\'epartement de Math\'ematiques, Facult\'e des Sciences et Techniques, Tanger, Morocco}\\
{\normalsize \em ${}^2$Laboratoire de Physique des Hautes Energies et Astrophysique } \\
{\normalsize \em  D\'epartement de Physiques,Facult\'e des Sciences
  Semlalia,Marrakech, Morocco} \\
{\normalsize \em ${}^3$ Facult\'e Polydisciplinaire, Universit\'e Cadi Ayyad, Sidi Bouzid, Safi-Morocco}\\
{\normalsize \em ${}^4$ Instituto de Fisica de Cantabria (CSIC-UC), Santander, Spain} \\
{\normalsize \em ${}^5$ Universit\'e Montpellier 2, Laboratoire Charles Coulomb UMR 5221,}\\
{\normalsize \em F-34095 Montpellier, France}\\
{\normalsize \em ${}^6$ CNRS, Laboratoire Charles Coulomb UMR 5221,  F-34095 Montpellier, France}\\
{\normalsize \em ${}^7$ Universit\'e Montpellier 2, Laboratoire Univers \& Particules de Montpellier UMR 5299,} \\
{\normalsize \em F-34095 Montpellier, France}\\
{\normalsize \em ${}^8$ CNRS/IN2P3, Laboratoire Univers \& Particules de Montpellier UMR 5299,}\\
{\normalsize \em  F-34095 Montpellier, France}\\
}
\setcounter{figure}{0}
\maketitle
\thispagestyle{empty}
\begin{abstract}
The Standard Model Higgs sector, extended by one  weak gauge triplet of scalar fields 
with a very small vacuum expectation value, is a very promising setting to account
for neutrino masses through the so-called type II seesaw mechanism.
 In this paper we consider the general renormalizable doublet/triplet 
Higgs potential of this model. We perform a detailed study of its main dynamical features that depend on five 
dimensionless couplings and two mass parameters after spontaneous symmetry breaking, and highlight the implications 
for the Higgs phenomenology. In particular, we determine
{\sl i)} the complete set of tree-level unitarity constraints on the couplings of the potential  
and {\sl ii)} the exact tree-level {\sl all directions} boundedness from below constraints on these couplings.
When combined, these constraints delineate precisely the theoretically allowed parameter space domain within our
perturbative approximation.
Among the seven physical Higgs states of this model,
the mass of the lighter (heavier) ${\mathcal{CP}}_{even}$  state $h^0$ ($H^0$) will always satisfy a 
theoretical {\sl upper (lower) bound} 
that is reached for a 
critical value  $\mu_c$ of $\mu$ (the mass parameter controlling triple couplings among the doublet/triplet Higgses). 
Saturating the unitarity bounds we find
an upper bound $m_{h^0} < {\cal O}(0.7 - 1 {\rm TeV})$, while the upper bound for the remaining Higgses lies in
the several tens of TeV. However, the actual masses can be much lighter. 
We identify two regimes corresponding to 
$\mu \gtrsim \mu_c$ and $\mu \lesssim \mu_c$. In the first regime the 
Higgs sector is typically very heavy and only
$h^0$ that becomes SM-like could be accessible to the LHC. In contrast, in the second regime, 
somewhat overlooked in the literature, most 
of the Higgs sector is light. In particular, the heaviest state $H^0$ becomes SM-like, the lighter states
being the ${\mathcal{CP}}_{odd}$ Higgs, the (doubly) charged Higgses  and a decoupled $h^0$, possibly 
leading to a distinctive phenomenology at the colliders. 

\vspace{.5cm}  
\noindent
{ ${}^*${\sl corresponding author}}



\end{abstract}

\newpage

\section{Introduction}
One of the major goals of the LHC is to uncover the mechanism underlying the
electroweak symmetry breaking and thereby the origin of the weak gauge boson
and fermion masses. 
Moreover, observation of neutrino oscillations has shown that neutrinos are massive (for a review see for instance
\cite{Altarelli:2010fk} and references therein). Such masses do not necessarily require 
physics beyond the standard model (SM), since one can accommodate a (Dirac) mass through a Yukawa coupling
assuming a right-handed  neutrino similarly to the other massive fermions. However, the introduction of such a 
right-handed state, whose only role is to allow for non-zero neutrino masses 
while being neutral under all the SM interactions, 
might seem rather mysterious. Furthermore, in contrast with the other right-handed fermion 
states of the SM, the right-handed neutrino allows also for a Majorana mass that is invariant under the SM gauge group 
but violates lepton number. These features make plausible the existence of new flavor physics beyond the SM
associated with the neutrino sector. Probably, one of the most attractive aspects is the ability to induce naturally 
the tiny  neutrino masses from this new flavor physics sector \cite{Weinberg:1979sa}. 
The celebrated seesaw mechanism \cite{GellMann:1979mm, Yanagida:1979as, Mohapatra:1979ia} relating directly
the smallness of the neutrino masses to the presence of a large new scale $\Lambda$ through $m_{\nu} \sim v^2/\Lambda$,
when $\Lambda \gg v$ where $v$ denotes the electroweak scale, is realized in a grand unified context (GUT)
comprising right-handed neutrinos and often dubbed {\sl type I seesaw}. It can also be achieved without right-handed
neutrinos through an extended Higgs sector including an $SU(2)_L$ triplet scalar field, {\sl type II seesaw}
\cite{Konetschny:1977bn, Cheng:1980qt, Lazarides:1980nt, Schechter:1980gr, Mohapatra:1980yp}, 
or by including two extra matter multiplets in the adjoint of $SU(2)_L$, {\sl type III seesaw} \cite{Foot:1988aq},
or a hybrid type mixture of {\sl type I} and {\sl type III}  \cite{Ma:1998dn, Bajc:2006ia, Perez:2007rm, Perez:2007iw}.

If such extended sectors are too heavy to be directly accessible to TeV scale experiments, they could still
be indirectly probed through distinctive low energy effective operators in the neutrino sector \cite{Abada:2007ux}.
In the present paper we will rather focus on the possibility of accessing directly the Higgs sector {\sl per se}
 of the {\sl type II} scenario, studying general dynamical constraints which originate from the potential that 
couples the Higgs doublet and the Higgs triplet.
However, given the present theoretical uncertainties, we do not  commit  to any specific GUT or flavor physics 
scenarios beyond the SM. In particular mass parameters such as $\mu$ and $M_{\Delta}$ will not necessarily take large 
GUT scale values, even though such a configuration is included in the analysis. We will even consider regimes with very 
small $\mu$ ($\ll G_F^{-1/2}$). As noted in \cite{Perez:2008ha}, such a small $\mu$ makes all the Higgs sector
accessible to the LHC. Here we carry out a complete study, taking into account the full set of
renormalizable operators present in the potential.  The aim is to exhibit the various possible
regimes  consistent with the {\sl dynamical constraints} dictated by the potential and their consequences on the phenomenology
of the extended Higgs sector. Most of these operators are often neglected in the existing phenomenological studies
of {\sl type II seesaw}, based on the fact that after spontaneous symmetry breaking their effects are suppressed 
by the small Higgs triplet vacuum expectation value (VEV), $v_t$,  when compared to the electroweak scale. This is, 
however, not justified when studying the small $\mu$ regimes just mentioned, where $\mu$ can be of order $v_t$.
In this case the detailed dynamics leads to an interesting structure of the Higgs sector.

  The paper is organized as follows: in section 2, we present the ingredients of the model, 
the physical Higgs states and mass spectrum, as well as a parameterization of the potential parameters in terms of the physical
masses. In section 3, we discuss some of the phenomenological and theoretical constraints on the parameters related
to precision measurements, the absence of tachyonic Higgs modes, as well as the presence of false vacua.     
In section 4, we provide a thorough study of the boundedness from below of the potential and establish for the first 
time simple necessary and sufficient conditions on the couplings that are valid for {\sl all field directions}. The unitarity
constraints are analyzed in detail in section 5, through the study of all the scalar scattering channels.
In section 6, we combine in an analytical compact form the constraints obtained in sections 4 and 5. 
Section 7 presents the behavior of the  ${\mathcal{CP}}_{even}$ Higgs masses as functions of the potential parameters, 
highlighting theoretical upper and lower mass bounds and identifying different regimes that give better insight into the overall
 Higgs sector phenomenology, as well as the determination of unitarity mass bounds on the lightest Higgs.
Section 8 is devoted to a short review of the salient features of the Higgs phenomenology at the colliders
as well as to specific illustrations of our results. We conclude in section 9 and give some technical details in
the appendices.



\setcounter{equation}{0}
\section{The model}
\label{sec:themodel}

We start by recalling the scalar potential and the main properties 
of the Higgs physical eigenstates after EWSB as well as the 
corresponding eigenmasses and mixing angles. We give the expressions without neglecting any of the couplings 
appearing in Eq.~(\ref{eq:Vpot}) nor making any specific assumption about the magnitudes of $\mu, m_H^2$  
and $M_{\Delta}^2$ which would originate from the unknown underlying high energy theory. The results of this
section fix the notations and will serve for the completely model-independent analysis carried out in the subsequent 
sections.

\subsection{The Higgs potential}
\label{sec:thehiggspot}
The scalar sector consists of the standard Higgs weak doublet  $H$ and a colorless scalar field
$\Delta$ transforming as a triplet under the $SU(2)_L$ gauge group with hypercharge $Y_\Delta=2$, so that 
$H \sim (1, 2, 1)$ and  $\Delta \sim (1, 3, 2)$ under the $SU(3)_c \times SU(2)_L \times U(1)_Y$.  

Under a general gauge transformation ${\cal U}(x)$, $H$ and $\Delta$ transform as
$H \to {\cal U}(x) H$  and $\Delta \to {\cal U}(x) \Delta {\cal U}^\dag(x)$.
One can then write the most general renormalizable and gauge invariant Lagrangian of this scalar sector as follows:


\begin{eqnarray}
\mathcal{L} &=&
(D_\mu{H})^\dagger(D^\mu{H})+Tr(D_\mu{\Delta})^\dagger(D^\mu{\Delta}) -V(H, \Delta) + \mathcal{L}_{\rm Yukawa}
\label{eq:DTHM}
\end{eqnarray}

\noindent
where the covariant derivatives are defined by 
\begin{eqnarray}
D_\mu{H} &=& \partial_\mu{H}+igT^a{W}^a_\mu{H}+i\frac{g'}{2}B_\mu{H} \label{eq:covd1}\\
D_\mu{\Delta} &=&
  \partial_\mu{\Delta}+ig[T^a{W}^a_\mu,\Delta]+ig' \frac{Y_\Delta}{2} B_\mu{\Delta} \label{eq:covd2}
\end{eqnarray}

\noindent
(${W}^a_\mu$, $g$), and ($B_\mu$, $g'$) denoting respectively the $SU(2)_L$ and $U(1)_Y$ gauge fields and couplings
and $T^a \equiv \sigma^a/2$, with $\sigma^a$ ($a=1, 2, 3$)  the Pauli matrices.
The potential $V(H, \Delta)$ is given by,


\begin{eqnarray}
V(H, \Delta) &=& -m_H^2{H^\dagger{H}}+\frac{\lambda}{4}(H^\dagger{H})^2+M_\Delta^2Tr(\Delta^{\dagger}{\Delta})
+[\mu(H^T{i}\sigma^2\Delta^{\dagger}H)+{\rm h.c.}] \nonumber\\
&&+\lambda_1(H^\dagger{H})Tr(\Delta^{\dagger}{\Delta})+\lambda_2(Tr\Delta^{\dagger}{\Delta})^2
+\lambda_3Tr(\Delta^{\dagger}{\Delta})^2 +\lambda_4{H^\dagger\Delta\Delta^{\dagger}H}
\label{eq:Vpot}
\end{eqnarray}

\noindent
where $Tr$ is the trace over $2\times2$ matrices. $\mathcal{L}_{\rm Yukawa}$ contains all the Yukawa sector
of the SM plus one extra Yukawa term that leads after spontaneous symmetry breaking to (Majorana) mass terms for
the neutrinos, without requiring right-handed neutrino states,

\begin{equation}
\mathcal{L}_{\rm Yukawa} \supset  - Y_{\nu} L^T C \otimes i \sigma^2 \Delta L  + {\rm h.c.} \label{eq:yukawa}
\end{equation}

\noindent
where $L$ denotes $SU(2)_L$ doublets of left-handed leptons, $Y_{\nu}$ denotes neutrino Yukawa couplings,  
$C$ the charge conjugation operator, and we have suppressed
falvor indices for simplicity. Although part of the type II seesaw model,  we will refer to 
the above model Eq.~(\ref{eq:DTHM}) as the doublet-triplet-Higgs-Model (DTHM) since in this paper we are mainly 
interested in the scalar sector, bringing up only occasionally the content of the Yukawa sector $\mathcal{L}_{\rm Yukawa}$
and the related neutrino masses issue.  

Defining the electric charge as usual, $Q= I_3 + \frac{Y}{2}$ where $I$ denotes the isospin,  
we write the two Higgs multiplets in components as,  
\begin{eqnarray}
\Delta &=\left(
\begin{array}{cc}
\delta^+/\sqrt{2} & \delta^{++} \\
\delta^0 & -\delta^+/\sqrt{2}\\
\end{array}
\right)~~~~{\rm and}~~~~H=\left(
                    \begin{array}{c}
                      \phi^+ \\
                      \phi^0 \\
                    \end{array}
                  \right)
\end{eqnarray}

\noindent
where we have used for convenience the $2 \times 2$ traceless matrix representation for the triplet.\footnote{Note 
that the electric charge assignments for the upper and lower component fields are only conventional and can be 
interchanged by taking $Y_\Delta = -2, Y_H=-1$, entailing an exchange of the upper and lower components
of the fermion weak doublets, without affecting the physical content. This seemingly
trivial statement is important to keep in mind when discussing possible electric charge breaking minima of the 
potential.}  

The potential defined in Eq.~(\ref{eq:Vpot}) exhausts all possible gauge invariant renormalizable operators.
For instance a term of the form 
$ \lambda_5 H^\dagger \Delta^\dagger \Delta H$, which would be legitimate to add if $\Delta$ contained a singlet
component, can actually be projected on the $\lambda_1$ and $\lambda_4$ operators appearing in Eq.~(\ref{eq:Vpot})
thanks to the identity
$H^\dagger \Delta^\dagger \Delta H + H^\dagger \Delta \Delta^\dagger H=
H^\dagger H Tr(\Delta^\dagger \Delta)  $ which is valid because $\Delta$ is a traceless $2\times2$ matrix. 
This simply amounts to 
redefining  $\lambda_1$ and $\lambda_4$ such as $\lambda_1 + \lambda_5 \to \lambda_1,\; 
\lambda_4 - \lambda_5 \to \lambda_4$. The potential thus depends on 
five independent dimensionless couplings  $\lambda$ and $\lambda_i,  (i=1,...4)$ and three mass parameters, 
$m_H^2, M_\Delta^2$ and $\mu$. 
In the present paper we will assume all these parameters to be real 
valued. Indeed, apart from the $\mu$ term, all other operators in $V$ are self-conjugate so that, 
by hermiticity of the potential, only the real parts of the $\lambda$'s and the $m_H^2, M_\Delta^2$ mass parameters 
will be relevant. As for $\mu$, the only parameter that can pick up a would-be CP phase, this phase is unphysical
and can always be absorbed in a redefinition of the
fields $H$ and $\Delta$. One thus concludes that the DTHM Lagrangian is CP conserving (see also the discussion
in \cite{Dey:2008jm}).
Moreover, $V$ depends on five complex (or ten real) scalar 
fields.
 

Assuming that spontaneous electroweak symmetry breaking (EWSB) is taking place at some electrically 
neutral point in the field space, and denoting the corresponding VEVs by

\begin{eqnarray}
\langle \Delta \rangle &=\left(
\begin{array}{cc}
0 & 0 \\
v_t/\sqrt{2} & 0\\
\end{array}
\right)~~~~{\rm and}~~~~\langle H \rangle =\left(
                    \begin{array}{c}
                      0 \\
                       v_d/\sqrt{2}\\
                    \end{array}
                  \right)
\label{eq:VEVs}
\end{eqnarray}

\noindent
one finds after minimization of the potential Eq.(\ref{eq:Vpot}) 
the following necessary conditions :

\begin{eqnarray}
  M_\Delta^2 &=& \frac{2\mu{v_d^2}-\sqrt{2}(\lambda_1+
\lambda_4)v_d^2v_t-2\sqrt{2}(\lambda_2+\lambda_3)v_t^3}{2\sqrt{2}v_t} \label{eq:ewsb1}\\
  m_H^2 &=& \frac{\lambda{v_d^2}}{4}-\sqrt{2}\mu{v_t}+\frac{(\lambda_1+\lambda_4)}{2}v_t^2  \label{eq:ewsb2}
\end{eqnarray}


\noindent
Even though, as we noted above, CP symmetry is realized at the level of the Lagrangian, there remains in principle the 
possibility for a spontaneous breakdown of this symmetry, an issue which we do not address in this paper. 
We can thus choose in the sequel $v_d$ and $v_t$ to be real valued; that is, we consider only CP conserving vacua 
for which complex valued $v_d$ and/or $v_t$ can always be rotated simultaneously to real values
through some unphysical phase redefinition of the fields.   

These equations, to which we will refer as the EWSB conditions, ensure that the vacuum corresponds to an extremum of
the potential, [that is $\partial V/\partial \eta_i |_{\Delta = \langle \Delta \rangle , H= \langle H \rangle}
=0$ for each of the ten real-valued field components denoted here by $\eta_i, (i=1, ...10)$], but one would still need 
to check that this extremum is indeed a stable, albeit local, minimum. 
The corresponding extra conditions are nothing else but the absence of tachyonic modes in the Higgs sector,  
to be considered in a later section. We just anticipate here that the latter conditions will enforce the
signs of $\mu$ and $v_t$ to be identical. We can thus choose in the sequel $v_t >0$, $\mu >0$ without
loss of generality. Furthermore, the two free parameters $m_H^2$ and $M_\Delta^2$ can now be traded for $v_d$ and $v_t$ through
Eqs.~(\ref{eq:ewsb1}, \ref{eq:ewsb2}). In the rest of the paper we will take the eight parameters of the 
potential as being $\lambda$, $\lambda_i,  (i=1,...4)$, $\mu, v_d$ and $v_t$; requiring the correct
electroweak scale will put the further constraint $v \equiv \sqrt{v_d^2 + 2 v_t^2}  = 246$GeV on  
$v_d, v_t$, reducing this set of free parameters down to seven.


Let us also note that the above EWSB conditions  will not necessarily imply that the gauge symmetric vacuum 
({\sl i.e.} at $\eta_i=0$) is unstable. Indeed the latter instability  
requires that $M_{\Delta}^2 < 0$ and/or $m_H^2 > 0$ which are not guaranteed by Eqs.~(\ref{eq:ewsb1}, \ref{eq:ewsb2}).
Even more so, regimes with large $\mu$ will lead through the EWSB conditions to a very narrow gauge
symmetric local minimum(!) so that metastability issues might have to be considered.
[More comments about the structure of the vacua of the model will be deferred to section \ref{sec:vacuum}.] 

On the other edge of the spectrum, very small values of $\mu$ could be favored if one
requires the lepton number not to be strongly violated. Indeed, the $\mu$ term in Eq.(\ref{eq:Vpot}) is 
the only source of lepton number violation at the Lagrangian level and {\sl before} spontaneous EWSB. 
If this term is absent 
the Yukawa term Eq.~(\ref{eq:yukawa}) together with the other standard Yukawa terms 
imply a conserved lepton number (where the $\Delta$ and $H$ Higgs fields  carry respectively the lepton numbers
 $l_{\Delta}=-2$ and $l_H=0$).\footnote{The processes mediated by Eq.~(\ref{eq:yukawa}) and involving Higgs
triplet decay or exchange are sometimes misleadingly dubbed 'lepton number violating'. One can check that the net 
overall lepton number of any process, comprising such decays or exchange, is conserved. This global symmetry will be violated only spontaneously when
$\Delta$ acquires a VEV, that is when the Majorana mass is induced from (\ref{eq:yukawa}).} Then, from the lepton number assignment for $H$ and 
$\Delta$ it follows that the $\mu$ term violates lepton number by two units. However, this violation is soft since 
the $\mu$-induced lepton number violating processes (corresponding either to loop suppressed $2 \to 2$ processes or to propagator suppressed  
multi-particle processes) will have to involve both the standard and neutrino 
Yukawa couplings .  These features suggest that if the two seemingly independent sources of lepton number 
violation, namely the $\Delta$ VEV and $\mu$, are assumed to have a common origin  such as some spontaneous symmetry 
breaking of an underlying flavor theory, then it is natural to expect $\mu = {\cal O}(v_t)$ up to possible 
Yukawa coupling factors.

\subsection{Higgs masses and mixing angles}

The $10 \times 10$ squared mass matrix

\begin{equation}
 {\mathcal M}^2=\frac{1}{2} \frac{\partial^2 V}{\partial \eta_i^2} |_{\Delta = \langle \Delta \rangle , H= \langle H \rangle}
\end{equation}

\noindent
can be recast, using Eqs.(\ref{eq:ewsb1}, \ref{eq:ewsb2}), 
in a block diagonal form of one doubly-degenerate eigenvalue $m_{H^{\pm\pm}}^2$ and four 
$2 \times 2$ matrices denoted in the following by ${\mathcal{M}}_{\pm}^2, {\mathcal{M}}_{{\mathcal{CP}}_{even}}^2$ and 
${\mathcal{M}}_{{\mathcal{CP}}_{odd}}^2$.

\subsection*{Mass of the doubly-charged field:}
The double eigenvalue  $m_{H^{\pm\pm}}^2$ corresponds to the doubly charged eigenstate $\delta^{\pm \pm}$
and could also be obtained directly by collecting all the coefficients of  $ \delta^{++}\delta^{--}$
in the potential. It reads
 
\begin{eqnarray}
m_{H^{\pm\pm}}^2=\frac{\sqrt{2}\mu{v_d^2}- \lambda_4v_d^2v_t-2\lambda_3v_t^3}{2v_t}  \label{eq:mHpmpm}
\end{eqnarray}

\noindent
From now on we will denote the doubly charged mass eigenstates $\delta^{\pm\pm}$ by $H^{\pm\pm}$.

\subsection*{Mass of the singly-charged field:}

The mass-squared matrix for the singly charged field is:

$$
{\mathcal{M}}_{\pm}^2= (\sqrt{2}\mu-\frac{ \lambda_4 v_t}{2} )\left(
  \begin{array}{cc}
  v_t & -v_d/\sqrt{2}\\
  -v_d/\sqrt{2} & v_d^2/2v_t\\
  \end{array}
\right)
$$

This matrix is diagonalized by the following matrix
$\mathcal{R}_{\beta^{'}}$, given by :
\begin{eqnarray}
{\mathcal{R}}_{\beta^{'}} &=& \left(
\begin{array}{cc}
\cos\beta^{'} & -\sin\beta^{'} \\
\sin\beta^{'} & \cos\beta^{'} \\
\end{array}
\right)
\label{eq:rotamatbetaprime}
\end{eqnarray}
where $\beta^{'}$ is a rotation angle. Among the two eigenvalues of ${\mathcal{M}}_{\pm}^2$, one is zero and 
corresponds to the charged Goldstone boson $G^\pm$ while the other corresponds to the singly charged Higgs boson $H^\pm$
and is given by 
\begin{eqnarray}
m_{H^\pm}^2=\frac{(v_d^2+2 v_t^2)[2\sqrt{2}\mu- \lambda_4 v_t]}{4v_t}
\label{eq:mHpm}
\end{eqnarray}
The  mass eigenstates $H^\pm$ and $G^\pm$ are rotated from the Lagrangian 
fields $\phi^{\pm}, \delta^{\pm}$ and defined by

\begin{eqnarray}
  G^\pm &=& \cos\beta^{'} \phi^{\pm}+ \sin\beta^{'} \delta^{\pm} \\
  H^\pm &=& -\sin\beta^{'} \phi^{\pm}+ \cos\beta^{'}\delta^{\pm}
\end{eqnarray}
The diagonalization of ${\mathcal{M}}_{\pm}^2$  leads to the following relations involving the rotation angle
 $\beta^{'}$:

\begin{eqnarray}
\frac{v_d^2}{2v_t}[\sqrt{2}\mu-\frac{ \lambda_4 v_t}{2}]&=&\cos^2\!\beta^{'}M_{H^\pm}^2 \label{eq:sbp1}\\
\frac{v_d}{\sqrt{2}}[\sqrt{2}\mu-\frac{ \lambda_4 v_t}{2}]&=&\frac{\sin2{\beta^{'}}}{2}M_{H^\pm}^2 \label{eq:sbp2} \\
v_t[\sqrt{2}\mu-\frac{ \lambda_4 v_t}{2}]&=& \sin^2\!\beta^{'}M_{H^\pm}^2 \label{eq:sbp3}
\end{eqnarray}

\noindent
These equations lead to a unique solution for $\sin\beta^{'}, \cos\beta^{'}$ up to a global sign ambiguity. Indeed,
Eq.~(\ref{eq:sbp1}) implies $\sqrt{2}\mu-\frac{ \lambda_4 v_t}{2} > 0$ in order not have a tachyonic $H^\pm$ state and
given our convention of $v_t > 0$. Then it follows from Eq.~(\ref{eq:sbp2}) that $\sin\beta^{'}$ and $\cos\beta^{'}$
should have the same sign. One finds

\begin{eqnarray}
\sin\beta^{'}&=& \epsilon_{\beta^{'}} \frac{\sqrt{2}v_t}{\sqrt{v_d^2+2v_t^2}} 
\qquad , \qquad
\cos\beta^{'}= \epsilon_{\beta^{'}} \frac{v_d}{\sqrt{v_d^2+2v_t^2}} \label{eq:scbprime}
\end{eqnarray}

\noindent
with a sign freedom $\epsilon_{\beta^{'}} = \pm 1$, and

\begin{eqnarray}
  \tan{\beta^{'}}&=&\sqrt{2}\frac{v_t}{v_d} \label{eq:tanbp}
\end{eqnarray}


\subsection*{Mass of the neutral fields:}

The neutral scalar and pseudoscalar mass matrices read:

\begin{eqnarray}
{\mathcal{M}}_{{\mathcal{CP}}_{even}}^2=\left(
                                                  \begin{array}{cc}
                                                   A & B \\
                                                   B & C  \end{array}
\right)~~~~{\rm and}~~~~{\mathcal{M}}_{{\mathcal{CP}}_{odd}}^2=\sqrt{2}\mu\left(
                                    \begin{array}{cc}
                                    2v_t & -v_d \\
                                    -v_d  & v_d^2/2v_t \\
                                    \end{array}
\right)
                                                  \end{eqnarray}
where
\begin{eqnarray}
  A &=& \frac{\lambda}{2}{v_d^2}, \; \;
  B =v_d ( -\sqrt{2}\mu+(\lambda_1+\lambda_4)v_t) , \; \; 
  C = \frac{\sqrt{2}\mu{v_d^2}+4(\lambda_2+\lambda_3)v_t^3}{2v_t} \label{eq:ABC}
\end{eqnarray}\\
These symmetric matrices are diagonalized by the following two orthogonal matrices :


\begin{eqnarray}
{\mathcal{R}}_\alpha = \left(
\begin{array}{cc}
\cos\alpha & -\sin\alpha \\
\sin\alpha & \cos\alpha \\
\end{array}
\right)~~~~{\rm and}~~~~{\mathcal{R}}_\beta = \left(
\begin{array}{cc}
\cos\beta & -\sin\beta \\
\sin \beta & \cos\beta \\
\end{array}
\right)
\label{eq:rotamatalphabeta}
\end{eqnarray}

\noindent
where $\alpha$, $\beta$ denote the rotation angles respectively in the
${\mathcal{CP}}_{even}$ and ${\mathcal{CP}}_{odd}$ sectors.\footnote{Hereafter, we will use the
shorthand notations, $s_x  \equiv \sin x $ and $c_x \equiv \cos x$, for all three angles $\alpha$, $\beta$, $\beta'$.}
Upon diagonalization of ${\mathcal{M}}_{{\mathcal{CP}}_{even}}^2$ 
one obtains two massive even-parity physical states  $h^0$ and $H^0$ defined by

\begin{eqnarray}
h^0 &=& ~~c_\alpha \, h + s_\alpha \, \xi^0  \\
H^0 &=& - s_\alpha \, h + c_\alpha \, \xi^0
\end{eqnarray}

\noindent
where $h$ and $\xi^0$ are the real parts of the $\phi^0$ and $\delta^0$ fields shifted by their VEV values,

\begin{eqnarray}
\phi^0=\frac{1}{\sqrt{2}} (v_d+h+iZ_1)~~~~{\rm and}~~~~\delta^0=\frac{1}{\sqrt{2}} (v_t+\xi^0+iZ_2).
\label{eq:shiftedfields}
\end{eqnarray}

\noindent

The masses are given by the eigenvalues of  ${\mathcal{M}}^2_{{\mathcal{CP}}_{even}}$
as follows,
 
\begin{eqnarray}
m_{h^0}^2&=&\frac{1}{2}[A+C-\sqrt{(A-C)^2+4B^2}] \label{eq:mh0}\\
m_{H^0}^2&=&\frac{1}{2}[A+C+\sqrt{(A-C)^2+4B^2}] \label{eq:mH0}
\end{eqnarray}

\noindent
so that $m_{H^0}>m_{h^0}$. Note that the lighter state $h^0$ is not necessarily the lightest of the Higgs sector
(see section \ref{sec:higgsbounds}). 

On the other hand, ${\mathcal{M}}^2_{{\mathcal{CP}}_{odd}}$ leads to one massive
physical state $A^0$ and one massless Goldstone boson $G^0$ defined by: 

\begin{eqnarray}
A^0 &=& - s_\beta \, Z_1 + c_\beta \, Z_2  \\
G^0 &=&  ~~c_\beta \,  Z_1 + s_\beta \, Z_2
\end{eqnarray}

\noindent
with masses
\begin{eqnarray}
  m_{A}^2 &=& \frac{\mu(v_d^2+4v_t^2)}{\sqrt{2}v_t}  
\label{eq:mA0}
\end{eqnarray}

\noindent
Knowing the above eigenmasses one can then determine the rotation angles $\alpha$ and $\beta$,
 which control the field content of the physical states, from the following diagonalization conditions: 

\begin{itemize}

\item[1.]\underline{${\mathcal{CP}}_{even}$}:
\begin{eqnarray}
  C &=& s_\alpha^2m_{h^0}^2+c_\alpha^2m_{H^0}^2 
\label{eq:sa1}\\
  B &=& \frac{\sin2\alpha}{2}(m_{h^0}^2-m_{H^0}^2) \label{eq:sa2}\\
  A &=& c_\alpha^2m_{h^0}^2+s_\alpha^2m_{H^0}^2  \label{eq:sa3}
\end{eqnarray}
\item[2.]\underline{${\mathcal{CP}}_{odd}$}:
\begin{eqnarray}
2\sqrt{2}\mu{v_t}&=& s_\beta^2m_{A}^2 \label{eq:sb1}\\
\sqrt{2}\mu{v_d} &=&\frac{\sin2\beta}{2}m_{A}^2 \label{eq:sb2}\\
\frac{\mu{v_d^2}}{\sqrt{2}v_t}&=& c_\beta^2m_{A}^2 \label{eq:sb3}
\end{eqnarray}
\end{itemize} 

\noindent
Of course Eq.~(\ref{eq:sa1}) should be equivalent to Eq.~(\ref{eq:sa3}) upon use of $s_\alpha^2 + c_\alpha^2=1$
and Eqs.~(\ref{eq:mh0}, \ref{eq:mH0}), and similarly for Eqs.~(\ref{eq:sb1})  and (\ref{eq:sb3}).
Furthermore, $s_\alpha, c_\alpha, s_\beta, c_\beta$ will all be determined up to a global sign. There is however
a difference between the two sectors. In the ${\mathcal{CP}}_{odd}$ sector $s_\beta$ and $c_\beta$ must have the same sign as can be seen from
Eq.(\ref{eq:sb2}) and the fact that $\mu >0$ 
(the latter being due to the absence of tachyonic $A^0$ state,Eq.(\ref{eq:sb1})). 
One then obtains unambiguously

\begin{equation}
  \tan\beta=\frac{2v_t}{v_d}~~~~{\rm and}~~~~\tan2\beta = \frac{4v_t{v_d}}{v_d^2-4v_t^2} \label{eq:tanb}
\end{equation}

\noindent
from Eqs.~(\ref{eq:sb1}, \ref{eq:sb3}), and

\begin{equation}
s_\beta = \epsilon_{\beta} \frac{2 v_t}{\sqrt{v_d^2 + 4 v_t^2}}~,~~~~
c_\beta = \epsilon_{\beta} \frac{v_d}{\sqrt{v_d^2 + 4 v_t^2}}
\end{equation}

\noindent
with a sign freedom $\epsilon_{\beta} = \pm 1$. 

In contrast, the relative sign between $s_\alpha$ and $c_\alpha$ in the ${\mathcal{CP}}_{even}$ sector depends on the values of $\mu$ as can be seen from
Eqs.(\ref{eq:sa2}, \ref{eq:ABC}). While they will have the same sign and $\tan \alpha > 0$  for most of the allowed 
$\mu$ and $\lambda_1, \lambda_4$ ranges, there will be a small but interesting domain of small $\mu$ 
values and  $\tan \alpha < 0$ which we discuss in detail in section \ref{sec:higgsbounds}. One obtains from
Eqs.~(\ref{eq:sa1} - \ref{eq:sa3})


\begin{eqnarray}
s_\alpha &=&    -\frac{\epsilon_{\alpha} {\epsilon}}{\sqrt{2}}(1 + \frac{(A - C)}{\sqrt{(A - C)^2 + 4 B^2}})^{1/2} 
\label{eq:sa}\\
c_\alpha &=&    \frac{\epsilon_{\alpha}}{\sqrt{2}}(1 - \frac{(A - C)}{\sqrt{ (A - C)^2 + 4 B^2  }})^{1/2}
\label{eq:ca}
\end{eqnarray}

\noindent
where $\epsilon_{\alpha}= \pm 1$ and ${\epsilon} \equiv {\rm sign}[B]$, and 

\begin{equation}
  \tan2\alpha = \frac{2 B}{ A - C}
\end{equation}


\noindent
Let us finally note that the
angles $\beta$ and $\beta^{'}$ are correlated since they depend exclusively on $v_d$ and $v_t$. For instance one always has

\begin{eqnarray}
\tan\beta&=&\sqrt{2}\tan{\beta^{'}} \label{eq:bbprime}
\end{eqnarray}

\noindent
as can be seen from Eqs.(\ref{eq:tanbp}, \ref{eq:tanb}).

\subsection*{Lagrangian parameters from physical masses and couplings:}

The full experimental determination of the DTHM would require not only evidence for the neutral and (doubly) charged
Higgs states but also the experimental determination of the masses and couplings of these states among themselves as 
well as to the gauge and matter sectors of the model. Crucial tests would then be driven by the predicted 
correlations among these measurable quantities. For instance one can easily express the Lagrangian parameters $\mu$ and the $\lambda$'s
in terms of the physical Higgs masses and the mixing angle $\alpha$ as well as the VEV's $v_d, v_t$, using equations
(\ref{eq:mA0}, \ref{eq:mHpm}, \ref{eq:mHpmpm}) and (\ref{eq:sa1} - \ref{eq:sa3}). One finds

\begin{eqnarray}
\lambda_1 &=& -\frac{2}{v_d^2+4v_t^2}.m_{A}^2+\frac{4}{v_d^2+2v_t^2}.m_{H^\pm}^2+\frac{\sin2\alpha}{2v_d{v_t}}.(m_{h^0}^2-m_{H^0}^2) \label{eq:lambda1}\\
\lambda_2&=&\frac{1}{v_t^2}\{\frac{s_{\alpha}^2m_{h^0}^2+c_{\alpha}^2m_{H^0}^2}{2}+\frac{1}{2}.\frac{v_d^2}{v_d^2+4v_t^2}.m_{A}^2-\frac{2v_d^2}{v_d^2+2v_t^2}.m_{H^\pm}^2+m_{H^{\pm\pm}}^2\}\\
\lambda_3 &=& \frac{1}{v_t^2}\{\frac{-v_d^2}{v_d^2+4v_t^2}.m_{A}^2+\frac{2v_d^2}{v_d^2+2v_t^2}.m_{H^\pm}^2-m_{H^{\pm\pm}}^2\}\\
\lambda_4 &=& \frac{4}{v_d^2+4v_t^2}.m_{A}^2-\frac{4}{v_d^2+2v_t^2}.m_{H^\pm}^2 \\
\lambda &=& \frac{2}{v_d^2}\{c_{\alpha}^2m_{h^0}^2+s_{\alpha}^2m_{H^0}^2\} \\
\mu     &=& \frac{\sqrt{2} v_t}{v_d^2+4v_t^2}.m_{A}^2 \label{eq:mu}
\end{eqnarray}

\noindent
The remaining two Lagrangian parameters $m_H^2$ and $M_\Delta^2$ are then related to the physical
parameters through the EWSB conditions Eqs.~(\ref{eq:ewsb1}, \ref{eq:ewsb2}). To complete the determination
in terms of physical quantities one should further extract the mixing angle $\alpha$ from the measurement
of some couplings (see also Appendix C) and $v_d$ and $v_t$ from the W (or Z) masses. Using equations 
(\ref{eq:mZ}, \ref{eq:mW}, \ref{eq:tanb}) one finds 

\begin{eqnarray}
v_d^2=\frac{1}{(1+\frac{1}{2} \tan^2 \beta)} \frac{\sin^2\theta_W \; M_W^2}{\pi \alpha_{QED} }
~,~~~~v_t^2=\frac{ \tan^2 \!\beta}{(1+ \frac{1}{2} \tan^2 \beta)} \frac{\sin^2\theta_W \; M_W^2}{4 \pi \alpha_{QED} }
\label{eq:vdvt1}
\end{eqnarray}

\noindent
or

\begin{eqnarray}
v_d^2=\frac{1}{(1+ \tan^2 \beta)} \frac{\sin 2\theta_W \; M_Z^2}{2 \pi \alpha_{QED} }
~,~~~~v_t^2=\frac{ \tan^2 \!\beta}{(1+  \tan^2 \beta)} \frac{\sin 2 \theta_W \; M_Z^2}{8 \pi \alpha_{QED} }
\label{eq:vdvt2}
\end{eqnarray}

\noindent
or

\begin{eqnarray}
v_d^2 = \frac{\sin^2\theta_W}{\pi \alpha_{QED} }(2 M_W^2 - \cos^2\theta_W M_Z^2)  
~,~~~~v_t^2 =  \frac{\sin^2\theta_W}{2 \pi \alpha_{QED} }( \cos^2\theta_W M_Z^2 -M_W^2 )  \label{eq:vdvt3}
\end{eqnarray}

\noindent
Using any of the above equations to substitute for $v_d, v_t$ in Eqs.~(\ref{eq:lambda1}, \ref{eq:mu}) 
allows us to obtain the Lagrangian parameters solely in terms of experimentally measurable quantities. 
Although Eqs.~(\ref{eq:vdvt1})-(\ref{eq:vdvt3}) are theoretically trivially equivalent, they involve different sets of
experimental observables and can thus lead to non equivalent reconstruction strategies depending on the achieved
accuracies in the measurement of these observables. Similarly, trading $\tan\beta$ for $\tan\beta^{'}$ through
Eq.~(\ref{eq:bbprime}) can be useful, depending on which of the two quantities is experimentally better determined
through some coupling measurements. We should also note that Eqs.~(\ref{eq:lambda1})-(\ref{eq:vdvt3})
not only allow to reconstruct the Lagrangian parameters from the measurable Higgs masses, $\alpha$, $\beta$, $M_Z$
and/or $M_W$, but can also serve  as consistency checks among observable quantities for the model when the 
$\lambda$'s and $\mu$ are determined independently through the measurement of couplings in the purely Higgs 
sector  (see also Appendix C). Finally, as can be seen from Eq.(\ref{eq:vdvt3}), the magnitude of $v_t$
entails the deviation of the $\rho$ parameter from its SM tree-level value, a point we will discuss
further in the following section. 




\section{Miscellaneous constraints}

\subsection{Constraints from electroweak precision measurements}

In the Standard Model the custodial symmetry ensures that the $\rho$ parameter,
$\rho\equiv \frac{M_W^2}{M_Z^2\cos^2\theta_W}$ is equal to $1$ at tree level. In the DTHM one obtains the $Z$ and $W$ gauge boson masses
readily from Eq.(\ref{eq:VEVs}) and the kinetic terms in Eq.(\ref{eq:DTHM}) as

\begin{eqnarray}
M_Z^2&=&\frac{(g^2+{g'}^2)(v_d^2+4v_t^2)}{4}   \label{eq:mZ}
     =\frac{g^2(v_d^2+4v_t^2)}{4\cos^2\theta_W} \\
M_W^2&=&\frac{g^2(v_d^2+2v_t^2)}{4}  \label{eq:mW}
\end{eqnarray}

\noindent
whence the modified form of the $\rho$ parameter:
\begin{eqnarray}
\rho&=&\frac{v_d^2+2v_t^2}{v_d^2+4v_t^2} \neq 1
\end{eqnarray}

\noindent
and actually $\rho < 1 $ at the tree-level.  
Since we are interested in the limit $v_t \ll v_d$ we rewrite

\begin{equation}
\rho \simeq 1 - 2 \frac{v_t^2}{v_d^2} = 1 + \delta \rho
\end{equation}

\noindent
with $\delta \rho = - 2 \frac{v_t^2}{v_d^2}  < 0$ and $\sqrt{v_d^2 + 2 v_t^2} =246$ GeV. 
Thus the model will remain viable as far as the 
experimentally driven values of $\delta \rho$
are compatible with a negative number. 
The implication for the DTHM has already been studied in the literature
\cite{Perez:2008ha}. Here we only discuss briefly this point taking into account the latest updates of the electroweak
observables fits as reported by the PDG \cite{Amsler:2008zzb}. 
One should compare the theoretical value with the 
experimental value after having subtracted from the latter all 
the known standard model contributions to the 
$\rho$ parameter. The quoted number after this
subtraction, $\rho_0 = 1.0008^{+0.0017}_{ -0.0007}$ 
obtained from a global fit including the direct
search limits on the standard Higgs boson, is not compatible with a 
negative $\delta \rho$ and would exclude
the DTHM. However at the 2 $\sigma$ level, one obtains 
$\rho_0 = 1.0004^{+0.0029}_{-0.0011}$ \cite{Amsler:2008zzb}, which is 
again compatible with $\delta \rho <0$. Moreover, 
relaxing the Higgs direct limit leads to 
$\rho_0 = 1.0008^{+0.0017}_{-0.0010}$, 
again compatible with $\delta \rho <0$. From the last two numbers one gets
an upper bound on $v_t$ of order $2.5-4.6$ GeV. 
In the present study we will thus be contented  by the conservative assumption that an upper bound of $2.5$ GeV 
guarantees consistency with the experimental constraints.  
We should note, though,  that the tree-level DTHM value of 
$\delta \rho$ being of order $10^{-4}$, it is legitimate to ask
about the effects of radiative corrections to this quantity 
within the DTHM. As far as we know, radiative corrections to 
$\delta \rho$ are not available in the literature in the 
case of  $Y_{\Delta}=2$ that concerns us here, 
while several studies have been dedicated to
this question in the framework of a $Y_{\Delta}=0$ triplet Higgs 
 \cite{Chen:2006pb,Blank:1997qa,Chankowski:2006hs,Chen:2008jg},  
In \cite{Chen:2006pb}, it has been shown that the tree level bound on the
triplet vev could be pushed to higher values by one-loop radiative corrections.
Whether this happens also in our case is still to be investigated and deserves a study that is
out of the scope of the present paper, including  for that matter all other LEP/SLC SM observables. 
\subsection{Absence of tachyonic modes}
\label{sec:tachyon}

From Eq.~(\ref{eq:mA0}), the requirement that $m_{A}^2$ should be positive implies  
$\mu v_t > 0$. The same positivity requirement in the singly charged and doubly
charged sectors, Eqs.~(\ref{eq:mHpm}, \ref{eq:mHpmpm}),  together with our phase convention $v_t >0$ discussed 
in section \ref{sec:themodel}, lead  to the following bounds on $\mu$:

\begin{eqnarray}
\mu&>&0  \label{eq:tachyonodd} \\
\mu &>& \frac{\lambda_4 v_t}{2\sqrt{2}} \label{eq:tachyoncharge1}\\
\mu &>& \frac{\lambda_4 v_t}{\sqrt{2}} + \sqrt{2} \frac{ \lambda_3 v_t^3}{v_d^2} \label{eq:tachyoncharge2}
\end{eqnarray}


\noindent
The tachyonless condition in the ${\mathcal{CP}}_{even}$ sector, Eqs.~(\ref{eq:mh0}, \ref{eq:mH0}),  
is somewhat more involved and reads 
\begin{eqnarray}
&\sqrt{2} \mu v_d^2 + \lambda  v_d^2 v_t + 4 (\lambda_2 + \lambda_3) v_t^3  > 0 & \label{eq:posh0H0-0}\\
&-8 \mu^2 v_t + \sqrt{2} \mu (\lambda v_d^2 + 8 (\lambda_1 + \lambda_4) v_t^2)
     + 4 ( \lambda (\lambda_2 + \lambda_3) - (\lambda_1 + \lambda_4)^2) v_t^3  \label{eq:posh0H0-1}
    > 0 &
\end{eqnarray}
The first of these two equations is actually always satisfied as a consequence of Eq.~(\ref{eq:tachyonodd}) and 
the boundedness from below conditions for the potential (see section \ref{sec:boundedness} and Eq.~(\ref{eq:BFBgen1})). 
The second equation, quadratic in $\mu$, will lead to new constraints on $\mu$ in the form of an allowed range
 
\begin{equation}
\mu_{-} < \mu < \mu_{+} \label{eq:posh0H0}
\end{equation}
 
\noindent
The full expressions of $\mu_{\pm}$ and a discussion of their real-valuedness are given in appendix A. 
Here we discuss their behavior in the regime $v_t \ll v_d$. In this case one finds a vanishingly small 
$\mu_{-}$ given by
\begin{equation}
\displaystyle\mu_{-} = 
( (\lambda_1 + \lambda_4)^2 -\lambda (\lambda_2 + \lambda_3) ) \, \frac{2 \sqrt{2}}{\lambda} \frac{v_t^3}{v_d^2}  
+ {\cal O}(v_t^4) \label{eq:muminusapprox}
\end{equation}

\noindent
and a large $\mu_{+}$ given by

\begin{equation} 
\displaystyle \mu_{+} =  \frac{\lambda}{4 \sqrt{2}} 
\frac{v_d^2}{v_t} + \sqrt{2} (\lambda_1 + \lambda_4) v_t +{\cal O}(v_t^2).
\label{eq:muplusapprox}
\end{equation}

\noindent
Depending on the signs and magnitudes of the $\lambda$'s, one of the lower bounds (\ref{eq:tachyonodd} -
\ref{eq:tachyoncharge2}) or $\mu_{-}$ will overwhelm the others. Moreover, these no-tachyon bounds will have 
eventually to be amended by taking into account the existing experimental exclusion limits. This is straightforward for
$A^0, H^\pm$ and $H^{\pm \pm}$. We thus define for later reference

\begin{eqnarray}
\mu_{\rm min} &=& \max \; \left[\begin{matrix} \displaystyle  \frac{\sqrt{2} \, v_t}{v_d^2 + 4 v_t^2} \, ({m_A^2})_{\rm exp},
\cr \displaystyle \frac{\lambda_4 v_t}{2\sqrt{2}} 
+ \frac{\sqrt{2} \, v_t }{v_d^2 + 2 v_t^2} \, ({m_{H^\pm}^2})_{\rm exp}, \cr \displaystyle
\frac{\lambda_4 v_t}{\sqrt{2}} + \sqrt{2} \frac{ \lambda_3 v_t^3}{v_d^2} 
+ \frac{\sqrt{2} \, v_t }{v_d^2} \, (m_{H^{\pm \pm}}^2)_{\rm exp}\end{matrix}\right] \label{eq:mumin}
\end{eqnarray}

\noindent
where $({m_{A}})_{\rm exp}, ({m_{H^\pm}})_{\rm exp}, ({m_{H^{\pm \pm}}})_{\rm exp}$ denote some experimental 
lower exclusion limits for the Higgs masses. Eqs.(\ref{eq:tachyonodd} - \ref{eq:tachyoncharge2})
are then replaced by 

\begin{equation}
\mu > \mu_{\rm min} \label{eq:mumumin}
\end{equation}

\noindent 
in order for the masses to satisfy these exclusion limits. 
Similar modifications on $\mu_{\pm}$ taking into account experimental exclusion
limits in the ${\mathcal{CP}}_{even}$ sector are more involved and will be differed to section  
\ref{sec:higgsbounds} after having
established the theoretical upper (lower) bounds on the $h^0$ ($H^0$) masses.
Furthermore, the upper bound $\mu_{+}$ will be instrumental in determining
the maximally allowed values of the six Higgs masses $m_{H^0}, m_{A}, m_{H^\pm}, m_{H^{\pm \pm}}$ as we will see
in section \ref{sec:higgsbounds}.  

\subsection{The vacuum structure}
\label{sec:vacuum} 


Obviously, violation of any of the constraints discussed in the previous subsection is a signal that the would-be 
electroweak vacuum is not a minimum (but rather a saddle-point or a local maximum) for the given set of values 
$\lambda, \lambda_i, v_d, v_t$ when $\mu$ is either very small or very large. However, since Eqs.(\ref{eq:ewsb1}, \ref{eq:ewsb2})
are non-linear in $v_d, v_t$, it could still be possible to find a different set of values $v_d', v_t'$, for the
same input values of $m_H^2, M_{\Delta}^2$, where the true electroweak minimum is obtained at a lower point of the 
potential than the previous one. More generally, and depending on the values of the parameters
of the potential, one expects on top of the electroweak minimum a rich structure of extrema that can affect
the interpretation and viability of this minimum and thus possibly lead to additional constraints
on these parameters. A complete study of such extrema can be very involved since the potential
depends on ten independent real fields. Here we only provide a partial
qualitative discussion. 

Upon use of Eqs.(\ref{eq:VEVs}, \ref{eq:ewsb1}, \ref{eq:ewsb2}) in Eq.(\ref{eq:Vpot}) one readily finds that the value
of the potential at the electroweak minimum, $\langle V \rangle_{\rm EWSB}$, is given by:

\begin{equation}
\langle V \rangle_{\rm EWSB} = -\frac{1}{16}(\lambda v_d^4 + 
4 (\lambda_2 + \lambda_3) v_t^4 + 4 v_d^2 v_t ((\lambda_1 + \lambda_4) v_t- \sqrt{2} \mu  ))
\end{equation}

\noindent
Since the potential vanishes at the gauge invariant origin of the field space, $V_{H=0, \Delta =0}=0$, then
spontaneous electroweak symmetry breaking would be energetically disfavored if  
$\langle V \rangle_{\rm EWSB} >0$.\footnote{We should, however, keep in mind the possibility
that a long-lived metastable vacuum could still be physically acceptable, even when $\langle V \rangle_{\rm EWSB} >0$,
thus altering our constraints; these issues are not addressed further in the present paper.}
One can thus require as a first approximation  the naive bound on $\mu$ 
\begin{equation}
\displaystyle \mu < \mu_{\rm max} \equiv 
\frac{\lambda}{4 \sqrt{2}} \frac{v_d^2}{v_t} + (\lambda_1 + \lambda_4) \frac{v_t}{\sqrt{2}} +{\cal O}(v_t^2) 
\label{eq:mumax}
\end{equation}

\noindent 
to ensure $V_{\rm EWSB} < 0$. As can be seen from Eq.(\ref{eq:muplusapprox}) one has either 
$\mu_{\rm max} < \mu_{+}$ or $\mu_{\rm max} > \mu_{+}$ depending on the sign of $\lambda_1 + \lambda_4$. But for
all practical purposes  $\mu_{\rm max} \simeq \mu_{+}$ in the regime $v_t/v_d \ll 1$,  so that the proviso stated above
concerning the relevance of the tachyonless conditions is weakened for the upper bound $\mu_{+}$ which can be replaced
by  $\mu_{\rm max}$. There is, however, yet another critical value of $\mu$. As mentioned at the end of section 
\ref{sec:thehiggspot}, $M_\Delta^2$ and $-m_H^2$ can be both positive for sufficiently large values of $\mu$ thus
making the gauge invariant point $H=0, \Delta=0$ a local minimum. This happens when $\mu > \mu_H$ where

\begin{equation}
\mu_H = \frac{\lambda}{4 \sqrt{2}} v_d^2 + (\lambda_1 + \lambda_4) \frac{v_t}{2\sqrt{2}}.  
\label{eq:muH}                                          
\end{equation}

\noindent
If $\lambda_1 + \lambda_4 > 0$ then $\mu_H < \mu_{\rm max} < \mu_{+}$. To delineate 
some consistency constraints in this case, it would be necessary 
to look more closely at the decay rate from a metastable gauge invariant vacuum to the EWSB vacuum,
 if $\mu_H < \mu <\mu_{\rm max}$, and vice-versa, from a metastable EWSB vacuum to the gauge invariant vacuum when 
$\mu_{\rm max} < \mu < \mu_{+}$. Fortunately, however, these 
configurations altogether are already excluded if we take into account the experimental mass limits on the Standard Model 
Higgs. Indeed, as will be shown in  sections \ref{sec:higgsbounds} and \ref{sec:higgspheno}, the lightest 
${\mathcal{CP}}_{even}$ Higgs state $h^0$ becomes purely standard model-like for such large values of $\mu$, irrespective of the
values of the couplings $\lambda$, $\lambda_i$,  while $m_{h^0}$ becomes very small for these values
(e.g. $m_{h^0} = \sqrt{3 (\lambda_1 + \lambda_4)}  v_t$ for $\mu= \mu_H$) and thus experimentally excluded. 

Nonetheless, the structure of the potential Eq.(\ref{eq:Vpot}) is sufficiently rich to provide dangerous extrema
configurations which are not excluded by the above mentioned experimental limits. 
We exhibit  here, without much details, one example among a manifold of possibilities. There is an extremum
in the field space direction defined by  
$\Re \phi^0 = \Re \phi^+ \equiv \frac{v_d^c}{\sqrt{2}}$ and 
$\Re \delta^0  =  - \Re \delta^{++} \equiv \frac{v_t^c}{\sqrt{2}}$, and all other fields put to zero.

This requires

\begin{eqnarray}
\mu &=& -\frac{\lambda_4 v_t^c}{\sqrt{2}} \label{eq:cb1}\\
 m_H^2 &=& \frac{1}{2}(\lambda {v_d^c}^{\,2} + (2 \lambda_1 - \lambda_4) {v_t^c}^{\,2}) \label{eq:cb2}\\
 M_{\Delta}^2 &=& - \lambda_1 {v_d^c}^{\,2} - (2 \lambda_2 + \lambda_3) {v_t^c}^{\,2} \label{eq:cb3}
\end{eqnarray}

\noindent
Note that this direction, and thus the corresponding extremum, breaks spontaneously charge
conservation. We will refer to this extremum as charge breaking (CB). Furthermore, in contrast with the EWSB point, 
Eqs.(\ref{eq:ewsb1}, \ref{eq:ewsb2}), here $\mu$ is not a free parameter. We can then seek for  a region in parameter
space where this CB extremum coexists with an EWSB minimum, and check what happens at the gauge invariant extremum point
as well. Requiring Eqs.(\ref{eq:ewsb1}, \ref{eq:ewsb2}, \ref{eq:cb1} - \ref{eq:cb3}) to be simultaneously satisfied
leads to correlations among $v_d, v_t, v_d^c, v_t^c$. These lead in turn to constraints  on the  $\lambda, \lambda_i$ 
parameter space in order for all these vev's to be real valued (modulo gauge transformations), together with the 
immediate constraint $\lambda_4 v_t^c < 0$ originating from Eq.(\ref{eq:cb1}) and $\mu >0$.\footnote{Working in the 
regime $v_t \ll v_d, |v_t^c|, |v_d^c|$ and keeping only terms ${\cal O} (v_t)$,
the constraint in the $(\lambda, \lambda_i)$ space satisfying all these requirements is found to be
$\lambda_1 < \frac{1}{4}(\lambda_4 - \sqrt{8 \lambda (2 \lambda_2 + \lambda_3) + 17 \lambda_4^2})$. Note
that $\lambda >0$ and $2 \lambda_2 + \lambda_3 > 0$ for a bounded from below potential (see section 
\ref{sec:boundedness}).} The ensuing correlations allow to write $m_{h^0}^2, \langle V \rangle_{\rm EWSB}$
and $\langle V \rangle_{\rm CB}$ (the value of the potential at the CB extremum) in the following form:


\begin{eqnarray}
m_{h^0}^2 &=& (\lambda (2 \lambda_2 + \lambda_3) + 2 \lambda_4^2 + \lambda_1 (\lambda_4 - 2 \lambda_1)) 
\frac{v_t v_t^c}{\lambda_4} + {\cal O} (v_t^2) \nonumber \\ 
&=& 2 m_H^2  + {\cal O} (v_t^2) \label{eq:cb4}\\
\langle V \rangle_{\rm EWSB} &=&-(\lambda (2 \lambda_2 + \lambda_3) + 2 \lambda_4^2 + 
\lambda_1 (\lambda_4 - 2 \lambda_1)) \times \nonumber \\
       && ~~~~~~~~(\lambda (2 \lambda_2 + \lambda_3) + \lambda_1 (-2 \lambda_1 + \lambda_4)) 
 \frac{v_t^2 {v_t^c}^{\,2}}{4 \lambda \lambda_4^2}
 + {\cal O} (v_t^3) \label{eq:cb5} \\
\langle V \rangle_{\rm CB} &=& (4 \lambda_1^2 - 2 \lambda (2 \lambda_2 + \lambda_3) 
- \lambda_4^2) \frac{{v_t^c}^{\, 4}}{4 \lambda}  +  {\cal O} (v_t^2) \label{eq:cb6}
\end{eqnarray}



\noindent
Various interesting conclusions can be drawn from the above equations.  As can be seen from
Eq.(\ref{eq:cb4}), a physical $h^0$, i.e. $m_{h^0}^2 > 0$, implies a positive $m_H^2$ and thus an {\sl unstable}
gauge invariant point at the origin of the fields $(H=0, \Delta =0)$. Furthermore, in the consistent 
$(\lambda, \lambda_i)$ domain 
(given in footnote 2) $m_{h^0}^2$ is indeed positive and, furthermore, one finds from Eq.(\ref{eq:cb5}) that 
$\langle V \rangle_{\rm EWSB} <0$. The EWSB vacuum is thus energetically favored over the gauge symmetry 
preserving one which lies at $V=0$. It then remains to compare the EWSB point with the CB point.
Close inspection of Eq.(\ref{eq:cb6}) shows that $\langle V \rangle_{\rm CB} > 0$ in all the $(\lambda, \lambda_i)$ 
domain given in footnote 2, if and only if $\lambda_4 < 0$, in which case the EWSB is energetically favored over the CB. 
However, if $\lambda_4 >0$ (and thus $v_t^c <0$), there are regions in the  $(\lambda, \lambda_i)$ 
consistent domain where $\langle V \rangle_{\rm CB} < 0$, provided that 
$4 \lambda_4^2 < \lambda (2 \lambda_2 + \lambda_3)$. 
Moreover, the potential at this CB point becomes
much deeper than at the EWSB point since we are in the regime $v_t \ll |v_t^c|$.
This is a dangerous configuration since it makes the EWSB vacuum potentially very short-lived due to tunneling 
effects \cite{Coleman:1977py, Callan:1977pt}. We stress here that this EWSB point is a true local
minimum in this configuration, i.e. there are no tachyonic Higgs states which could have signaled its non-relevance
beforehand. [This is easily seen from the fact that  $h^0$ is non-tachyonic and is the lightest Higgs
state when $\mu \sim |v_t^c| \gg v_t$; see also section \ref{sec:higgsbounds}.] 
Even more so, the potential is bounded from below, as can be shown by comparing the 
corresponding $(\lambda, \lambda_1)$ domain with the boundedness from below constraints that we will derive
 in the following section. We have thus exhibited 
an example of a configuration where $\mu$ can be very large, consistent with the experimental $h^0$ mass limit,
and {\sl a fortiori} with all the non-tachyon constraints, corresponding locally to an acceptable EWSB  
vacuum, but still non-viable due to the existence of lower (charge breaking) points akin to what happens
in two-Higgs-doublet models (see for instance \cite{Sher:1988mj}).

We end this section with a general comment concerning the neutrino mass see-saw mechanism.
The common lore is to assume a GUT origin for $\mu$ and
$M_{\Delta}$, and taking  $\mu \sim M_{\Delta} \sim {\cal O}(M_{\rm GUT})$ leads through
Eq.(\ref{eq:ewsb1}) naturally to a tiny $v_t$. However, as noted in the introduction we do not
commit in the present study to specific high energy physics scenarios, so that  $M_{\Delta}$ and/or $\mu$ could be 
smaller than a hypothetical GUT scale. It is then interesting to note that even in this case a kind of see-saw 
is actually still at work model-independently due to the dynamics of the potential. This is simply due to the form
of the $\mu$ upper bound $\mu_{+}$, Eq.(\ref{eq:muplusapprox}): the larger is $\mu$ the smaller should be $v_t$ in
order to avoid a tachyonic $h^0$. For instance,  taking $\lambda \simeq 0.5$  and $\mu_{+} \simeq 2 \times 10^{12}$GeV 
leads to $v_t \simeq 1$eV and $M_{\Delta} \simeq 10^{13}$GeV.






\setcounter{equation}{0}
\section{Boundedness of the potential}
 \label{sec:boundedness}
A necessary condition for the stability of the vacuum comes from requiring that the potential given
in Eq.~(\ref{eq:Vpot}) be bounded from below when the scalar fields become large in any direction
of the field space.  The constraints ensuring boundedness 
from below (BFB) of the DTHM potential have been  studied in the literature so far only partially
(see e.g. \cite{Dey:2008jm}), and at the tree-level.  
It would thus be somewhat premature to invoke possible quantum modifications of these constraints before 
fully settling first the tree-level issue. This section is devoted to this issue and aims at
deriving, at the tree-level, the complete {\sl necessary  and sufficient} BFB conditions valid for 
{\sl all} directions in field space.\footnote{We will thus not address in this paper the 
possibility that loop corrections could lift the potential in some otherwise unbounded from below directions, nor the 
issues related to metastability of the vacuum which could relax some of the constraints. See also
section \ref{sec:vacuum}.}

Obviously, at large field values the potential Eq.~(\ref{eq:Vpot})
is generically dominated by the part containing the terms that are quartic in the fields, 

\begin{equation}
V^{(4)}(H, \Delta) =\frac{\lambda}{4}(H^\dagger{H})^2
+\lambda_1(H^\dagger{H})Tr(\Delta^{\dagger}{\Delta})+\lambda_2(Tr\Delta^{\dagger}{\Delta})^2
+\lambda_3Tr(\Delta^{\dagger}{\Delta})^2
+\lambda_4{H^\dagger\Delta\Delta^{\dagger}H} 
\label{eq:Vquartic}
\end{equation}

\noindent 
The study of $V^{(4)}(H, \Delta)$ will thus be sufficient to obtain the main constraints. 
To obtain BFB conditions it is common in the 
literature to pick up specific field directions or to put some of the couplings
to zero. Consider for instance the two following cases:

\begin{enumerate}
\item in the absence of any coupling
between doublet and triplet Higgs bosons, i.e. $\lambda_1 = \lambda_4=0$,  it is obvious that 

\begin{equation}
\lambda >0 \;\;{\rm \&}\;\; \lambda_2 >0 \;\;{\rm \&}\;\; \lambda_3 >0 
\end{equation}

\noindent
will ensure that the potential is bounded from below. 

\item if one picks up the field space directions where only  
the electrically neutral components are non vanishing,
one finds
\begin{equation}
V^{(4)}_0
=\frac{\lambda}{4} |\phi^0|^4 +(\lambda_2 +\lambda_3) |\delta^0|^4
           + (\lambda_1 + \lambda_4) |\phi^0|^2 |\delta^0|^2 
\label{eq:VquarticNeut}
\end{equation}

\noindent
In order for the potential to be bounded from below in this sub-space, $V^{(4)}_0$ should be positive for
any values of $|\phi^0|$ and $|\delta^0|$ including when one or the other is vanishing. The latter
cases imply the necessary conditions  $\lambda >0$ and  $\lambda_2+\lambda_3 > 0$. It is then possible
to rewrite Eq.(\ref{eq:VquarticNeut}) in the form

\begin{eqnarray} V^{(4)}_0=
 [\frac{\sqrt{\lambda}}{2} |\phi^0|^2 -\sqrt{\lambda_2+\lambda_3} |\delta^0|^2]^2
 +(\lambda_1+\lambda_4+\sqrt{\lambda(\lambda_2+\lambda_3)})|\phi^0|^2 |\delta^0|^2 \label{eq:Pot2dir8}
\end{eqnarray}

\noindent
Since the first term is non-negative and vanishes in the
direction \newline $|\phi^0|^2/|\delta^0|^2=2\sqrt{(\lambda_2+\lambda_3)/\lambda}$,
then the necessary and sufficient conditions for the BFB of the potential in {\sl this} direction are
\begin{eqnarray}
  \lambda &>& 0\nonumber \\
  \lambda_2+\lambda_3 &>& 0 \label{eq:BFB2dir8} \\
   \lambda_1+\lambda_4+\sqrt{\lambda(\lambda_2+\lambda_3)} &>& 0\nonumber
\end{eqnarray}
\end{enumerate}

\noindent
As it will become clear later on in this section, the conditions in case 1 are sufficient but not
necessary, even for this special case. Furthermore, while the conditions in case 2 are necessary and sufficient 
for the corresponding direction, they obviously remain necessary for the general potential but it is 
{\sl a priori} not clear whether  they can be sufficient. By looking at other special directions in $2$-field and 
$3$-field directions  we will show that they are generally not sufficient. Before doing so, let us first point out a 
more convenient method to obtain positivity constraints like Eq.(\ref{eq:BFB2dir8}) directly from 
Eq.(\ref{eq:VquarticNeut}) rather than writing it first in the form of Eq.(\ref{eq:Pot2dir8}).  
The potential Eq.(\ref{eq:VquarticNeut}) can be cast in the form

\begin{equation}
 V(\chi) = a + b\chi^2+c \chi^4 \label{eq:Vtype}
\end{equation}

\noindent
by the change of variable $\chi= |\phi^0|/|\delta^0|$. Since $\chi$ is by definition real valued and the moduli 
$|\phi^0|$ and $|\delta^0|$ can have any value, then the problem of finding the necessary and  sufficient BFB
conditions for Eq.(\ref{eq:VquarticNeut}) is equivalent to finding the conditions on $a, b, c$ such that
$V(\chi) > 0$ for any $\chi \in [0, \infty[$. Since $V(\chi)$ has no linear or cubic terms in $\chi$ it is easy
to find these conditions  by studying $V(\chi)$ as a bi-quadratic function:

\begin{eqnarray}
 a&>&0  \nonumber \\
c&>&0 \label{eq:poscond} \\
b+2 \sqrt{a c} &>& 0 \nonumber
\end{eqnarray}

\noindent
Applied to Eq.(\ref{eq:VquarticNeut}) these conditions reproduce immediately Eq.~(\ref{eq:BFB2dir8}). We can now
easily study other field directions. For instance, the direction where only $\delta^{++}$ and $\phi^0$ are 
non vanishing yields

\begin{equation}
V=  (\lambda_2 + \lambda_3)\,|\delta^{++}|^4 + \lambda_1 \,|\delta^{++}|^2\,|\phi^0|^2 + 
    \frac{\lambda}{4}\,|\phi^0|^4 
\end{equation}

\noindent
for which the BFB constraints are readily obtained from Eq.(\ref{eq:poscond}) as

\begin{equation}
\lambda > 0 \;\;{\rm \&}\;\; \lambda_2 + \lambda_3 >0 \;\;{\rm \&}\;\; \lambda_1 + \sqrt{\lambda  (\lambda_2 + \lambda_3)} >0 .
\end{equation}

\noindent
Similarly, if we consider the field direction with non-vanishing $\delta^+$ and $\phi^+$,
then

\begin{equation}
V= (\lambda_2 + \frac{\lambda_3}{2})\,|\delta^+|^4 + 
      (\lambda_1 + \frac{\lambda_4}{2} )\,|\delta^+|^2\,|\phi^+|^2 + \frac{\lambda}{4}\,|\phi^+|^4
\end{equation}

\noindent
and the corresponding BFB conditions read

\begin{equation}
 \lambda > 0 \;\;{\rm \&}\;\; \lambda_2 + \frac{\lambda_3}{2} >0 \;\;{\rm \&}\;\;
\lambda_1 + \frac{\lambda_4}{2} + \sqrt{\lambda (\lambda_2 + \frac{\lambda_3}{2})} >0.
\end{equation}

\noindent
It is then obvious that these two sets of conditions are neither equivalent nor contained in the conditions
of Eq.(\ref{eq:BFB2dir8}). This shows that the BFB conditions derived only from the neutral direction 
Eq.(\ref{eq:VquarticNeut}) are neither necessary nor sufficient to ensure boundedness from below of the full potential
Eq.(\ref{eq:Vpot}). In Appendix B we have listed the potentials for all the field directions with only 
two non-vanishing fields together with the corresponding BFB conditions. Adding these conditions we come closer to
the real sufficient and necessary conditions. But one can get more conditions by going now to field directions
where $3$ fields are non-vanishing.  We give the exhaustive list of all these $3$-field directions
potentials in Appendix B. In these more complicated configurations,  an iteration of the method described above 
allowed  to treat all of them, although the results become somewhat complicated and not so telling.
For instance the $3$-field direction with non-vanishing $\phi^0, \phi^+, \delta^+$, see Eq.(\ref{eq:threedir9})
yields some of the simplest  BFB conditions

\begin{eqnarray}
   \lambda>0\land 2 \lambda_2+\lambda_3>0\land \sqrt{\lambda (4
    \lambda_2+2 \lambda_3)}+2 \lambda_1+\lambda_4>0\land \nonumber \\
\Bigl(2 \lambda (2
    \lambda_2+\lambda_3)>(2 \lambda_1+\lambda_4)^2\lor 2
    \lambda_1+\lambda_4>0\Bigr)
\end{eqnarray}

\noindent 
where $\land, \lor$ stand respectively for the logical AND, OR. These conditions are obtained by first defining
the reduced variables $\chi_1 = |\phi^+|/|\phi^0|, \chi_2=|\delta^+|/|\phi^0|$, and then using iteratively
the constraints Eqs.~(\ref{eq:poscond}). By the same method we could obtain even more complicated BFB conditions
given in Eqs.~(\ref{eq:BFB3dir1} - \ref{eq:BFB3dir10}).  Analyzing them numerically we confirm that 
Eqs.~(\ref{eq:BFB2dir8}) are far from being the full story.
However, and despite their
apparently complicated structure the intersection of the regions they delineate in the space of the $\lambda$'s 
has a form similar to equations (\ref{eq:BFB2dir8}) and (\ref{eq:BFB2dir4}). Moreover, the true BFB conditions will
be obtained only if all field directions are taken into account, up to some arbitrary $SU(2) \times U(1)$ gauge
transformations, but in this case the method used so far is not tractable anymore. 

To proceed to the most general case, we adopt a different parameterization of the fields that will turn out to be
particularly convenient to entirely solve the problem. Without loss of generality we can define:

\begin{eqnarray}
 r &\equiv& \sqrt{H^\dagger{H} + Tr \Delta^{\dagger}{\Delta}} \\
 H^\dagger{H} &\equiv& r^2 \cos^2 \gamma  \\
 Tr(\Delta^{\dagger}{\Delta}) &\equiv& r^2 \sin^2 \gamma   \\
Tr (\Delta^{\dagger}{\Delta})^2/(Tr \Delta^{\dagger}{\Delta})^2 &\equiv& \zeta \\
 (H^\dagger\Delta\Delta^{\dagger}H)/(H^\dagger{H} Tr\Delta^{\dagger}{\Delta} ) &\equiv& \xi 
\end{eqnarray}.

\noindent
(where we adopted here a parameterization similar
to the one used in \cite{ElKaffas:2006nt} to study two-Higgs-doublet models, although for the latter models
the problem is not fully solved by such a parameterization).
Obviously, when $H$ and $\Delta$ scan all the field space, the radius $r$ scans the domain $[0, \infty[$ and
the angle $\gamma \in [0, \frac{\pi}{2}]$. Moreover, one can show 
that 

\begin{eqnarray}
 0 \le \xi \le 1  & {\rm and} &  \frac{1}{2} \le \zeta \le 1  
\label{eq:conditions}
\end{eqnarray}

\noindent
With this parameterization it is straightforward to cast $V^{(4)}(H, \Delta)$ in the following simple form,

\begin{equation}
V^{(4)}(r, \tan \gamma, \xi, \zeta)= \frac{r^4}{ 4 (1+ \tan^2 \gamma)^{2}} ( \lambda + 
4 (\lambda_1 + \xi \lambda_4) \tan^2 \gamma  + 4 (\lambda_2 + \zeta \lambda_3) \tan^4 \gamma)
\label{eq:V4general}
\end{equation}

\noindent
Due to the bi-quadratic dependence in $\tan \gamma$, one can indeed  
consider only the range  $0 \le \tan \gamma <  +\infty$ in accordance with the above stated range for $\gamma$. 
We have thus written $V^{(4)}$ in the form of Eq.~({\ref{eq:Vtype}).
Boundedness from below is then equivalent to requiring $V^{(4)} > 0$ {\sl for all} $\tan \gamma \in [0, \infty[$ 
and {\sl all} $\xi, \zeta$ satisfying Eq.(\ref{eq:conditions}). 
Now applying directly the conditions Eq.(\ref{eq:poscond}) one obtains 


\begin{equation}
\lambda > 0 \;\;{\rm \&}\;\; \lambda_2+ \zeta \lambda_3 > 0 \;\;
{\rm \&} \;\;\lambda_1+ \xi \lambda_4 + \sqrt{\lambda(\lambda_2+\zeta \lambda_3)} > 0 \; \;
\forall \zeta \in [\frac{1}{2}, 1], \forall \xi [0, 1]
\end{equation}

Due to the monotonic dependence in $\zeta$ and $\xi$, it is easy to show that these conditions can be rewritten as,

\begin{eqnarray}
&& \lambda > 0 \;\;{\rm \&}\;\; \lambda_2+\lambda_3 > 0  \;\;{\rm \&}  \;\;\lambda_2+\frac{\lambda_3}{2} > 0 
\label{eq:BFBgen1} \\
&& {\rm \&} \;\;\lambda_1+ \sqrt{\lambda(\lambda_2+\lambda_3)} > 0 \;\;{\rm \&}\;\;
\lambda_1+ \sqrt{\lambda(\lambda_2+\frac{\lambda_3}{2})} > 0  \label{eq:BFBgen2}\\
&& {\rm \&} \;\; \lambda_1+\lambda_4+\sqrt{\lambda(\lambda_2+\lambda_3)} > 0 \;\; {\rm \&} \;\; 
\lambda_1+\lambda_4+\sqrt{\lambda(\lambda_2+ \frac{\lambda_3}{2})} > 0  \label{eq:BFBgen3}
\end{eqnarray}

\noindent

We stress here that the above conditions ensure BFB for all possible directions in field space and thus provide
the most general {\sl all directions necessary and sufficient BFB conditions} that solve completely
the issue at the tree-level. Note that all the 2-field directions conditions given in Eqs.(\ref{eq:BFB2dir1} -
\ref{eq:BFB2dir10} ) are  special cases of the above conditions. We also checked numerically that this is 
the case for all the ten 3-field directions conditions Eqs.~(\ref{eq:BFB3dir1} - \ref{eq:BFB3dir10}).


\setcounter{equation}{0}
\section{Unitarity constraints}
 \label{sec:unitarity}
Constraints on the scalar potential parameters can be obtained by demanding that tree-level unitarity be
 preserved in a variety of scattering processes: scalar-scalar scattering,
gauge boson-gauge boson scattering and scalar-gauge boson scattering as was initially done for the
SM \cite{Appelquist:1971yj,Cornwall:1974km,Lee:1977eg}. The generalizations of such constraints to
various extended Higgs sector scenarios have been studied in the literature, see for instance \cite{Kanemura:1993hm,
Akeroyd:2000wc, Aoki:2007ah, Gogoladze:2008ak}. 
Here we treat such constraints in the DTHM at the tree-level, limiting ourselves to $2$-body scalar scattering
processes dominated by quartic interactions. This is justified by the fact that we are interested
in the leading unitarity constraints, that is in the limit where $\sqrt{s}$ is much larger than 
any other mass scale involved. In particular this means that we disregard here unitarity constraints that
would involve the $\mu$ parameter when the latter is very large. Indeed, this parameter contributes to the
scalar scattering processes through the cubic interactions entering the Feynman diagrams with scalar exchange
in the $s, t$ and $u$ channels. Furthermore, the ratio $\mu/v_t$ controls the size of the exchanged scalar masses
so that some of the aforementioned diagrams can be important in the vicinity of the resonance pole in the limit
of large $\sqrt{s} \sim \mu v_d/v_t$.

In order to derive the unitarity constraints on the scalar masses
we adopt the basis of unrotated states, corresponding to the fields before electroweak symmetry breaking.
The quartic scalar vertices have in this case a much simpler form than the complicated functions of $\lambda_i$,
$\alpha$ and $\beta$ obtained in the physical basis ($H^{\pm\pm}$, $H^\pm$, $G^\pm$, $h^0$, $H^0$, $A^0$
and $G^0$) of mass eigenstate fields.  The $S$-matrix for the physical fields
is related by a unitary transformation to the
$S$-matrix for the unrotated fields.
Close inspection shows that the full set of $2$-body scalar scattering processes leads to
a $ 35\times35$ $S$-matrix which can be decomposed into 7 block submatrices corresponding to mutually
 unmixed sets of channels with definite charge and CP states. One has the following submatrix dimensions,  
structured in terms of net electric charge in the initial/final states:   
${S}^{(1)}(6 \times 6)$, ${ S}^{(2)}(7\times 7)$ and ${ S}^{(3)}(2\times 2)$, corresponding to 
$0$-charge channels, ${ S}^{(4)}(10 \times 10)$ corresponding to the $1$-charge channels,  
${ S}^{(5)}(7 \times 7)$
corresponding to the $2$-charge channels, ${ S}^{(6)}(2 \times 2)$ corresponding to the $3$-charge channels
and  ${S}^{(7)}(1 \times 1)$ corresponding to the unique $4$-charge channel.
The corresponding $T$-matrix submatrices $T^{(1)}, ...,T^{(7)}$ 
--with a momentum conservation factor $(2 \pi)^4 \delta^4( \sum {\rm momenta})$ properly factored out--
are then easily extracted using the pure scalar quartic
interactions expressed in terms of the non-physical fields $\phi^\pm$, $\delta^\pm$, $\delta^{\pm\pm}$,
$h$, $\xi^{0}$, and $Z_i$(i=1,2) as listed in Appendix C.

One can then in principle extract the unitarity constraints on each component of the $T$-matrix through the
 unitarity equation which we write here in a shorthand form as 
 \begin{equation}
-i (T - T^\dag) \sim \int ``T {T^\dag}"  \label{eq:uniteq}
\end{equation} 
where $\int$ denotes symbolically the phase space integral over each intermediate state channel
(see for instance \cite{Pilkuhn:1979ps}). However, it proves more efficient 
to define a modified matrix in such a way that its diagonalized form still satisfies Eq.~(\ref{eq:uniteq}).
The usual unitarity bound on partial-wave amplitudes that is valid for {\sl elastic} scattering 
would then apply readily to all the eigenvalues, thus encoding indirectly the bounds on all the components of the 
$T$-matrix.\footnote{This, however, cannot be achieved in general by simply diagonalizing $T$, since on the right-hand 
side of Eq.~(\ref{eq:uniteq}) the phase space factor is not the same for all the $2$-particle channels, even in the 
high energy massless limit we are considering. It picks up a factor $1/2$ only for internal states with identical 
particles so as to avoid double counting. The right-hand side of Eq.~(\ref{eq:uniteq}) is thus not a proper matrix 
multiplication of $T$ by $T^\dag$, a fact emphasized by the quotation marks in the equation.} The proper redefinition 
is a $\tilde{T}$-matrix having the same entries as $T$ but with an extra $1/\sqrt{2}$ factor for each initial or final 
state channel having two identical particles. $\tilde{T}$ now satisfies Eq.~(\ref{eq:uniteq}) with the same phase 
space for {\sl all} channels and a true matrix multiplication of $\tilde{T}$ by $\tilde{T}^\dag$. Its diagonalized 
form will thus satisfy the same equation. Defining ${\cal M}_n \equiv i \tilde{T}^{(n)}$, with $n=1, \cdots, 7$,  
we give hereafter the resulting submatrices whose entries correspond  to the quartic couplings that mediate the 
$2 \to 2$ scalar processes. These submatrices are hermitian, thus the sought for eigenvalues will all be real-valued.

The first submatrix ${\cal M}_1$  corresponds to scattering whose
initial and final states are one of the following:
$(\phi^+\delta^-$,$\delta^+\phi^-$, $h Z_2$, $\xi^0 Z_1
$, $Z_1 Z_2$, $h \xi^0)$. With the help of Appendix C
one finds, 

\begin{eqnarray}
{\cal M}_1=\left(
\begin{array}{cccccc}
\displaystyle \lambda_1+\frac{\lambda_4}{2} &0 &
\displaystyle -\frac{i\lambda_4}{2\sqrt{2}} & \displaystyle \frac{i\lambda_4}{2\sqrt{2}} &
\displaystyle \frac{\lambda_4}{2\sqrt{2}}  & \displaystyle \frac{\lambda_4}{2\sqrt{2}}  \\
 0 & \displaystyle \lambda_1+\frac{\lambda_4}{2} &
\displaystyle \frac{i\lambda_4}{2\sqrt{2}} & \displaystyle -\frac{i\lambda_4}{2\sqrt{2}} &
\displaystyle \frac{\lambda_4}{2\sqrt{2}}  &  \displaystyle \frac{\lambda_4}{2\sqrt{2}} \\
\displaystyle \frac{i\lambda_4}{2\sqrt{2}} & \displaystyle -\frac{i\lambda_4}{2\sqrt{2}}  &
 \lambda_{14}^+ & 0 &
 0  &  0 \\
\displaystyle -\frac{i\lambda_4}{2\sqrt{2}}  & \displaystyle \frac{i\lambda_4}{2\sqrt{2}}  &
 0 & \lambda_{14}^+ &
 0  &  0 \\
\displaystyle \frac{\lambda_4}{2\sqrt{2}} & \displaystyle \frac{\lambda_4}{2\sqrt{2}}  & 0 &0
 & \lambda_{14}^+ & 0 \\
\displaystyle \frac{\lambda_4}{2\sqrt{2}}  & \displaystyle \frac{\lambda_4}{2\sqrt{2}}  & 0 &0
 & 0 & \lambda_{14}^+
\end{array}\right)
\end{eqnarray}
$\ $
\\
where $\lambda_{ij}^{\pm}=\lambda_i\pm \lambda_j$. 
We find that ${\cal M}_1$ has the following
$3$ double eigenvalues:
\begin{eqnarray}
  e_1 &=& \lambda_1+\lambda_4 \ \ \ \ \ \\
  e_2 &=& \lambda_1 \ \ \ \ \  \\
  e_3 &=& \frac{1}{2}(2\lambda_1 + 3\lambda_4 ) \ \ \ \ \ 
\end{eqnarray}

The second submatrix ${\cal M}_2$ corresponds to scattering with one of the following
initial and final states:
$(\phi^+\phi^-$, $\delta^+\delta^-$, $\delta^{++}\delta^{--}$, $\frac{Z_1 Z_1}{\sqrt{2}}$, $\frac{Z_2 Z_2}{\sqrt{2}}$, 
$\frac{h h}{\sqrt{2}}$,
$\frac{\xi^0 \xi^0}{\sqrt{2}})$, where the $\sqrt{2}$ accounts for
identical particle statistics. Again, with the help of Appendix C,
one finds that ${\cal M}_2$ is given by:

\begin{eqnarray}
{\cal M}_2=\left(
\begin{array}{ccccccc}
\lambda & \displaystyle \frac{\tilde{\lambda}_{14}}{2} & \lambda_{14}^+ & \displaystyle \frac{\lambda}{2\sqrt{2}} & \displaystyle \frac{\lambda_1}{\sqrt{2}} & \displaystyle \frac{\lambda}{2\sqrt{2}} & \displaystyle \frac{\lambda_1}{\sqrt{2}} \\
\displaystyle \frac{\tilde{\lambda}_{14}}{2} & 2\tilde{\lambda}_{23} & 2\lambda_{23}^+ & \displaystyle \frac{\tilde{\lambda}_{14}}{2\sqrt{2}} & \sqrt{2}\lambda_{23}^+ & \displaystyle \frac{\tilde{\lambda}_{14}}{2\sqrt{2}} & \sqrt{2}\lambda_{23}^+ \\
\lambda_{14}^+ & 2\lambda_{23}^+ & 4\lambda_{23}^+ & \displaystyle \frac{\lambda_1}{\sqrt{2}} & \sqrt{2}\lambda_2 & \displaystyle \frac{\lambda_1}{\sqrt{2}} & \sqrt{2}\lambda_2 \\
\displaystyle \frac{\lambda}{2\sqrt{2}} & \displaystyle \frac{\tilde{\lambda}_{14}}{2\sqrt{2}} & \displaystyle \frac{\lambda_1}{\sqrt{2}} & \displaystyle \frac{3}{4}\lambda & \displaystyle \frac{\lambda_{14}^+}{2} & \displaystyle \frac{\lambda}{4} & \displaystyle \frac{\lambda_{14}^+}{2} \\
\displaystyle \frac{\lambda_1}{\sqrt{2}} & \sqrt{2}\lambda_{23}^+ & \sqrt{2}\lambda_2 & \displaystyle \frac{\lambda_{14}^+}{2} & 3\lambda_{23}^+ & \displaystyle \frac{\lambda_{14}^+}{2} & \lambda_{23}^+ \\
\displaystyle \frac{\lambda}{2\sqrt{2}} & \displaystyle \frac{\tilde{\lambda}_{14}}{2\sqrt{2}} & \displaystyle \frac{\lambda_1}{\sqrt{2}} & \displaystyle \frac{\lambda}{4} & \displaystyle \frac{\lambda_{14}^+}{2} & \displaystyle \frac{3\lambda}{4} & \displaystyle \frac{\lambda_{14}^+}{2}\\
\displaystyle \frac{\lambda_1}{\sqrt{2}} & \sqrt{2}\lambda_{23}^+ & \sqrt{2}\lambda_2 & \displaystyle \frac{\lambda_{14}^+}{2} & \lambda_{23}^+ & \displaystyle \frac{\lambda_{14}^+}{2} & 3\lambda_{23}^+ \\
\end{array}
\right)
\end{eqnarray}
$\ $ \\
where $\tilde{\lambda}_{14}=2\lambda_1+\lambda_4 $ and $\tilde{\lambda}_{23}=2\lambda_2+\lambda_3$.
Despite its apparently complicated structure, one can easily determine the seven eigenvalues of ${\cal M}_2$ 
as follows:

\begin{eqnarray}
& & f_{1} = \frac{\lambda}{2} \label{eq:f10}\\
& & f_{2} =2 \lambda_2\\
& & f_{3} =2 (\lambda_2+\lambda_3) \label{eq:f30}\\
& & a_{\pm}=\frac{1}{4}[ \lambda+4\lambda_2+8\lambda_3\pm\sqrt{(\lambda-4\lambda_2-8\lambda_3)^2+16 \lambda_4^2} \ ]\\
& & b_{\pm}=\frac{1}{4}[ 3\lambda+16\lambda_2+12\lambda_3\pm\sqrt{(3\lambda-16\lambda_2-12\lambda_3)^2+24(2\lambda_1+\lambda_4 )^2} \ ]
\end{eqnarray}

The third submatrix ${\cal M}_3$ corresponds to the basis
$(h Z_1 , \xi^0 Z_2)$ and is given by:

\begin{eqnarray}
{\cal M}_3=\left(
\begin{array}{cc}
\displaystyle \frac{\lambda}{2} & 0 \\
0 & 2\lambda_{23}^+ \\
\end{array}
\right)
\label{eq:f11}
\end{eqnarray}
$\ $ \\
with eigenvalues $k_1=f_1$ and $k_2=f_3$.

The $1$-charge channels occur for two-by-two body scattering
between the 10 charged states $ (h \phi^+ $,  $\xi^0 \phi^+ $,
$Z_1 \phi^+ $, $Z_2 \phi^+$, $h \delta^+$, $\xi^0 \delta^+$, $Z_1 \delta^+$,
$Z_2 \delta^+ $, $\delta^{++} \delta^-$, $\delta^{++} \phi^-)$.
The 10$\times$10 submatrix ${\cal M}_4$ obtained from the above
scattering processes is given by:

\begin{eqnarray}
{\cal M}_4=\left(
\begin{array}{cccccccccc}
\displaystyle \frac{\lambda}{2} & 0 & 0 & 0 & 0 & \displaystyle \frac{\lambda_4}{2\sqrt{2}} & 0 & \displaystyle \frac{-i \lambda_4}{2\sqrt{2}} & \displaystyle \frac{-\lambda_{4}}{2} & 0 \\
0 & \lambda_1 & 0 & 0 & \displaystyle \frac{\lambda_4}{2\sqrt{2}} & 0 & \displaystyle \frac{i\lambda_4}{2\sqrt{2}} & 0 & 0 & 0 \\
0 & 0 & \displaystyle \frac{\lambda}{2} & 0 & 0 & \displaystyle \frac{i\lambda_4}{2\sqrt{2}} & 0 & \displaystyle \frac{\lambda_4}{2\sqrt{2}} & \displaystyle \frac{-i\lambda_4}{2} & 0  \\
0 & 0 & 0 & \lambda_1 & \displaystyle \frac{-i \lambda_4}{2\sqrt{2}} & 0 & \displaystyle \frac{\lambda_4}{2\sqrt{2}} & 0 & 0 & 0 \\
0 & \displaystyle \frac{\lambda_4}{2\sqrt{2}} & 0 & \displaystyle \frac{i\lambda_4}{2\sqrt{2}} & \displaystyle \frac{1}{2}\tilde{\lambda}_{14} & 0 & 0 & 0 & 0 & \displaystyle \frac{-\lambda_{4}}{2} \\
\displaystyle \frac{\lambda_4}{2\sqrt{2}} & 0 & \displaystyle \frac{-i \lambda_4}{2\sqrt{2}} & 0 & 0 & 2\lambda_{23}^+ & 0 & 0 & -\sqrt{2}\lambda_3 & 0 \\
0 & \displaystyle \frac{-i \lambda_4}{2\sqrt{2}} & 0 & \displaystyle \frac{\lambda_4}{2\sqrt{2}} & 0 & 0 & \displaystyle \frac{1}{2}\tilde{\lambda}_{14} & 0 & 0 & \displaystyle \frac{-i \lambda_4}{2}\\
\displaystyle \frac{i\lambda_4}{2\sqrt{2}} & 0 & \displaystyle \frac{\lambda_4}{2\sqrt{2}} & 0 & 0 & 0 & 0 & 2\lambda_{23}^+ & -i\sqrt{2}\lambda_3 & 0\\
\displaystyle \frac{-\lambda_{4}}{2} & 0 & \displaystyle \frac{i\lambda_4}{2} & 0 & 0 & -\sqrt{2}\lambda_3 & 0 & i\sqrt{2}\lambda_3 & 2\lambda_{23}^+ & 0\\
0 & 0 & 0 & 0 & \displaystyle \frac{-\lambda_{4}}{2} & 0 & \displaystyle \frac{i\lambda_4}{2} & 0& 0 & \lambda_{14}^+ \\
\end{array}
\right)
\end{eqnarray}
$\ $\\
As one can see, this matrix contains many vanishing elements,
and the 10 eigenvalues are straightforward to obtain analytically. They read as follows:

\begin{eqnarray}
& & d_{1} = e_1\\
& & d_{2} = e_2 \ \  (twice)\\
& & d_{3} = e_3\\
& & d_{4} = f_1 \label{eq:f1}\\
& & d_{5} = f_2\\
& & d_{6} = f_3 \label{eq:f3}\\
& & d_{7} = \lambda_1-\frac{\lambda_4}{2}\\
& & d_{\pm}=a_{\pm}
\end{eqnarray}
The fifth submatrix ${\cal M}_5$ corresponds to scattering with
initial and final
states being one of the following $7$ sates:
$(\frac{\phi^+\phi^+}{\sqrt{2}}$,$\frac{\delta^+\delta^+}{\sqrt{2}}$,$\delta^+\phi^+$,$\delta^{++}\xi^0$, $\delta^{++} Z_2$,  $\delta^{++} Z_1 $,
$\delta^{++} h)$. It reads, 

\begin{eqnarray}
{\cal M}_5=\left(
\begin{array}{ccccccc}
\displaystyle{\frac{\lambda}{2}} & 0 & 0 & 0 & 0 & 0 & 0 \\
0 & \tilde{\lambda}_{23} & 0 & -\lambda_3 & -i \lambda_3 & 0 & 0 \\
0 & 0 & \displaystyle \frac{\tilde{\lambda}_{14}}{2} & 0 & 0 & \displaystyle \frac{-i \lambda_4}{2} & \displaystyle \frac{-\lambda_{4}}{2} \\
0 & -\lambda_3 & 0 & 2\lambda_2 & 0 & 0 & 0 \\
0 & i \lambda_3 & 0 & 0 & 2\lambda_2 & 0 & 0 \\
0 & 0 & \displaystyle \frac{i\lambda_4}{2} & 0 & 0 & \lambda_1 & 0 \\
0 & 0 & \displaystyle \frac{-\lambda_{4}}{2} & 0 & 0 & 0 & \lambda_1 \\
\end{array}
\right)
\end{eqnarray}
$\ $\\
and possesses the following 7 distinct eigenvalues:
\begin{eqnarray}
& & c_{1} = e_1\\
& & c_{2} = e_2\\
& & c_{3} = f_1  \label{eq:2f1}\\
& & c_{4} = f_2\\
& & c_{5} = f_3\\
& & c_{6} = d_{7}\\
& & c_{7} = 2 \lambda_2 - \lambda_3
\end{eqnarray}
There are also  triply-charged states.The submatrix ${\cal M}_6$ corresponding to this case generates the scattering 
with initial and final
states being one of the following $(\delta^{++}\phi^+$,$\delta^{++}\delta^+)$, and is given by :
\begin{eqnarray}
{\cal M}_6=\left(
\begin{array}{cc}
\lambda_{14}^+ & 0 \\
0 & 2\lambda_{23}^+ \\
\end{array}
\right)
\label{eq:f31}
\end{eqnarray}
with eigenvalues  $k_1=e_1$ and $k_2=f_3$. Finally, it is easy to check that there is just one quadruply-charged 
state $\frac{1}{\sqrt{2}} \delta^{++}\delta^{++} $ leading to 
\begin{equation}
{\cal M}_7 = f_3  \label{eq:2f3}
\end{equation}
with $f_3$ as eigenvalue.

From the usual expansion in terms of partial-wave amplitudes $a_J$, we write, 
following our notations, 
\begin{equation}
{\cal M}^{(k f)} = i \tilde{T}_{k f} = 16 i \pi \sum_{J\ge0} ( 2 J + 1) \, a^{(k f)}_J(s) \, P_J(\cos \theta)
\end{equation}
where ${\cal M}^{(k f)}$ denotes the entries of the ${\cal M}$ matrix, the subscripts $k$ and $f$ run
over all possible initial and final states of the above $35$-state basis, 
$\theta$ denotes the scattering angle of the corresponding processes and the $P_J$'s are the Legendre polynomials. 
Since we considered only the leading high energy (massless limit) contributions that are $s$ and $\theta$ independent, 
all the partial waves with $J\neq 0$ vanish and one is left with 

\begin{equation} a^{(k f)}_0 = - \frac{i}{16 \pi} {\cal M}^{(k f)} \label{eq:Unitconst} \end{equation}
for each channel. The S-matrix unitarity constraint for elastic scattering $|a^{(kk)}_0| \le 1$ 
(or alternatively $| Re(a^{(kk)}_0)| \le \frac{1}{2}$
\cite{Luscher:1988gc, Marciano:1989ns}) applies to the diagonal entries of ${\cal M}$. 
It encodes as well the constraints for non-elastic scattering, 
provided that it is applied to the eigenchannels of the $35$-state basis as noted previously. 
Thus, this constraint translates through Eq.~(\ref{eq:Unitconst}) directly to all the eigenvalues we determined above.
We defer to the next section, Eqs.~(\ref{eq:unit1} - \ref{eq:unit10}) the list of
all the resulting constraints.

\setcounter{equation}{0}
\section{Combined unitarity and potential stability constraints}
\label{sec:comb}

\noindent
Let us first recall all the constraints obtained in sections \ref{sec:boundedness} and \ref{sec:unitarity}.\\

{\sl \underline{BFB}:}
\begin{eqnarray}
&& \lambda \geq 0 \;\;{\rm \&}\;\; \lambda_2+\lambda_3 \geq 0  \;\;{\rm \&}  \;\;\lambda_2+\frac{\lambda_3}{2} \geq 0
\label{eq:bound1} \\
&& {\rm \&} \;\;\lambda_1+ \sqrt{\lambda(\lambda_2+\lambda_3)} \geq 0 \;\;{\rm \&}\;\;
\lambda_1+ \sqrt{\lambda(\lambda_2+\frac{\lambda_3}{2})} \geq 0  \label{eq:bound2} \\
&& {\rm \&} \;\; \lambda_1+\lambda_4+\sqrt{\lambda(\lambda_2+\lambda_3)} \geq 0 \;\; {\rm \&} \;\; 
\lambda_1+\lambda_4+\sqrt{\lambda(\lambda_2+ \frac{\lambda_3}{2})} \geq 0 \label{eq:bound3}
\end{eqnarray}

{\sl \underline{unitarity}:}
\begin{eqnarray}
&&|\lambda_1 + \lambda_4| \leq \kappa \pi  \label{eq:unit1} \\
&&|\lambda_1| \leq \kappa \pi \label{eq:unit2} \\
&&|2 \lambda_1 + 3 \lambda_4| \leq 2 \kappa \pi \label{eq:unit3} \\
&&|\lambda| \leq  2 {\kappa} \pi \label{eq:unit4} \\
&&|\lambda_2| \leq  \frac{\kappa}{2} \pi \label{eq:unit5} \\
&&|\lambda_2 + \lambda_3| \leq  \frac{\kappa}{2} \pi \label{eq:unit6} \\
&&|\lambda + 4 \lambda_2 + 8 \lambda_3 \pm \sqrt{(\lambda - 4 \lambda_2 - 8 \lambda_3)^2
+ 16 \lambda_4^2} \;| \leq  4 \kappa \pi \label{eq:unit7} \\
&&| 3 \lambda + 16 \lambda_2 + 12 \lambda_3 \pm \sqrt{(3 \lambda - 16 \lambda_2 - 12 \lambda_3)^2
+ 24 (2 \lambda_1 +\lambda_4)^2} \;| \leq  4 \kappa \pi \label{eq:unit8} \\
&&|2 \lambda_1 - \lambda_4| \leq 2 \kappa \pi \label{eq:unit9} \\
&&|2 \lambda_2 - \lambda_3| \leq  \kappa \pi \label{eq:unit10} 
\end{eqnarray}
where we introduced the parameter $\kappa$ that takes the
values $\kappa=16$ or $8$, depending on whether we choose $|a_0| \le 1$ or $| Re(a_0)| \le \frac{1}{2}$ as pointed out 
at the end of section \ref{sec:unitarity}.

\noindent
Working out analytically these two sets of BFB and unitarity constraints, one can reduce them to a more compact 
system where the allowed ranges for the $\lambda$'s are easily identified.  One can obtain a necessary domain 
for $\lambda, \lambda_2, \lambda_3$ that does not depend on $\lambda_1$ and $\lambda_4$, by considering
simultaneously Eqs.(\ref{eq:unit4} - \ref{eq:unit10}) together with Eq.~(\ref{eq:bound1}). It then turns out that 
 Eqs.(\ref{eq:unit5}, \ref{eq:unit6}) as well as the lower part of Eq.(\ref{eq:unit10})
are weaker than the actually allowed domains for $\lambda_2, \lambda_3$, and similarly Eq.(\ref{eq:unit4}) is weaker 
than the constraint on $\lambda$ coming from Eq.(\ref{eq:unit8}). We find,

\begin{eqnarray}
&& 0 \leq \lambda \leq \frac{2}{3} \kappa \pi  \label{eq:unit4new} \\
&& \lambda_2+\lambda_3 \geq 0  \;\;{\rm \&}  \;\;\lambda_2+\frac{\lambda_3}{2} \geq 0 \label{eq:bound}\\
&& \lambda_2 + 2 \lambda_3 \leq \frac{\kappa}{2} \pi \\
&& 4 \lambda_2 + 3 \lambda_3 \leq  \frac{\kappa}{2} \pi \\
&& 2 \lambda_2 - \lambda_3 \leq  \kappa \pi  \label{eq:unitnew}
\end{eqnarray}

\noindent
We stress here that the above
constraints define the largest possible domain for $\lambda, \lambda_2, \lambda_3$ for any set of allowed
values of $\lambda_1, \lambda_4$, although Eqs.(\ref{eq:unit7}, \ref{eq:unit8}) have been used
to determine this domain. It is noteworthy that the upper bound on $\lambda$ Eq.(\ref{eq:unit4new}) is reduced by a 
factor $3$ with respect to the naive expectation Eq.~(\ref{eq:unit4}). 
Studying further Eqs.(\ref{eq:unit7}, \ref{eq:unit8}), one can rewrite them
in the following simple form where the dependence on $\lambda_1, \lambda_4$ has been explicitly separated
from that on $\lambda, \lambda_2, \lambda_3$:

\begin{eqnarray}
&&  |\lambda_4| \leq 
 \min{\sqrt{(\lambda \pm 2 \kappa \pi) (\lambda_2 + 2 \lambda_3 \pm \frac{\kappa}{2} \pi)} } \label{eq:unit7new}\\
&& |2 \lambda_1 + \lambda_4| \leq \sqrt{2 (\lambda - \frac{2}{3} \kappa \pi) (4 \lambda_2 + 3 \lambda_3 
- \frac{\kappa}{2} \pi)} \label{eq:unit8new}
\end{eqnarray}

\noindent
where Eqs.(\ref{eq:unit4new}, \ref{eq:unitnew}) have been used in deriving 
Eq.(\ref{eq:unit8new})\footnote{ In writing Eq.~(\ref{eq:unit8new}) we relied on the  fact that
the minimum of $\sqrt{2(\lambda \pm \frac{2}{3} \kappa \pi) (4 \lambda_2 + 3 \lambda_3 
\pm \frac{\kappa}{2} \pi)}$ is given by $\sqrt{2
(\lambda - \frac{2}{3} \kappa \pi) (4 \lambda_2 + 3 \lambda_3 
- \frac{\kappa}{2} \pi)}$  in all the domain allowed by $\lambda, \lambda_2, \lambda_3$. In contrast,  
$\min{\sqrt{(\lambda \pm 2 \kappa \pi) (\lambda_2 + 2 \lambda_3 \pm \frac{\kappa}{2} \pi)} }$
appearing in Eq.~(\ref{eq:unit7new}) cannot be written in a closed form in this domain.}. Various comments are in 
order here. First, to obtain
the full domain for $\lambda_1, \lambda_4$, one has to add to Eqs.(\ref{eq:unit7new}, \ref{eq:unit8new})
equations (\ref{eq:unit1} - \ref{eq:unit3}, \ref{eq:unit9})  as well as Eqs.(\ref{eq:bound2}, \ref{eq:bound3}).
Thus for each set of values of $\lambda, \lambda_2, \lambda_3$, the allowed domain for $\lambda_1, \lambda_4$
is easily determined as the overlap of a set of linear bands, as illustrated in 
Fig.~1.

\begin{figure}[ht]
\label{fig:lam14}
\centering
\includegraphics[]{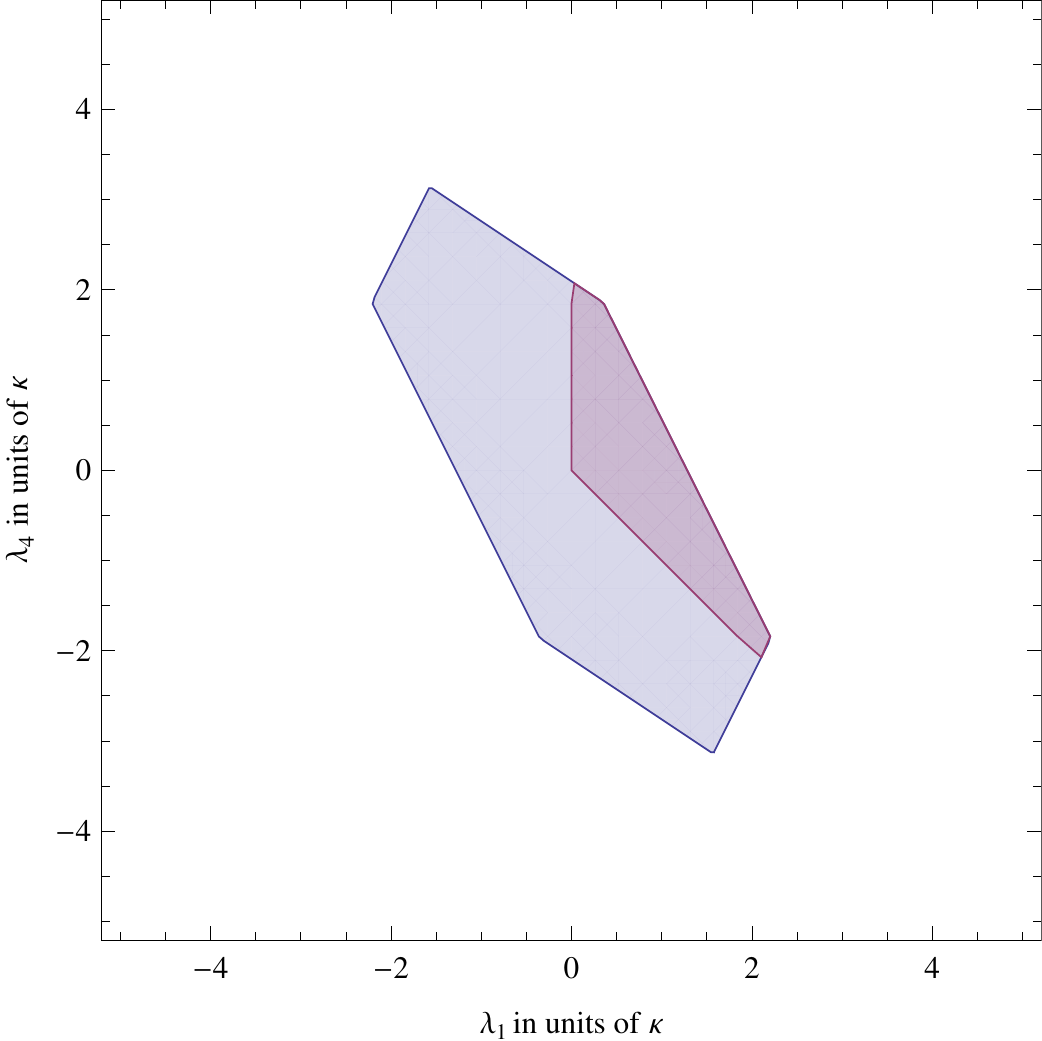}
\caption{ An illustration of a section of the $(\lambda_1, \lambda_4)$ domain (blue) in units of $\kappa$ as 
determined by Eqs.~(\ref{eq:unit1} - \ref{eq:unit3}, \ref{eq:unit9}, \ref{eq:unit7new}, \ref{eq:unit8new}), where we 
fixed $\lambda =\lambda_2=\lambda_3=0$. Adding the BFB constraints Eqs.(\ref{eq:bound2}, \ref{eq:bound3}) one obtains
the reduced domain shown (light purple).}
\end{figure}

\noindent
As stated earlier, Eqs.~(\ref{eq:bound} - \ref{eq:unitnew}) define the largest possible domain for  
$\lambda_2, \lambda_3$ allowed by the combined unitarity and BFB constraints. The reason is seen
from Eqs.~(\ref{eq:unit7new}, \ref{eq:unit8new}) that are the only extra constraints on $\lambda_2, \lambda_3$
depending on the actual values of $\lambda, \lambda_1, \lambda_4$. As one can easily check, these constraints
become trivially satisfied when $\lambda_1=\lambda_4=0$ and thus correspond to the case of largest 
$\lambda_2, \lambda_3$ domain.
For each set of non vanishing values  for $\lambda_1, \lambda_4$ the domain of $\lambda_2, \lambda_3$
given by Eqs.~(\ref{eq:bound} - \ref{eq:unitnew}) will be further reduced according
to Eqs.~(\ref{eq:unit7new}, \ref{eq:unit8new}). We illustrate the largest $(\lambda_2, \lambda_3)$ domain 
in 
Fig.~2.

\begin{figure}[ht]
\label{fig:lam23}
\centering
\includegraphics[]{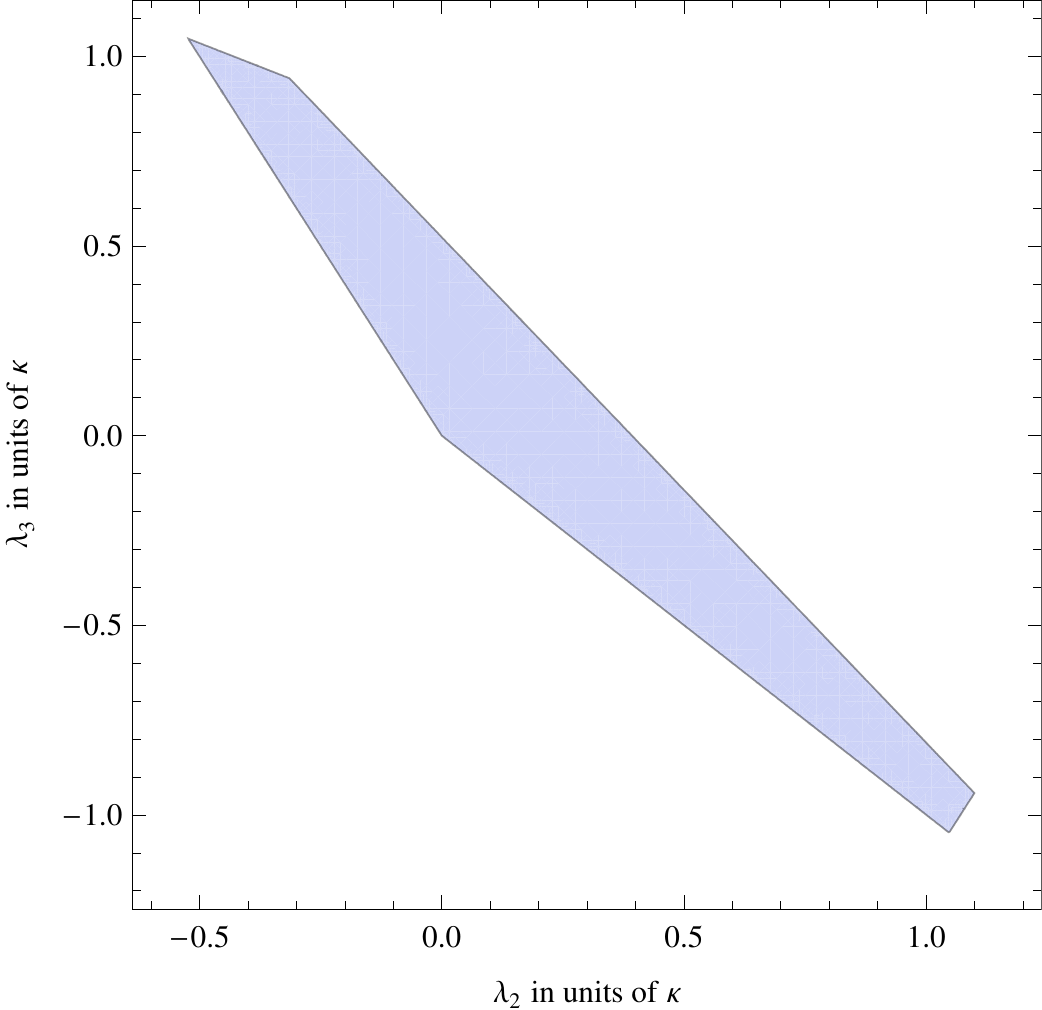}
\caption{ We illustrate here the {\sl largest} $\lambda_2-\lambda_3$ domain allowed by the combined 
unitarity and BFB constraints in units of $\kappa$. This domain corresponds to 
Eqs.~(\ref{eq:bound} - \ref{eq:unitnew}) and is attained for $\lambda_1= \lambda_4=0$ in which case
Eqs.~(\ref{eq:unit7new}, \ref{eq:unit8new}) are trivially satisfied. As can be seen from
Eqs.~(\ref{eq:unit7new}, \ref{eq:unit8new}), a smaller domain obtains as soon as $\lambda_1$ and/or $\lambda_4$
are non-zero, irrespecive of the value of $\lambda$  satisfying Eq.~(\ref{eq:unit4new}).}
\end{figure}

To summarize, the boundaries of the combined unitarity and general BFB 
domains for the five couplings are now given by the reduced set of equations
(\ref{eq:bound2} - \ref{eq:unit3},  \ref{eq:unit4new}- \ref{eq:unit8new}) which moreover have an analytically simpler 
form. In particular, one readily
finds from Eq.~(\ref{eq:unit8new}) that saturating the unitarity bound on 
$\lambda$, i.e. $\lambda = \frac{2}{3} \kappa \pi$, reduces the two-dimensional $(\lambda_1, \lambda_4)$ domain 
to the one-dimensional (straight line)  $\lambda_4 = -2 \lambda_1$. This, as well as other features, will be
useful in determining lower and upper bounds on the Higgs masses in the next section.

\setcounter{equation}{0}
\section{Higgs mass theoretical bounds} \label{sec:higgsbounds}

In this section we rely on the results of the previous sections to study the theoretically allowed ranges
of the Higgs masses when varying the $\lambda_i's$ and the $\mu$ parameter in their allowed
domains. Rather than assuming that $\mu$ is very large, i.e. of the order of the GUT scale together with
$\mu \simeq M_{\Delta}$, we will study all the phenomenologically allowed range.
We stress here that even very small values of $\mu$ are consistent with a tiny value of $v_t$ necessary 
for realistic  neutrino masses (and ${\cal O}(1)$ Yukawa couplings) provided that we take into account
consistently Eq.~(\ref{eq:ewsb1}).

Let us first describe qualitatively the generic behavior of the masses when $\mu$ is varied.
We will show that, as a function of $\mu$, the $h^0$ mass features a {\sl maximum} 
$m_{h^0}^{\rm max}$
for a specific value $\mu = \mu_c$. This maximum will translate into an  
upper bound on $m_{h^0}$ when the unitarity bound on $\lambda$ is saturated. Similarly, the $H^0$ mass
reaches a {\sl minimum} $m_{H^0}^{\rm min}$ at a nearby value which we momentarily denote also $\mu = \mu_c$ for the
sake of the qualitative discussion. 
In the range $\mu < \mu_c$,
$H^0$ is the heaviest among all the Higgses, decreasing very slowly with increasing $\mu$ towards its minimum
value $m_{H^0}^{\rm min}$, while $m_{h^0}$ increases very quickly with $\mu$ to 
$m_{h^0}^{\rm max}$. 
 The other Higgs masses can have various hierarchies and in particular the unusual one where the
$m_{H^{\pm \pm}}$ is the lightest state,  
$m_{H^{\pm \pm}} < m_{H^{\pm}} < m_A \simeq m_{h^0}$. In contrast, in the range 
$\mu_c < \mu < \mu_{\rm max}$, $m_{H^0}$  now increases quickly with $\mu$ while $m_{h^0}$ decreases very slowly from 
its maximal value.
This sharply different behavior of $m_{h^0}$ and $m_{H^0}$ below and above $\mu_c$ can be traced
back to the smallness of $v_t$. We illustrate numerically such a behavior in 
Fig.~\ref{fig:fig2} 
where the seemingly constant $m^2_{h^0}$ for $\mu > \mu_c$ and constant $m^2_{H^0}$ for
$\mu < \mu_c$ is an artifact of the very small ratio $v_t/v_d$.  In fact, $m^2_{h^0}$ is decreasing very
slowly to the right of $\mu_c$
and reaches zero when $\mu = \mu_{+}$, cf. section \ref{sec:tachyon} and Appendix A, while  $m^2_{H^0}$
starts off at $\mu = \mu_{\rm min}$ decreasing very slowly until its minimum value at $\mu= \mu_c$,
then increases very slowly between $\mu_c$  and approximately $ \mu = \bar{\mu} \simeq \lambda v_t/\sqrt{2}$, 
and increases very quickly afterwards.~\footnote{ \label{footnote:mubar} The precise value is $\bar{\mu}= v_t (\lambda v_d^2 + 4 (4 \lambda_1 
- \lambda_2 - \lambda_3 + 4 \lambda_4) v_t^2)/\sqrt{2} (v_d^2 + 16 v_t^2)$. In fact $\bar{\mu}$ is the common value 
of $\mu$ at which the slopes of the variations of 
 $m^2_{h^0}$ and $m^2_{H^0}$ as functions of $\mu$ experience a sudden change.}

More quantitatively, we find that there are two different values of $\mu_c$, that we dub 
$\mu_c^{(1)}$, $\mu_c^{(2)}$ that are uniquely determined in terms of $v_d, v_t$ and the $\lambda$'s. When one
of these two critical values corresponds to $m_{h^0}^{\rm max}$ the other will correspond to $m_{H^0}^{\rm min}$, 
and vice versa, depending on the sign of the following quantity:

\begin{equation}
{\cal V}_\lambda \equiv (-\lambda + \lambda_1 + \lambda_4) \, v_d^2 + 4\,(\lambda_2 + \lambda_3)\,v_t^2  \label{eq:mucregimes}
\end{equation}

\noindent
Moreover, it turns out that  at these extrema one of the two $h^0$ or $H^0$ states will correspond to a purely SM-like 
Higgs state, and this too is controlled by the sign of ${\cal V}_\lambda$. 
One can summarize the behavior analytically as follows: \\

\begin{itemize}

\item[{\sl i)}] \underline{${\cal V}_\lambda > 0 \label{eq:regime1} $}: \\

\noindent
In this case $m_{h^0}$ reaches a maximum given by 

\begin{equation}
{m_{h^0}^2}^{\rm max}_{~(1)} = m^2_{(1)} \equiv\frac{\lambda\,v_d^2}{2} \label{eq:h0max}
\end{equation}

\noindent
when $\mu$ takes the value

\begin{equation} \mu = \mu_c^{(1)} \equiv (\lambda_1 + \lambda_4) \frac{v_t}{\sqrt{2}} 
\label{eq:mc1}
\end{equation}

\noindent
and $m_{H^0}^2$ reaches a minimum given by 

\begin{eqnarray}
 {m_{H^0}^2}^{\rm min}_{~(1)} = m^2_{(2)}& \equiv &\frac{1}{
     2\,(v_d^2 + 16\,v_t^2)} (\lambda\,v_d^4 + 16\,v_t^2\,((\lambda_1 + \lambda_4) v_d^2 + 
4(\lambda_2 +\lambda_3) \,v_t^2)) \nonumber\\
 &=& \frac{\lambda\,v_d^2}{2} + {\cal O}(v_t^2)  \label{eq:H0min}
\end{eqnarray}

\noindent
when $\mu$ takes the value

\begin{eqnarray}
&&\mu= \mu_c^{(2)} \equiv \frac{v_t}{\sqrt{2}\,(v_d^2 + 16\,v_t^2)}((2\,\lambda - \lambda_1 - \lambda_4)\,v_d^2 + 8\,(2\,
\lambda_1  + 2\,\lambda_4 - \lambda_2 - \lambda_3)\,v_t^2). \nonumber \\ 
\label{eq:mc2}
\end{eqnarray}

\noindent
Expanding the Higgs masses squared around $\mu_c^{(1)}$ one finds 

\begin{eqnarray}
&&m_{h^0}^2 =  {m_{h^0}^2}^{\rm max}_{~(1)} - \frac{4  v_d^2}{{\cal V}_\lambda} \delta_{\mu 1}^2 
+ {\cal O}(\delta_{\mu 1}^3)\\
&&    \Delta_{H^0}^2 = (-\lambda +\lambda_1 + \lambda_4) \frac{v_d^2}{2} + 2\,(\lambda_2 + \lambda_3)\,v_t^2
+  \frac{v_d^2}{\sqrt{2} v_t} \delta_{\mu 1} + {\cal O}(\delta_{\mu 1}^2) \\
&&\Delta_{A^0}^2 = (-\lambda + \lambda_1 + \lambda_4) \frac{v_d^2}{2} 
+ 2 \,(\lambda_1 + \lambda_4) \,v_t^2 +  \frac{(v_d^2 + 4 v_t^2)}{\sqrt{2} v_t} \delta_{\mu 1} + {\cal O}(\delta_{\mu 1}^2)\\
&&\Delta_{H^\pm}^2 = (-\lambda +  \lambda_1 + \frac{\lambda_4}{2}) \frac{v_d^2}{2} + (2 \lambda_1 + \lambda_4) \,\frac{v_t^2}{2} + 
\frac{(v_d^2 + 2 v_t^2)}{\sqrt{2} v_t} \delta_{\mu 1} + {\cal O}(\delta_{\mu 1}^2)\\ 
&&\Delta_{H^{\pm \pm}}^2 = (- \lambda + \lambda_1) \frac{v_d^2}{2} - \lambda_3 v_t^2  
+  \frac{v_d^2}{\sqrt{2} v_t} \delta_{\mu 1} + {\cal O}(\delta_{\mu 1}^2) 
\end{eqnarray}

\noindent
where $\Delta_X^2 \equiv m_X^2 - {m_{h^0}^2}^{\rm max}$ denotes the various squared mass splittings 
from ${m_{h^0}^2}^{\rm max}$ and $\delta_{\mu 1} \equiv \mu - \mu_c^{(1)}$.\\

\item[{\sl ii)}] \underline{${\cal V}_\lambda < 0 \label{eq:regime2}$}: \\

\noindent
In this case the reversed configuration
occurs. $m_{h^0}$ reaches a maximum, given by 

\begin{equation}
{m_{h^0}^2}^{\rm max}_{~(2)}= m^2_{(2)}  \label{eq:h0maxp}
\end{equation}

\noindent
at $\mu = \mu_c^{(2)}$,
while $m_{H^0}$ reaches a minimum given by

\begin{equation}
{m_{H^0}^2}^{\rm min}_{~(2)}= m^2_{(1)}  \label{eq:H0minp}
\end{equation}

\noindent
at $\mu = \mu_c^{(1)}$, 
where $m^2_{(1)}, m^2_{(2)}, \mu_c^{(1)}, \mu_c^{(2)}$ are as defined in Eqs.(\ref{eq:h0max} - \ref{eq:mc2}),

Again, expanding around $\mu_c^{(2)}$ we find

\begin{eqnarray}
&&m_{h^0}^2 = {m_{h^0}^2}^{\rm max}_{~(2)} + \frac{4  v_d^2}{{\cal V}_\lambda} \delta_{\mu 2}^2 
+ {\cal O}(\delta_{\mu 2}^3) \end{eqnarray}

\noindent
and the squared mass splittings

\begin{eqnarray}
&&\Delta_{H^0}^2 = (\lambda-\lambda_1 - \lambda_4)\,\frac{v_d^2}{2} 
- 2 (\lambda_2 + \lambda_3)\,v_t^2  
           +  \frac{v_d^2}{\sqrt{2} v_t} \delta_{\mu 2} + {\cal O}(\delta_{\mu 2}^2) \end{eqnarray}
\begin{eqnarray}
&&\Delta_{A^0}^2 = \frac{v_d^2}{(v_d^2 + 16\,v_t^2)}\,( \; (\lambda - \lambda_1 - \lambda_4)\,\frac{v_d^2}{2} + 
         2\,(2\,(\lambda - \lambda_2 - \lambda_3)  - \lambda_1- \lambda_4)\,v_t^2 \;) \nonumber \\
&&~~~~~~~~~~~~~~~~~~~~~~~~~+ \frac{(v_d^2 + 4 v_t^2)}{\sqrt{2} v_t} \delta_{\mu 2} + {\cal O}(\delta_{\mu 2}^2, v_t^4/v_d^2) \\
&&\Delta_{H^\pm}^2 = \frac{v_d^2}{(v_d^2 + 16\,v_t^2)}\,((\lambda - \,\lambda_1 - \frac{3}{2}
\,\lambda_4)\,\frac{v_d^2}{2} + 
         (2\,\lambda -\lambda_1 - 4 \lambda_2 - 4 \lambda_3 -\frac{11}{2} \lambda_4)\,v_t^2) \nonumber \\
&&~~~~~~~~~~~~~~~~~~~~~~~~~+ \frac{(v_d^2 + 2 v_t^2)}{\sqrt{2} v_t} \delta_{\mu 2} + {\cal O}(\delta_{\mu 2}^2, v_t^4/v_d^2) \\
&&\Delta_{H^{\pm \pm}}^2 = \frac{v_d^2}{(v_d^2 + 16\,v_t^2)} 
((\lambda - \lambda_1 - 2\,\lambda_4)\,\frac{v_d^2}{2} - 
(4\,\lambda_2 + 5\,\lambda_3 + 8\, \lambda_4)\,v_t^2  \nonumber\\
&&~~~~~~~~~~~~~~~~~~~~~~~~~+ \frac{v_d^2}{\sqrt{2} v_t} \delta_{\mu 2} + {\cal O}(\delta_{\mu 2}^2, v_t^4/v_d^2) 
\end{eqnarray}

\noindent
where $\delta_{\mu 2} \equiv \mu - \mu_c^{(2)}$.

\end{itemize}

\noindent
Noting that $\mu_c^{(1)} - \mu_c^{(2)}$, $m^2_{(2)} - m^2_{(1)}$ and ${\cal V}_\lambda$ have the same sign,
and defining

\begin{eqnarray}
\mu_c^{\rm min} &\equiv& \min \{\mu_c^{(1)}, \mu_c^{(2)} \} \label{eq:mucmin} \\
\mu_c^{\rm max} &\equiv& \max \{\mu_c^{(1)}, \mu_c^{(2)} \}  \label{eq:mucmax}
\end{eqnarray}
one can recast the results of Eqs.(\ref{eq:h0max}, \ref{eq:H0min}, \ref{eq:h0maxp}, \ref{eq:H0minp}) in a more
compact form as

\begin{eqnarray}
{m_{h^0}^2}^{\rm max} &=& m_{h^0}^2(\mu = \mu_c^{\rm max}) = \min \{ m^2_{(1)}, m^2_{(2)} \} \label{eq:CPevenbounds1} \\
&&  \nonumber \\
{m_{H^0}^2}^{\rm min} &=& m_{H^0}^2(\mu = \mu_c^{\rm min}) = \max \{ m^2_{(1)}, m^2_{(2)} \} \label{eq:CPevenbounds2}
\end{eqnarray}

\noindent
with an implicit reference to the two regimes {\sl i)} and {\sl ii)} if one keeps in mind that 
$m^2_{(i)}$ is reached for $\mu = \mu_c^{(i)}$.\\

\noindent
\underline{\sl The mixing pattern}: 
for $\mu= \mu_c^{(1)}$,  $h^0$
and $H^0$ become pure doublet or triplet states, since in this case $B=0$ as can be seen from Eq.(\ref{eq:ABC}). 
However, a close inspection of Eq.(\ref{eq:sa})  shows that in the regime 
{\sl i)} (resp. {\sl ii)}) one has $s_\alpha=0$ (resp. $s_\alpha=1$) for this value of $\mu$. 
Thus, at $\mu= \mu_c^{(1)}$, $h^0$ becomes a pure SM-like Higgs in regime  {\sl i)}, but it is 
$H^0$ that becomes a pure SM-like Higgs in regime {\sl ii)}. The fact that the SM-like state is not always associated 
with the lightest ${\mathcal{CP}}_{even}$ state is important when discussing the Higgs phenomenology and 
interpretation of the experimental limits and is consistent with the fact that $m^2_{(1)}$ is 
indeed the SM Higgs squared mass, Eq.(\ref{eq:h0max}).  In fact, due to the smallness of $v_t/v_d$ 
the behavior of  the mixing angle $\alpha$ over the full range of the $\mu$ parameter follows closely the generic  
pattern discussed above: in both regimes 
{\sl i)} and {\sl ii)} one has essentially  
$s_\alpha \simeq \pm 1$  or $s_\alpha \simeq 0$ over most of the $\mu$ range, except for a very narrow region 
in the vicinity of $\bar{\mu}$ defined in footnote \ref{footnote:mubar} and satisfying  
\begin{equation} 
\bar{\mu} = \frac12 (\mu_c^{(1)}+ \mu_c^{(2)}) \label{eq:mubar} 
\end{equation}

\noindent
 where $|s_\alpha|$ changes quickly from  $\simeq 0$ to
$\simeq 1 $. 
The generic dominance of no-mixing regimes can be understood from the asymptotic behavior at 
small and large $\mu$ values, i.e. $ |\sin \alpha_{|\mu \to 0}| = 1 - 2 \frac{(\lambda_1 + \lambda_4)^2}{\lambda^2} 
(v_t^2/v_d^2) + {\cal O}(v_t^3)$ and 
$\sin \alpha_{| \mu \to \mu_{+}} = 2 (v_t/v_d) + {\cal O}(v_t^2)$, together with the fact that
$d s_\alpha/d \mu = {\cal O}(v_t^3)$. We illustrated this behavior in Fig.~\ref{fig:mixingangle} adopting the sign 
convention $\epsilon_\alpha= +1$. 
As seen in Fig.~\ref{fig:mixingangle}.b, $s_\alpha$ remains positive in all the 
$\mu$ range since $B <0$ ({\sl cf.} Eq.(\ref{eq:ABC}, \ref{eq:sa})). And in accordance with the asymptotic behavior, 
$s_\alpha$ tends to ${\cal O} (10^{-2})$ at large $\mu (> \bar{\mu})$ where $h^0$ is nearly SM-like, and to 
${\cal O} (1)$ at small $\mu (< \bar{\mu})$ where $H^0$ is nearly SM-like. [Note that in this numerical example
$\mu_c^{(1)}$ becomes negative and is never reached]. In contrast, for the regime illustrated
in Fig.~\ref{fig:mixingangle}.a, 
$s_\alpha$ remains negative for $\mu < \mu_c^{(1)}$, crosses zero at $\mu_c^{(1)}$ and again tends to a 
positive value ${\cal O} (10^{-2})$ for $\mu \gg \mu_c^{(1)}$.

The exact magnitude of $|s_\alpha|$ at the three critical values of $\mu$ can be summarized as follows:

\begin{eqnarray}
\begin{array}{rcccc}
&~~& {\cal V}_\lambda > 0:  &~~& {\cal V}_\lambda < 0:  \\
&&&&\\
|s_\alpha(\mu= \mu_c^{(1)})| = &~~&  0   &~,~&  1  \\
&&&& \\
|s_\alpha(\mu= \bar{\mu})| =  &~~& \displaystyle  \left(\frac{1}{2} - \frac{2 v_t}{\sqrt{v_d^2 + 16 v_t^2}} \right)^{1/2} 
&~,~&  \displaystyle \left(\frac{1}{2} + \frac{2 v_t}{\sqrt{v_d^2 + 16 v_t^2}} \right)^{1/2} \\
&&&&\\
= &~~&  \displaystyle \frac{1}{\sqrt{2}} - \sqrt{2} \frac{v_t}{v_d} + {\cal O}(\frac{v_t^2}{v_d^2}) 
&~,~& \displaystyle \frac{1}{\sqrt{2}} + \sqrt{2} \frac{v_t}{v_d} + {\cal O}(\frac{v_t^2}{v_d^2})  \\
&&&& \\
|s_\alpha(\mu= \mu_c^{(2)})| = &~~&  \displaystyle \frac{v_d}{\sqrt{v_d^2 + 16 v_t^2}} 
&~,~&  \displaystyle \frac{4 v_t}{\sqrt{v_d^2 + 16 v_t^2}}  \\
&&&&\\
= &~~&  \displaystyle 1 - 8 \frac{v_t^2}{v_d^2} + {\cal O}(\frac{v_t^3}{v_d^3})
&~,~& 4 \displaystyle \frac{v_t}{v_d}  + {\cal O}(\frac{v_t^3}{v_d^3}) \\
\end{array}
\end{eqnarray}

\noindent
Large mixing scenarios have been discussed previously
in \cite{Dey:2008jm,Akeroyd:2010je} while here we quantify more precisely the
regions where such a large mixing takes place.\\
For later analyses it is useful to characterize the $\mu$ range in the large $|s_\alpha|$ regime. 
One sees from the above equations that the size of this range is ${\cal O}(v_t)$.
As a first approximation one can characterize it by the interval $0< \mu < \mu_c^{\rm min}$, with
$\mu_c^{\rm min}$ given by Eq.~(\ref{eq:mucmin}). However, depending on the values
of $\lambda$ and $\lambda_1 + \lambda_4$, $|s_\alpha|$ can still be very close to 1 in the range
$\mu_c^{\rm min} < \mu < \bar{\mu}$, especially that $\mu_c^{\rm min}$ is not positive definite [it 
becomes negative when $\lambda_1 + \lambda_4 < 0$ or $2 \lambda - (\lambda_1 + \lambda_4) < 0$]. 
It is more sensible to base this characterization on the amount of deviation from the value $|s_\alpha|=1$.
Defining $\hat{\mu}$ in the vicinity of $\bar{\mu}$ in the form 
$\hat{\mu} \equiv \bar{\mu} - \delta {\scriptstyle \times} v_t$,
with $\delta$ strictly $> 0$, one finds 
$\displaystyle |s_\alpha(\hat{\mu})| = 1 - k(\delta) \frac{v_t^2}{v_d^2} + {\cal O}(\frac{v_t^3}{v_d^3})$.
For each given {\sl positive} value of $k$ there corresponds a value of $\hat{\mu}$ given by

\begin{eqnarray}
\hat{\mu}_{(\pm)}&=& (\frac{\lambda}{\sqrt{2}} - \frac{\lambda - \lambda_1 - \lambda_4}{\sqrt{2} \pm \sqrt{k}}) v_t +
{\cal O}(\frac{v_t^3}{v_d^2}) \label{eq:muhat}
\end{eqnarray}

\noindent
The two-fold ambiguity in this expression is resolved as follows: 
requiring consistently $\mu_c^{\rm min} \leq \hat{\mu} < \bar{\mu}$ to hold, one should take
 for ${\cal V}_\lambda > 0$,  
$\hat{\mu} = \hat{\mu}_{(-)}$ with $k\geq 8$, and for ${\cal V}_\lambda < 0$,  
$\hat{\mu} = \hat{\mu}_{(+)}$ for any $k \geq 0$.\footnote{Strictly speaking, in the case  ${\cal V}_\lambda > 0$
one can still choose $k$ in the interval $2 < k < 8$ if ${\cal V}_\lambda$ is sufficiently close
to zero so that to ensure that $\mu_c^{\rm min} \leq \hat{\mu}$. In practice these details will not be important,
since one does not expect an experimental sensitivity to the deviation from $|s_\alpha|=1$ to be better than a few
percent. A deviation of $1 \%$, with $v_t=1$GeV, puts the value of $k$ already around 600 !}   
In particular $\hat{\mu}$ reproduces respectively $\mu_c^{(1)}$ and $\mu_c^{(2)}$ 
for the special values $k=0$ and $k=8$ as expected, while $\bar{\mu}$ cannot be reached for any finite value
of $k$ [consistently with the fact that $|s_\alpha(\hat{\mu})| \simeq 1$ and $|s_\alpha(\hat{\mu})| \simeq 1/\sqrt{2}$
are not perturbatively close to each other in terms of powers of $v_t/v_d$].

With the above prescription one can 
characterize the $\mu$ range in the large $|s_\alpha|$ regime by $0 < \mu < \hat{\mu}(k)$, where $k$ can now be
interpreted as triggering the experimental sensitivity to the deviation of $|s_\alpha|$ from its maximal value
$|s_\alpha| = 1$.
Equation (\ref{eq:muhat}) shows that the lower the sensitivity to  large $|s_\alpha|$ (i.e. the larger $k$), the lower
the sensitivity of the size of the $\mu$ domain to $\lambda_1 + \lambda_4$. We will come back to the above issues
in the phenomenological discussion of section \ref{sec:higgspheno}. \\


\begin{figure}[t]
\begin{picture}(200,200)
\begin{minipage}[b]{0.5\linewidth}
\includegraphics[scale=.8, width=7cm, height=7cm]{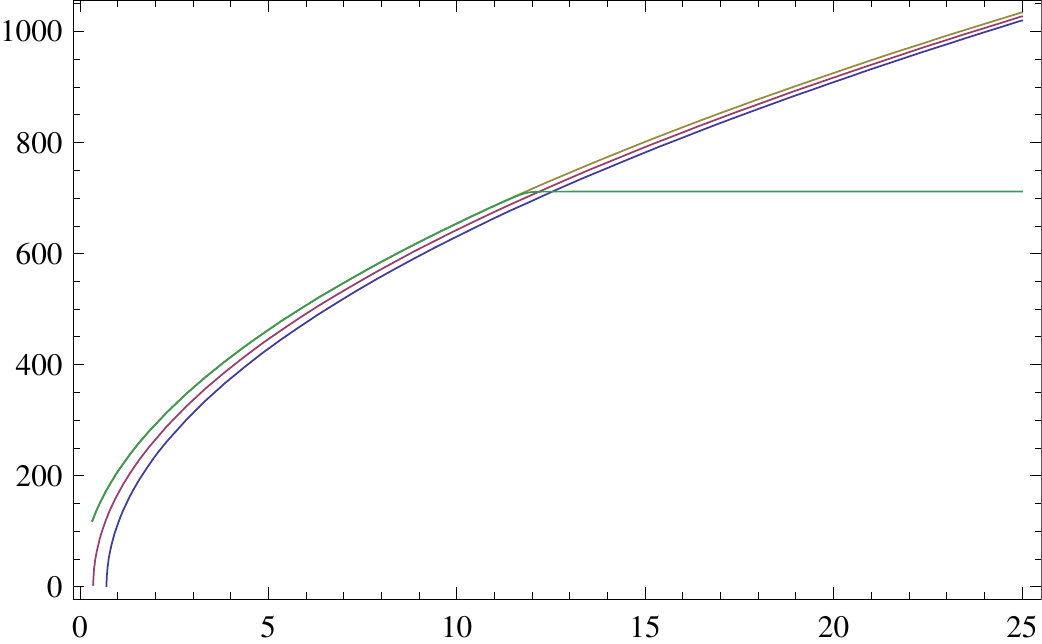}
\put(-220,50){\makebox(0,0)[bl]{\rotatebox{90}{{$m_{h^0, A^0, H^\pm, H^{\pm\pm}}$ \;{(GeV)}}}}}
\put(-120, -10){$\mu$ (GeV)}
\put(-180, 70){\makebox(0,0)[bl]{\rotatebox{45}{{${}^{h^0, A^0}$}}}}
\put(-180, 50){${}_{H^\pm}$}
\put(-175, 30){${}_{H^{\pm\pm}}$}
\put(-70, 130){${}_{h^0}$}
\put(-45, 180){${}_{A^0}$}
\put(-50, 170){${}_{H^\pm}$}
\put(-65, 155){${}_{H^{\pm\pm}}$}
\end{minipage}
\hspace{0.5cm}
\begin{minipage}[b]{0.5\linewidth}
\centering
\includegraphics[scale=.78, width=7cm, height=7cm]{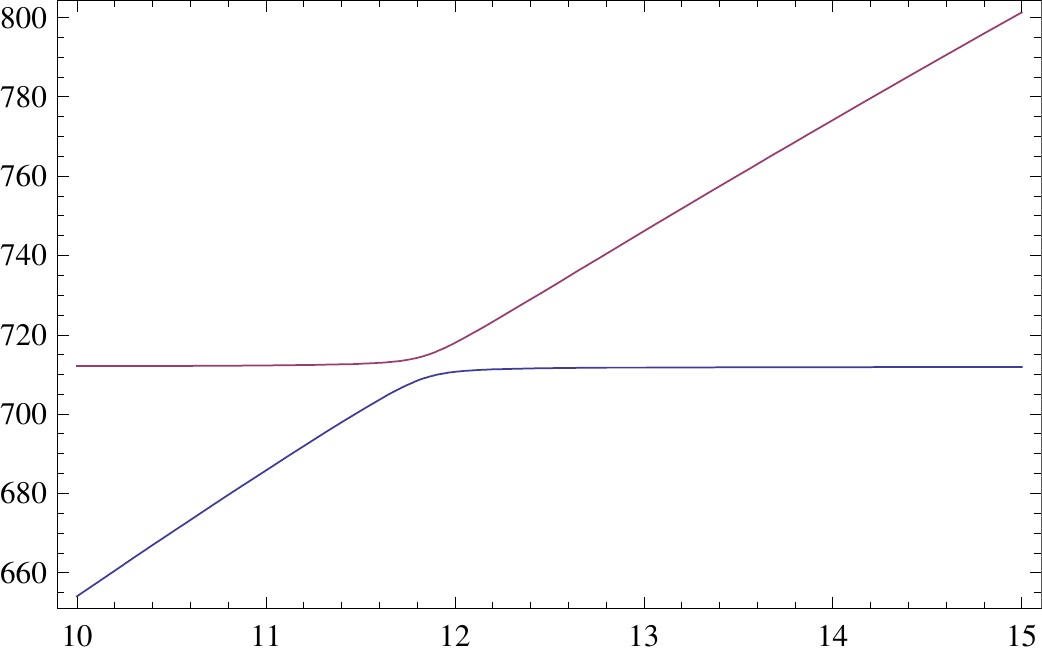}
\put(-220,70){\makebox(0,0)[bl]{\rotatebox{90}{{$m_{h^0, H^0}$ \;{(GeV)}}}}}
\put(-120, -10){$\mu$ (GeV)}
\put(-65, 155){${}_{H^{0}}$}
\put(-125, 70){${}_{h^{0}}$}
\end{minipage}
\end{picture}
\caption{Illustration of the regime ${\cal V}_\lambda < 0$ with 
$\lambda = \frac{16 \pi}{3},\lambda_2 = 10^{-1},\lambda_3 = 2 \times 10^{-1}, \lambda_1= -\frac{1}{2},
\lambda_4 = 1,v_t = 1 ~{\rm GeV},  
v= 246 ~{\rm GeV}, v_d=\sqrt{v^2-2 v_t^2}, \kappa=8$, leading to $\mu_c^{(2)} \simeq 23$ GeV. see Eq.~(\ref{eq:mc2}).
}
\label{fig:fig2}
\end{figure}
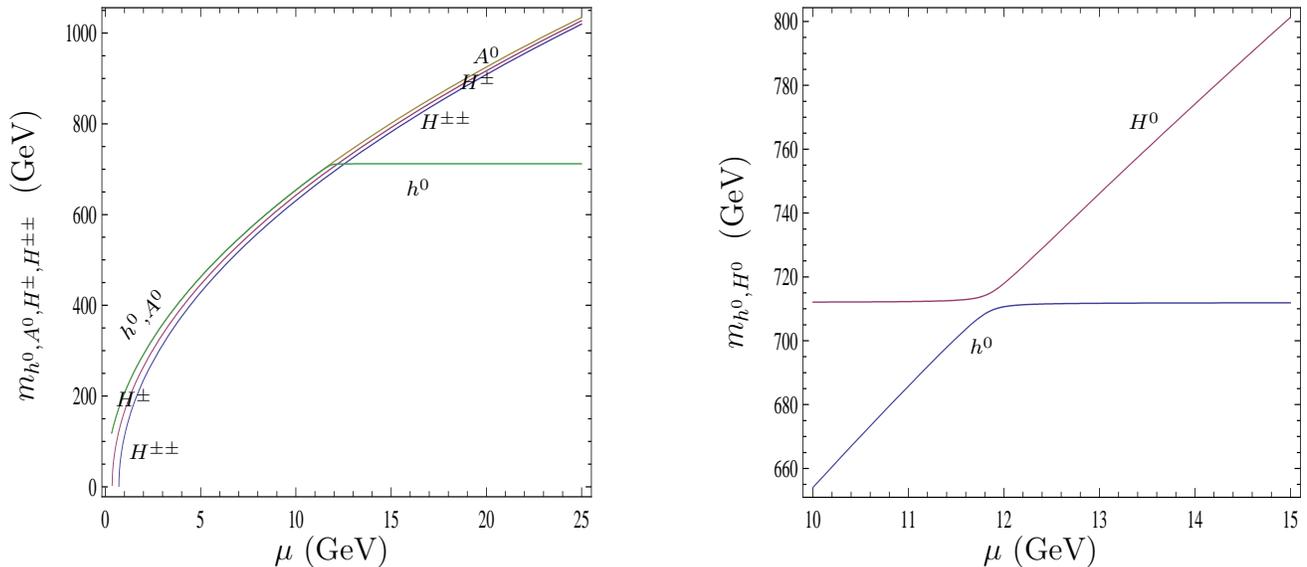

\begin{figure}[t]
\begin{picture}(200,200)
\begin{minipage}[b]{0.5\linewidth}
\includegraphics[scale=.72, width=6cm, height=6cm]{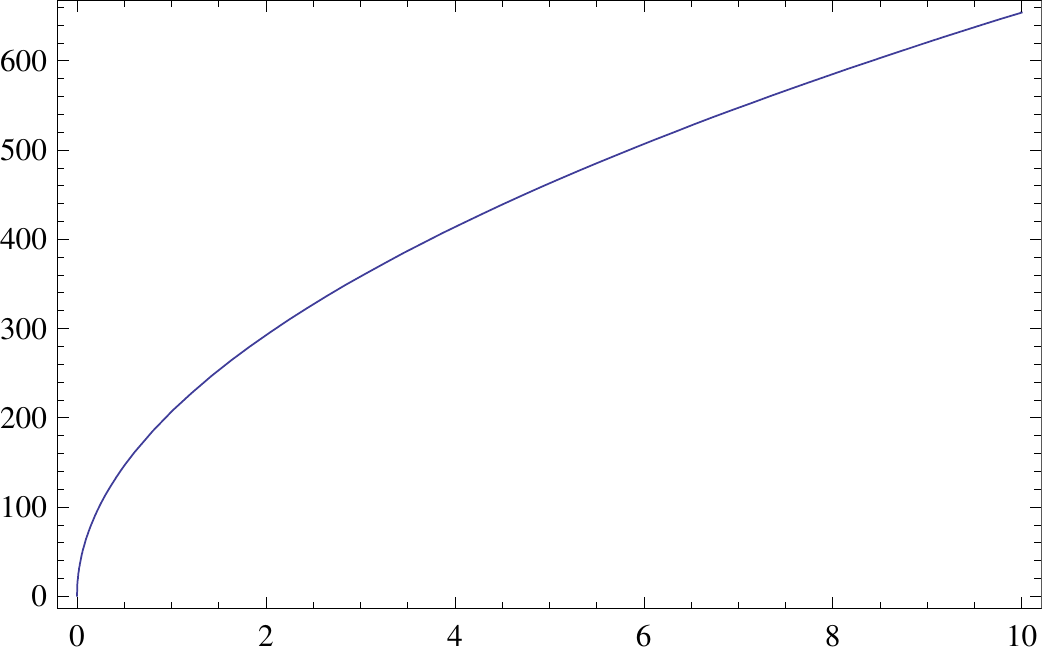}
\put(-190,50){\makebox(0,0)[bl]{\rotatebox{90}{{$m_{h^0}$ \;{(GeV)}}}}}
\put(-120, -10){$\mu$ (GeV)}
\end{minipage}
\hspace{0.5cm}
\begin{minipage}[b]{0.5\linewidth}
\centering
\includegraphics[scale=.60, width=6cm, height=6cm]{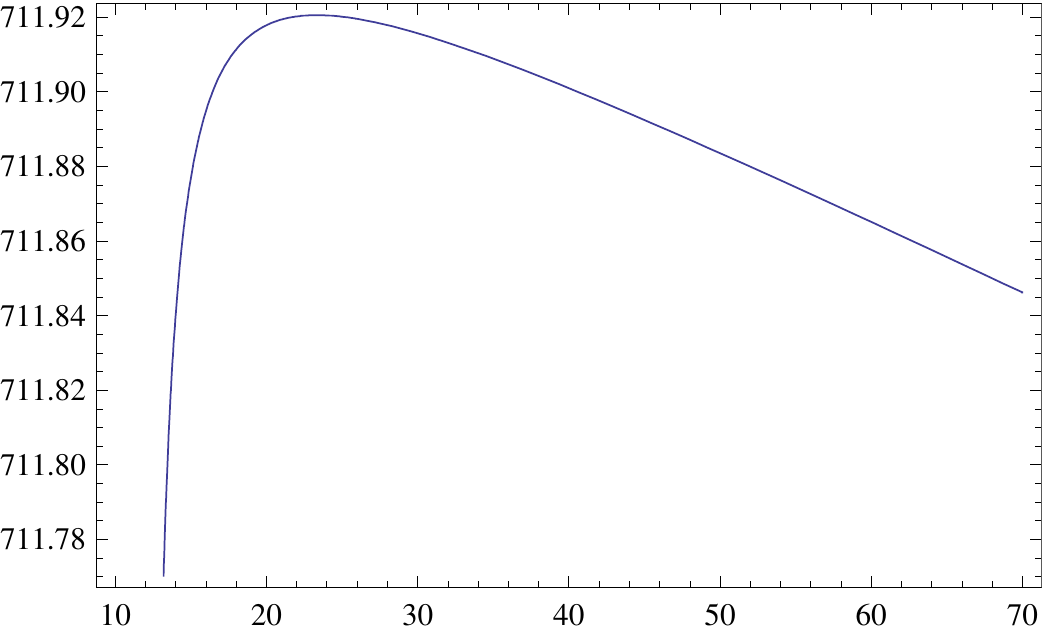}
\put(-190,50){\makebox(0,0)[bl]{\rotatebox{90}{{$m_{h^0}$ \;{(GeV)}}}}}
\put(-120, -10){$\mu$ (GeV)}
\end{minipage}
\end{picture}
\caption{Zoom on the variation of $m_{h^0}$ with $\mu$ in the vicinity of $\mu_c^{(2)}$; 
$\lambda = \frac{16 \pi}{3},\lambda_2 = 10^{-1},\lambda_3 = 2 \times 10^{-1}, \lambda_1= -\frac{1}{2},
\lambda_4 = 1,v_t = 1 ~{\rm GeV},  
v= 246 ~{\rm GeV}, v_d=\sqrt{v^2-2 v_t^2}, \kappa=8$, leading to $\mu_c^{(2)} \simeq 23$ GeV. see Eq.~(\ref{eq:mc2}).}
\label{fig:fig1}
\end{figure}

\noindent
\underline{\sl Unitarity bounds}: relying on the above properties we can now easily derive the theoretical upper 
bounds on the various Higgs masses. 
From Eq.~(\ref{eq:h0max}), and using the maximal value allowed by the tree-level unitarity
constraint for $\lambda$, Eq.~(\ref{eq:unit4new}), and $v_d \simeq 246$GeV, we determine an upper bound on
$m_{h^0}$,

\begin{eqnarray}
 m_{h^0} &\lesssim&   712 \, {\rm GeV} \; ({\rm for} \; \kappa=8) \\
         &\lesssim&   1 \, {\rm TeV} \; ({\rm for} \; \kappa=16)  \label{eq:unitarity8}
\end{eqnarray}

\noindent
If Eq.~(\ref{eq:h0maxp}) is used instead, then the saturation of unitarity and BFB bounds on $\lambda_1 + \lambda_4$
should be also considered. However, due to the smallness of $v_t/v_d$, this would lead to only a few GeV change in the 
above upper bounds. As far as $m_{H^0}$ is concerned, the above bounds are essentially the minimally allowed values,  
as obvious from Eq.~(\ref{eq:CPevenbounds1}, \ref{eq:CPevenbounds2}), in the unitarity saturation limit. To obtain its theoretical upper
bound as well as those of the other Higgs masses, one should rather take $\mu$ at its maximally allowed value,
$\mu_{\rm max} \simeq \mu_{+} \simeq
\frac{\lambda}{4 \sqrt{2}} \frac{v_d^2}{v_t}$,
since all these masses increase monotonically with $\mu$. For instance, with the set of parameters chosen in 
Figs.~\ref{fig:fig2}, \ref{fig:fig1}
and $v_t=1$ GeV one finds the upper bounds

\begin{eqnarray}
 m_{H^{\pm \pm}} \simeq m_{H^{\pm}} \simeq m_A \simeq m_{H^0} &\simeq& 88 \,{\rm TeV} \; ({\rm for} \; \kappa=8) \\
 &\simeq& 124 \, {\rm TeV} \;  ({\rm for} \; \kappa=16 )
\end{eqnarray}

\noindent
which are not phenomenologically compelling. Actually, somewhat lower bounds are obtained when taking into
account the experimental exclusion limits on a light Higgs $m_{h^0}$ but remain too high to be useful.
In contrast,  phenomenologically interesting 
scenarios with light charged, doubly charged, ${\mathcal{CP}}_{odd}$ and ${\mathcal{CP}}_{even}$ Higgses are possible 
for small values of $\mu$. For instance as illustrated in 
Figs.~\ref{fig:fig2}, \ref{fig:fig1},
such a light spectrum occurs when $\mu \ll \mu_c^{(2)} \simeq 23$ GeV. More generally, the analytical expressions 
given above  for the mass splittings show that in the vicinity of $\mu_c$ and in particular for
$\mu < \mu_c$ the neutral ${\mathcal{CP}}_{even}$ $h^0$ is not necessarily the lightest Higgs.\footnote{We have kept in these expressions
subleading terms of ${\cal O}(v_t^2)$ in order to handle as well the small parts of the $\lambda_i$'s parameter space 
where the leading ${\cal O}(v_d^2)$ are suppressed.}  The detailed patterns will depend
on the actual values of the $\lambda$'s and will be studied more thoroughly in the next section, but one can 
already see some generic features in regimes {\sl i)} and {\sl ii)} at $\mu \simeq \mu_c$.  
In regime {\sl i)} where $-\lambda + \lambda_1 + \lambda_4 >0$ one expects $H^{\pm \pm}$ to become the lightest Higgs 
if $-\lambda + \lambda_1 < 0$, that is when $  \lambda_1   <\lambda < \lambda_1 + \lambda_4$.
Similarly, in regime {\sl ii)} where typically $\lambda - \lambda_1 - \lambda_4 >0$ one again expects $H^{\pm \pm}$
to be the lightest Higgs when $  \lambda_1 +\lambda_4  <\lambda < \lambda_1 + 2 \lambda_4$. More generally,
a close inspection of Eqs.(\ref{eq:mHpmpm}, \ref{eq:mh0}) shows that $m_{H^{\pm\pm}} < m_{h^0}$ 
when $\mu < \mu^\star= (\lambda + \lambda_4) v_t/\sqrt{2} +  {\cal O}(v_t^3/v_d^2)$,  
and only if $\lambda_4>0$.\footnote{
Although this expression of $\mu^\star$ is well-defined for $\lambda_4<0$, one finds that
the splitting $m_{H^{\pm\pm}}^2 - m_{h^0}^2$ is negative only in the domain defined 
by $\mu < (\lambda + 2 \lambda_4) v_t/\sqrt{2} + {\cal O}(v_t^3/v_d^2)$  and 
$-\lambda_4 v_d^2/(4 \sqrt{2} v_t)  + {\cal O}(v_t) <  \mu <  (\lambda + \lambda_4) v_t/\sqrt{2} +  
{\cal O}(v_t^3/v_d^2)$, which is clearly non-empty only for $\lambda_4>0$.} Furthermore, it immediately follows
from Eqs.(\ref{eq:mHpm}, \ref{eq:mA0}) that $m_{H^{\pm\pm}} < m_{H^{\pm}} < m_A$ when $\lambda_4>0$ 
so that the necessary and sufficient condition for $H^{\pm\pm} $ to be the lightest Higgs is
\begin{equation}
\mu < \mu^\star \;\; {\rm with} \;\; \lambda_4>0. 
\label{eq:H++lightest}
\end{equation}

\noindent
\underline{\sl Phenomenological bounds}: in order to prepare for a phenomenological study, we discussed in 
section \ref{sec:tachyon} the modification on the tachyonic
bounds of $\mu$ when experimental exclusion limits are available for $m_{A^0}, m_{H^\pm}$ and $m_{H^{\pm \pm}}$, 
{\sl cf.} Eqs.(\ref{eq:mumin}, \ref{eq:mumumin}). Here we address the same question concerning $m_{h^0}$ and $m_{H^0}$. 
Given some experimental exclusion {\sl lower} bounds  $(m_{h^0})_{\rm exp}$, (resp. $(m_{H^0})_{\rm exp})$ there corresponds
two values $\mu_{\pm}^{h^0}$ (resp. $\mu_{\pm}^{H^0}$), namely

\begin{eqnarray}
\displaystyle
\mu_{\pm}^{h^0} &=& \frac{1}{{8\sqrt2 \upsilon_t}}(\lambda \upsilon_d^2+8(\lambda_1+\lambda_4 )\upsilon_t^2  
-2 (m_{h^0}^2)_{\rm exp}  \nonumber \\
&&\pm \; 2 \, [(m^2_{(1)} - (m_{h^0}^2)_{\rm exp}) 
(m^2_{(2)} - (m_{h^0}^2)_{\rm exp})]^{\frac12}(1 + 16 \frac{\upsilon_t^2}{\upsilon_d^2})^{\frac12}) 
\label{eq:solmuposexp}
\end{eqnarray}

\noindent
(and similarly for $\mu_{\pm}^{H^0}$ with  $(m_{h^0})_{\rm exp}$ replaced by $(m_{H^0})_{\rm exp}$ ),
for which $m_{h^0}$ reaches $(m_{h^0})_{\rm exp}$   (resp.  $m_{H^0}$ reaches $(m_{H^0})_{\rm exp}$). 
Note that in the limit of no experimental bounds, i.e.  $(m_{h^0, H^0}^2)_{\rm exp} \to 0$, Eq.(\ref{eq:solmuposexp})
gives back  Eq.(\ref{eq:solmupos}).  
Furthermore, relying on the fact that $m_{h^0}$ has a maximum and $m_{H^0}$ has a minimum as functions of $\mu$, {\sl cf.} 
Eq.~(\ref{eq:CPevenbounds1}, \ref{eq:CPevenbounds2}),  the phenomenological bounds read  

\begin{eqnarray}
\mu_{-}^{h^0} \leq \mu \leq \mu_{+}^{h^0}  \; \; &{\rm assuming}& \; \; (m_{h^0})_{\rm exp} \leq m_{h^0}^{\rm max} 
\nonumber \\
& {\rm and} & \label{eq:solmuposexpTOT} \\
\mu \leq \mu_{-}^{H^0}  \; {\rm or} \;  \mu_{+}^{H^0} \leq \mu   
\; \; &{\rm assuming}& \; \; (m_{H^0})_{\rm exp} \ge m_{H^0}^{\rm min} \nonumber 
\end{eqnarray}

\noindent
Obviously $(m_{h^0})_{\rm exp} > m_{h^0}^{\rm max}$ would be an inconsistent assumption, while
$(m_{H^0})_{\rm exp} < m_{H^0}^{\rm min}$ would be an empty assumption not leading to any constraint
as far as $\mu$ is concerned. 


In summary, the experimental lower bounds on the various Higgs masses will typically constrain the $\mu$
parameter to lie in a finite domain defined by the combination of Eqs.(\ref{eq:mumumin}, \ref{eq:solmuposexpTOT}).




\setcounter{equation}{0}

\section{Higgs phenomenology} 
\label{sec:higgspheno}

Although previous studies in the literature assumed typically the triplet mass 
$M_{\Delta}$ and the mass parameter $\mu$ to be much larger than the electroweak scale, $M_\Delta \gg v_d$, 
attention has been paid more recently to the possibility 
of  having  $M_{\Delta}, \mu \lesssim 1$ TeV where the Higgses of 
the DTHM might be accessible at the Tevatron and the LHC 
\cite{Perez:2008ha,Akeroyd:2007zv,delAguila:2008cj,Akeroyd:2009nu,
Fukuyama:2009xk,Akeroyd:2009hb,Petcov:2009zr, Fukuyama:2010mz, Akeroyd:2010ip}. 
In this spirit, the results obtained in the previous sections help defining educated strategies to extract
constraints on the physical Higgs masses and model parameters from experimental data, rather than performing merely 
blind (and CPU time consuming) scans on these parameters. The existing experimental exclusion limits on the SM 
Higgs particle are readily translated into constraints on the DTHM in the parameter space region where
$h^0$ becomes SM-like, i.e. when the mixing between the doublet and the triplet
is very small. However, even when far from this region, existing exclusion limits for an extended Higgs sector 
(such as in two Higgs doublets models or in the minimal supersymmetric extension of the SM) can also be partially 
adapted to $h^0$, $H^0$, $A^0$ and $H^\pm$,  while of course $H^{\pm \pm}$ has a distinctive experimental search.  
 
In this section we give a quick overview of the Higgs sector phenomenology and experimental searches
(for an extended overview on the phenomenology of triplet models see Ref.~\cite{Accomando:2006ga}),
followed by a preliminary analysis using our results.
A detailed study taking into account all present-day experimental limits lies out of the scope of this paper and 
will be presented elsewhere.


\subsection*{Doubly charged Higgs}
Observation of the doubly charged Higgs $H^{\pm\pm}$ would 
signal unambiguously physics beyond Higgs doublets, let alone physics beyond the SM Higgs sector.
Owing to charge conservation, it is obvious that  $H^{\pm\pm}$ 
cannot couple to a pair of quarks, therefore, its possible decay modes are:

\begin{itemize}
\item[i)] same sign charged lepton pair $H^{\pm\pm}\to l^\pm l^\pm$ 
          that proceed via lepton number violating coupling, 
\item[ii)] a pair of $W^\pm$ gauge bosons $H^{\pm\pm}\to W^\pm W^\pm $,
\item[iii)] $H^{\pm\pm}\to W^\pm H^\pm $,
\item[iv)] a pair of charged Higgs bosons $H^{\pm\pm}\to H^\pm H^\pm $.
\end{itemize}
We emphasize also that the doubly charged Higgs couples 
to the photon and  to the Z boson through gauge couplings
Eqs.~(\ref{eq:coup14}, \ref{eq:coup15}), while its 
couplings to a pair of $W^{\pm}$  is  
proportional to the triplet vev $v_t$, see Eqs.~(\ref{eq:coup12}).
Therefore the decay channel ii) will be suppressed for $v_t \ll v_d$. 
The decay channel iv) will also be suppressed for small $v_t$ as
can be seen from the form of the coupling of  $H^{\pm\pm}$ to a pair
charged Higgses $H^\pm$, Eq.~(\ref{eq:coup13}). Indeed, one has
$\cos\beta^{'} \simeq 1$ and $\sin\beta^{'} \sim v_t/v_d$ from Eq.~(\ref{eq:scbprime}),
and furthermore, the $\mu  \sin^2\beta^{'}$ is also of order $v_t$ due to the
$\mu$ upper bound $\mu_{+}  \sim  v_d^2/v_t$, {\sl viz} Eq.~(\ref{eq:solmupos}).
In contrast, the coupling $H^{\pm\pm} W^\pm H^\pm $ is proportional to the gauge
coupling and has no suppression factors. The decay channel ii) will thus contribute
substantially if kinematically open. 
Depending on the size of the Yukawa couplings of the leptons, the doubly 
charged Higgs can decay dominantly either to a pair of leptons or to $W^{\pm}$ and $H^{\pm}$,
and subdominantly to a pair of $W^{\pm}$ and/or a pair of $H^{\pm}$ if kinematically  allowed.

In $e^+e^-$ collisions, the doubly charged Higgs can be pair produced through 
$\gamma$ and $Z$ s-channel\footnote{the t-channel mediated by a lepton is in
general suppressed by the small Yukawa coupling} 
$e^+e^-\to \gamma^*,Z^*\to H^{\pm\pm} H^{\mp\mp}$ 
\cite{Akeroyd:1995ci,Gunion:1996pq,Gunion:1989ci,Chen:2008jh}. 
One can have also access
to the associate production of $H^\pm$ with $W^\mp$ 
through s-channel $Z$ exchange  
\cite{Ghosh:1996jg,Cheung:1994rp,Godbole:1994np}.
If the $e^-e^-$ option is
available at ILC, then the doubly charged Higgs can be produced in 
$W^\pm W^\pm$ fusion through $e^-e^-\to W^{-*}W^{-*}\to e^-e^-H^{++}$. 
Even if $H^{\pm\pm}W^\mp W^\mp$ has a $v_t$ suppression, the rate 
for $W^\pm W^\pm$ fusion could be substantial especially at higher energies
options for $e^-e^-$ \cite{Gunion:1996pq}.\\

At the Tevatron or the LHC, the two production 
mechanisms with potentially large cross sections are 
$p\bar{p}/pp\to \gamma^*,Z^*\to 
H^{\pm\pm} H^{\mp\mp}X$ or a single production through WW fusion 
$p\bar{p}/pp\to W^{\pm *}W^{\pm *}\to H^{\pm\pm} X$ 
\cite{Huitu:1996su,Azuelos:2005uc}. 
The latter process as well as the s-channel 
$p\bar{p}/pp\to W^{\pm *}\to W^{\mp}H^{\pm\pm}$
depend on the coupling $H^{\pm\pm}W^\mp W^\mp$ which is proportional to
the triplet vev. 
However, the suppression due to the small value of $v_t$ is somewhat 
compensated by the fact that $W^\pm W^\pm$ fusion could be 
substantial at high energy.
Those processes have to be supplemented by the associate production 
of singly and doubly charged Higgs bosons 
$p\bar{p}/pp\to H^{\pm\pm} H^{\mp}X$ which could have a comparable 
cross section to $p\bar{p}/pp\to  H^{\pm\pm} H^{\mp\mp}X$ 
\cite{Barger:1982cy}, \cite{Akeroyd:2005gt}.

Such doubly charged Higgs have been subject to many experimental searches.
At LEP-II, the experiments L3, OPAL and Delphi 
\cite{Achard:2003mv},\cite{Abbiendi:2001cr}, \cite{Abdallah:2002qj} performed a 
search for doubly charged Higgs boson assuming that $H^{\pm\pm}$ decay 
dominantly to a pair of leptons $H^{\pm\pm}\to l^\pm l^\pm$.
Four leptons final states have been analyzed at L3, OPAL and Delphi.
L3 performed a search for the six possibilities: $ee$, $\mu\mu$, $\mu e$, 
$\mu\tau$, $e\tau$ and $\tau\tau$.
No excess has been found and lower limits in the range 95-100 GeV 
 at 95\% confidence level on the doubly charged Higgs boson mass are derived.
Those lower limits depend on the doubly charged Higgs decay modes.
For example, if $H^{\pm\pm} \to e^\pm e^\pm$ is the dominant decay,
 then the lower limit is  100 GeV while if $H^{\pm\pm} \to \mu^\pm \tau^\pm$ 
is the dominant decay then the lower limit is about 95 GeV.\\
At the Tevatron, D$\emptyset$\cite{Abazov:2004au},\cite{:2008iy} 
and CDF \cite{Acosta:2004uj}, \cite{Aaltonen:2008ip} have searched for 
 $p\bar{p}\to \gamma^*,Z^*\to 
H^{\pm\pm} H^{\mp\mp}X$ with $H^{\pm\pm}\to l^\pm l^\pm$.
D$\emptyset$ measurement \cite{Abazov:2004au}
represents the first doubly charged Higgs 
search with the decay $H^{\pm\pm}\to \mu^\pm \mu^\pm$.
 Note that D$\emptyset$ search was limited to 
$H^{\pm\pm}\to \mu^\pm \mu^\pm$ that is an almost background free signal,  
while CDF explored the three final states
$e^\pm e^\pm, \mu^\pm \mu^\pm$ and $e^\pm \mu^\pm$.
Both D$\emptyset$ and CDF excluded a doubly charged Higgs with a mass
in the range $100\to 150$ GeV. We stress that all those bounds assume
 a 100\% branching ratio for $H^{\pm\pm}\to l^\pm l^\pm$ decay, while in
 realistic cases one can easily find scenarios where 
$H^{\pm\pm}\to l^\pm l^\pm$ is suppressed while 
$H^{\pm\pm}\to W^{\pm *} W^{\pm *}$ is substantial 
\cite{Perez:2008ha,Akeroyd:2007zv,Garayoa:2007fw, Kadastik:2007yd}
which could invalidate partially the CDF and D$\emptyset$ limits.
However, the LHC has the capability to extend the 
above limits up to a mass about 1 TeV for high luminosity 
option \cite{Perez:2008ha,delAguila:2008cj,Azuelos:2005uc,Rommerskirchen:2007jv}. Observation of 
doubly charged Higgs bosons at the LHC and measurement of its 
leptonic branching ratios will shed also some light on the neutrino mass pattern 
\cite{Perez:2008ha,Akeroyd:2007zv,
Chen:2008jh,Akeroyd:2005gt,Garayoa:2007fw,Kadastik:2007yd,Chun:2003ej,Ren:2008yi,Petcov:2009zr}.

Finally, indirect limits on the mass and the bileptonic couplings of the 
doubly charged Higgs boson can be extracted from  low energy lepton flavor violating  processes, such as $\mu\to e\gamma$, $\mu \to 3e$, $\tau \to 3l$,... 
 (see for instance \cite{Akeroyd:2009nu,Fukuyama:2010mz}).

\subsection*{Singly charged Higgs}
Let us now discuss briefly the couplings of the singly 
charged Higgs and its  decay modes. 
The charged Higgs coupling to lepton and neutrino is proportional
to $\approx m_\nu/v_t\approx Y_\nu$ \cite{Perez:2008ha} which could be of the 
order ${\cal{O}}(1)$ if $v_t$ is very small. 
Similarly, the charged Higgs coupling to a pair of quark $u$ and $d$ 
is proportional to $\tan\beta'$ which is suppressed by $v_t/v_d$ 
\cite{Perez:2008ha}. 
In the case of  $H^-\bar{t}b$, this coupling could enjoy some 
enhancement from Yukawa coupling of the top quark. 
The suppression of the coupling $H^-\bar{t}b$ has three consequences:

\begin{itemize}
\item Given the suppression factor of the order $v_t/v_d$ 
for $H^-\bar{t}b$, the charged Higgs mass can not be subject to 
$b\to s\gamma$ constraint, similarly to the two Higgs doublet 
model type I  where the coupling is suppressed by $1/\tan\beta$.

\item Some of the conventional mechanisms for charged Higgs production at
  Hadron colliders such as $bg\to tH^+$ and $gg\to tbH^+$ will be suppressed.

\item Since the charged Higgs search at the Tevatron is based on the top decay
  $t\to H^+b$, given the suppression of $H^+t\bar{b}$ coupling the branching
  ratio of $t\to b H^+$ would also be suppressed. One concludes then that  
the CDF limit does not apply in this case.
\end{itemize}

Besides those processes which are suppressed, 
one can still produce charged Higgses through the 
associate production of singly and doubly charged Higgs 
$pp/p\bar{p}\to W^*\to H^{\pm \pm}H^\mp$ 
\cite{Barger:1982cy,Akeroyd:2005gt} with a spectacular signature 
from $H^{\pm \pm}\to l^\pm l^\pm$. Other mechanisms are:
the Drell-Yan process 
$pp/p\bar{p}\to \gamma^*, Z^*\to H^\pm H^\mp$,  the 
associate production of charged Higgs and neutral Higgs 
$pp/p\bar{p}\to W^*\to H^{\pm}h^0$, 
$pp/p\bar{p}\to W^*\to H^{\pm}H^0$, 
$pp/p\bar{p}\to W^*\to H^{\pm}A^0$ and the associate production of
charged Higgs with W gauge boson $pp/p\bar{p}\to Z^*\to W^{\pm}H^\mp$.
Note that among the latter processes, the ones with $W^{\pm}H^\mp$
or $H^\mp h^0$ final states are suppressed by a $v_t/v_d$ factor
as compared to the Drell-Yan and the two other associate production processes that
are controlled by gauge couplings, cf. Eqs.~(\ref{eq:coup7}, \ref{eq:coup10}) and
Eqs.~(\ref{eq:coup8}, \ref{eq:coup9}).

If the charged Higgs decays dominantly to leptons (for small $v_t$) 
we can apply the LEP mass lower bounds that are of the order of 80 GeV 
\cite{Achard:2003gt}, \cite{Abdallah:2003wd}.
For large $v_t$, i.e. much larger than the neutrino masses but still well below the electroweak scale, 
 the dominant decay is either $H^+\to t\bar{b}$ or one of the
bosonic decays $H^+\to W^+Z$, $H^+\to W^+h^0/W^+A^0$ . For the first two decay modes there has been no
explicit search neither at LEP nor at the Tevatron, while for the $H^+\to W^+A^0$ decay (and possibly 
for $H^+\to W^+ h^0/$ if $h^0$ decays similarly to $A^0$), one can use the LEPII search performed
 in the framework of two Higgs doublet models. In this case the charged Higgs mass limit is again of the order 
of 80 GeV \cite{Abdallah:2003wd}.

\subsection*{Neutral Higgses}
The lighter ${\mathcal{CP}}_{even}$ Higgs boson $h^0$ is fully
dominated by the doublet component (i.e the mixing $|s_{\alpha}| \ll 1$) when $\mu > \bar{\mu}$, 
as discussed in section \ref{sec:higgsbounds} and illustrated on Fig.~\ref{fig:mixingangle}.
In this case the coupling of $h^0$ to a pair of neutrinos is suppressed 
being proportional to $s_\alpha$. 
Such Higgs will completely mimic the SM Higgs boson and then the
LEP and the recent Tevatron limits  would apply. In this scenario of very small
mixing, the other neutral Higgses $H^0$ and $A^0$ would be fermiophobic to 
all charged leptons and quarks 
 but their coupling to a pair of neutrinos that
is proportional to $\cos\alpha Y_\nu\approx Y_\nu=m_\nu/v_t$ 
could be enhanced for small $v_t$. Then the dominant decay
mode for $H^0$ and $A^0$, for small $v_t$, would be a 
pair of neutrinos \cite{Perez:2008ha}.

Note that $A^0$, being ${\mathcal{CP}}_{odd}$, does not couple to a pair of gauge bosons
while the couplings $H^0ZZ$ and $H^0WW$ in the small mixing case
 are suppressed by $v_t/v_d$, Eqs.(\ref{eq:coup2}, \ref{eq:coup4}). 
Thus the $W$ and $Z$ Higgsstrahlung productions of $H^0$ and $A^0$
are expected to be small.
Furthermore, while the ${H^0 A^0 Z}$ vertex is controlled by the gauge coupling, Eq.~(\ref{eq:coup6}),
$h^0A^0Z$ has an extra $v_t/v_d$ suppression, Eq.~(\ref{eq:coup5}). 
This implies that in the small mixing case, one can still produce $A^0$ and $H^0$
 through the Drell-Yan process $e^+e^-/pp/p\bar{p}\to Z^* \to H^0A^0$.
For very small $v_t$, $A$ and $H^0$ would decay essentially into a pair of neutrinos.
At LEP, the signal would then be a photon (from initial state
radiation) and missing energy in the  final state.  
A lower bound  on $m_H$ and $m_A$ of the order of 55 GeV can be extracted in this case from LEPII data,
assuming mass degeneracy between $A^0$ and $H^0$ \cite{Datta:1999nc}.
[In the non degenerate case the lower bound translates into 
$m_H+m_A\geq 110$ GeV.] Increasing $v_t$ well above the neutrino masses decreases significantly 
$H^0/A^0\to \nu\nu$, and  
the decay channels  $H^0\to b\bar{b}$, $A^0\to b\bar{b}$,  as well as
 $H^0\to ZZ$, $A^0\to Zh^0$ if open, see Eqs.~(\ref{eq:coup2}, \ref{eq:coup5}),  
can become dominant. If $H^0\to b\bar{b}$, $A^0\to b\bar{b}$ dominate, the LEPII Higgs search 
through $e^+e^-\to H^0A^0$ in the two Higgs doublet Model can apply to the DTHM in this
case, and the limit is roughly $m_H+m_A\ge 185$ GeV \cite{Abdallah:2004wy}.
There is however a distinctive feature in the DTHM related to the $H^0 h^0 h^0$ coupling,
Eq.~(\ref{eq:coup21}), the latter becoming substantial for increasing $\mu$ and thus for heavier $H^0$.
The $H^0\to h^0h^0$ decay mode would then be important in both the large and small (with respect to the neutrino
masses) $v_t$ regimes. 

In the case of maximal mixing $|s_\alpha| \approx 1$ which occurs for 
 $\mu < \bar{\mu}$, see Fig.~\ref{fig:mixingangle}, the roles of $h^0$ and $H^0$ are interchanged. 
$H^0$ is fully doublet and $h^0$ is fully triplet. Taking into account this  interchange, the previous discussion 
applies here to $H^0$. However, since $h^0$ remains the lighter Higgs which can be now far from SM-like, one expects 
weaker experimental constraints on its mass than the ones quoted above.  

\begin{figure}[t!] 
\begin{picture}(320,260)
\put(-18,-30){\mbox{\psfig{file=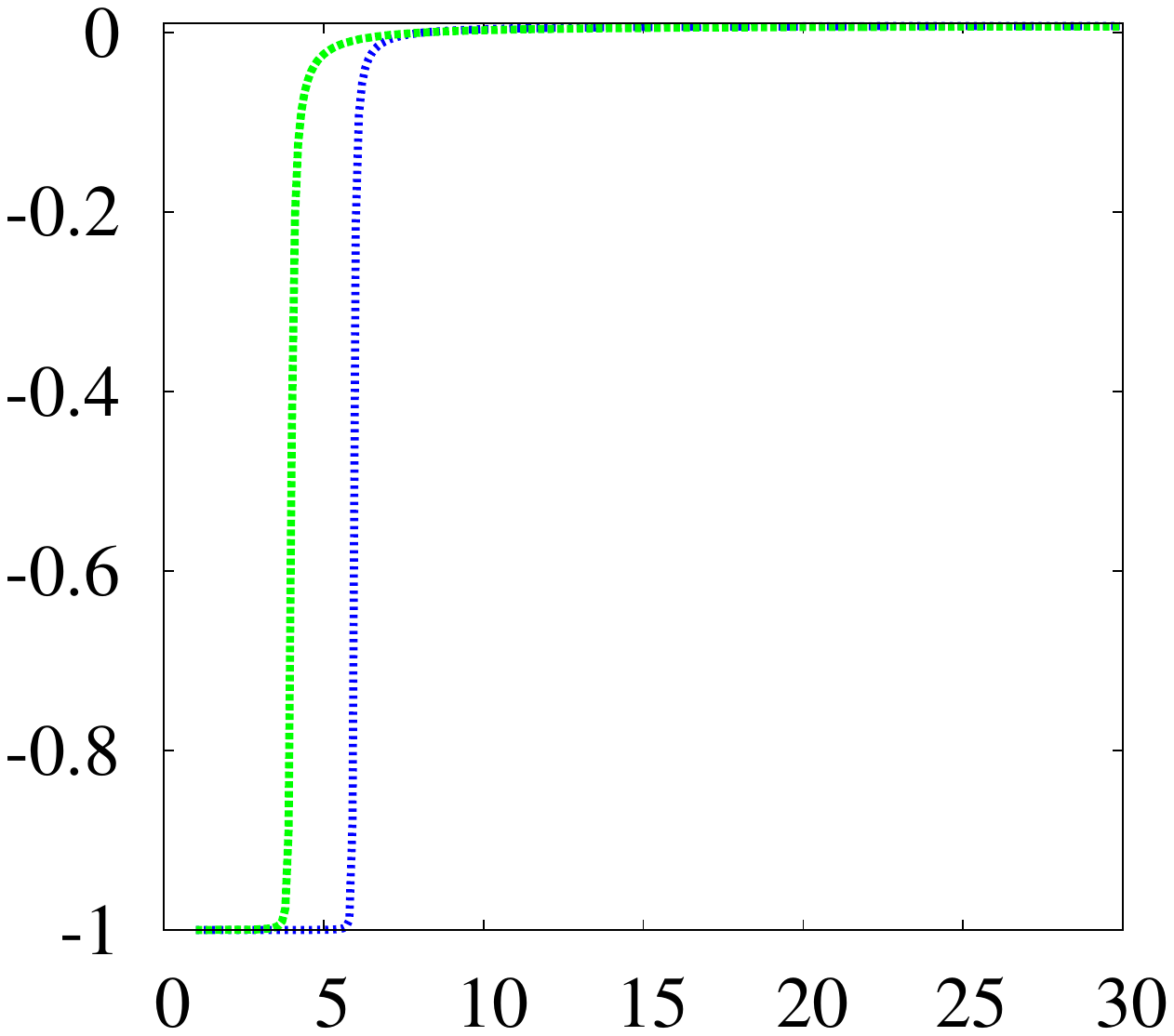,height=6.8in,width=4.6in}}}
\put(20,95){\makebox(0,0)[bl]{\rotatebox{90}{\large{$\sin \alpha$ }}}}
\put(120,-10){\makebox(0,0)[bl]{\large{$\mu$ (GeV)}}}
\put(105,160){\makebox(0,0)[bl]{{${\lambda} = \frac{\lambda_{\mbox{\tiny{max}}}}{2}$ }}}
\put(65,100){\makebox(0,0)[bl]{\large{$\frac{\lambda_{\mbox{\tiny{max}}}}{3}$ }}}
\put(120,40){\makebox(0,0)[bl]{\large{${\cal V_{\lambda}} > 0, \, \lambda_4 = 10$ }}}
\put(195,185){\makebox(0,0)[bl]{(a)}}
\put(250,0){\mbox{\psfig{file=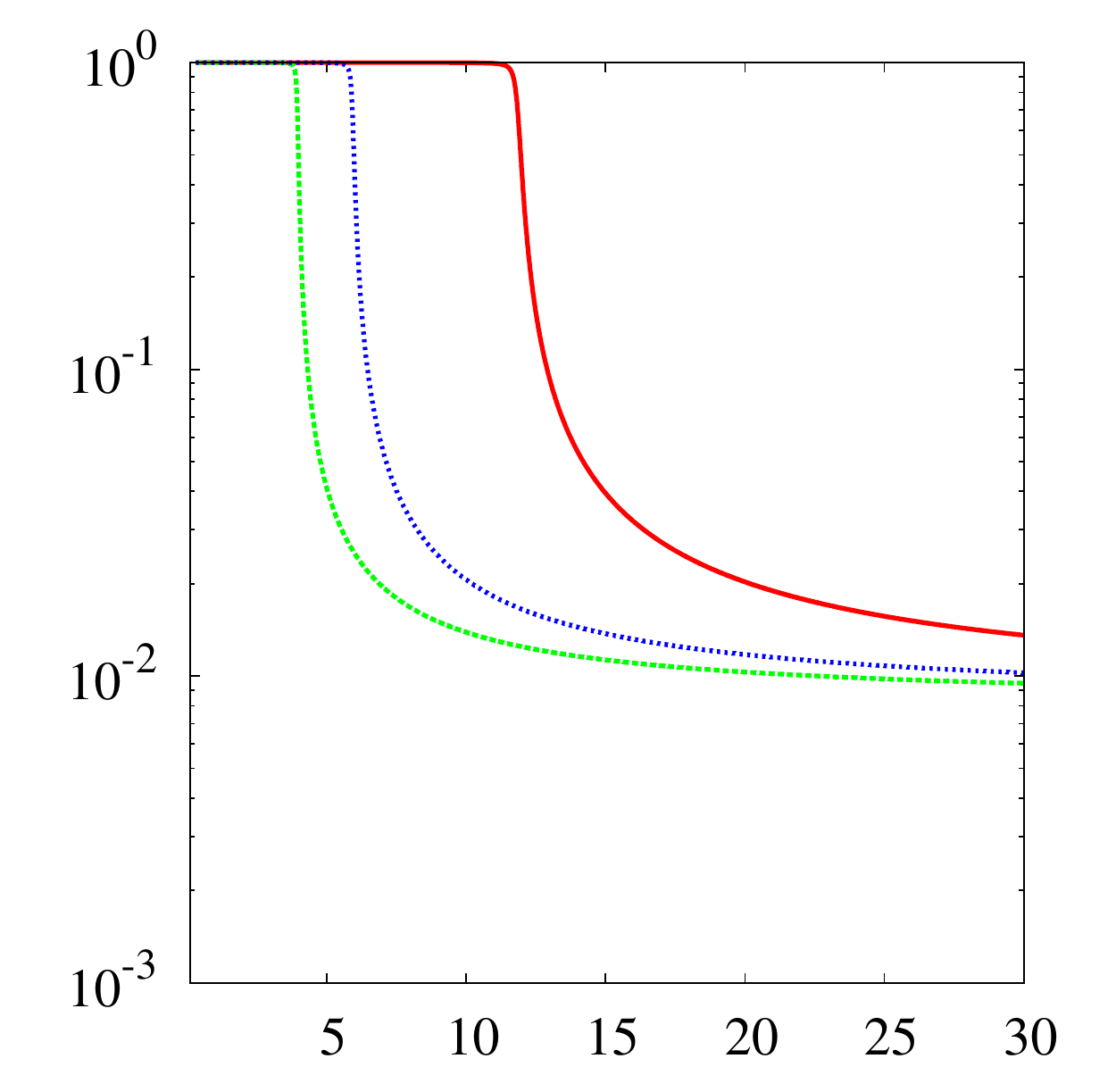,height=3.0in,width=3.0in}}}
\put(340,-10){\makebox(0,0)[bl]{\large{$\mu $ (GeV)}}}
\put(250,95){\makebox(0,0)[bl]{\rotatebox{90}{\large{$\sin \alpha$ }}}}
\put(355,180){\makebox(0,0)[bl]{{${\lambda} = \lambda_{\mbox{\tiny{max}}}$ }}}
\put(320,180){\makebox(0,0)[bl]{\large{$\frac{\lambda_{\mbox{\tiny{max}}}}{2}$ }}}
\put(290,100){\makebox(0,0)[bl]{\large{$\frac{\lambda_{\mbox{\tiny{max}}}}{3}$ }}}
\put(340,40){\makebox(0,0)[bl]{\large{${\cal V_{\lambda}} < 0, \, \lambda_4 = -1$ }}}
\put(420,185){\makebox(0,0)[bl]{(b)}}
\end{picture}
\caption{ 
The mixing angle as a function of $\mu$, in the regimes ${\cal V_{\lambda}} > 0$ (a) and 
${\cal V_{\lambda}} < 0$ (b);
the other parameters are given by $v_t = 1$ GeV, 
$\lambda_{max} = 16\pi/3$, $\lambda_{2} = \lambda_{3} = 0.1$, 
$\lambda_{1} = 0.5 $ and $\epsilon_{\alpha} = +$. The log scale in (b) shows the asymptotic values at large $\mu$.
The same asymptotic values apply in (a); see text for further discussion.}
\label{fig:mixingangle}
\end{figure}

\subsection*{Top decay into charged Higgs}
A light charged Higgs of the 
order $100\to 200$ GeV is still allowed by theoretical constraints
as well as by experimental search. If the charged Higgs satisfies 
$m_{H\pm} \leq m_t -m_b$, one could ask whether the decay 
$t\to bH^+$ can have a significant branching ratio to be 
observed at the LHC.
As mentioned before, the coupling 
$H^+tb$ has a $v_t/v_d$ suppression and the branching ratio for $t\to bH^+$ is
expected to be small. We perform a systematic scan over the DTHM parameters space
 looking for charged Higgs masses that allow the $t\to bH^+$ decay to be open.
In Fig.~\ref{fig:ttobH} we show the branching ratio for $t\to bH^+$
where we included $t\to b W^+$ and $t\to s W^+$ decay channels and 
the QCD corrections. It is obvious that a large
effect on $t\to bH^+$ would appear for the largest possible values of $v_t$ that are allowed
by EW precision constraints and the theoretical constraints. 
Indeed for a triplet vev $v_t$ in the range $0.1\to 3.5$ GeV and a
charged Higgs mass less than 165 GeV, one finds Br$(t\to bH^+)$
in the range $10^{-5}\to 10^{-4}$. 

\begin{figure}[t] 
\begin{picture}(320,260)
\put(-10,0){\mbox{\psfig{file=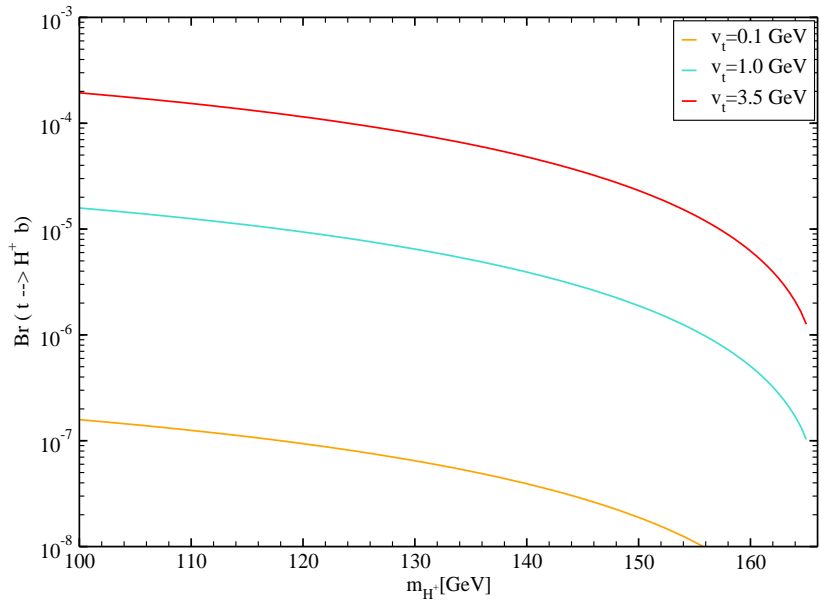,width=5.in,height=3.96in}}}
\end{picture}
\caption{ Branching ratio for $t\to bH^+$ as a function of charged Higgs mass
 for three values of the triplet vev $v_t$.}
\label{fig:ttobH}
\end{figure}

However, it is well known that the LHC will act as a top factory. 
With low luminosity $10 {\rm fb}^{-1}$, 8 million $t\bar{t}$ pairs 
per experiment per year will be produced. This number will increase
by one order of magnitude with the high luminosity option.
Therefore, the properties of top quarks can be examined 
with significant precision at LHC. For instance, it has been shown that for top decays through flavor changing neutral
processes, it is possible to reach  
${\rm Br}(t\to c H^0) \leq 4.5\times 10^{-5}$ at the LHC
\cite{AguilarSaavedra:2000aj}. 
For $t\to b H^+$, no such studies are available. But it is clear that
if we let one top decay to $b W$ and the other one decay to $b H$ 
with a branching ratio in the range $10^{-5}\to 10^{-4}$, 
this would lead to $800\to 8000$ raw $bW^+\bar{b}H^-$ (or $\bar{b} W^- b H^+$events 
in the case of high  luminosity option which may be enough to extract
 charged Higgs and measure its coupling to the top. Note also that high sensitivity to the charged
or neutral Higgses of top decays through loop induced flavor changing neutral currents, can also be attained at the 
ILC  \cite{Beneke:2000hk,AguilarSaavedra:2000db, AguilarSaavedra:2004wm}.\\  

\subsection*{DTHM spectrum and theoretical constraints}

We illustrate, in Figs.~\ref{fig:correlmu-1}.a and \ref{fig:correlmu-2}.a,  
the correlations among $\mu$, $\sin \alpha$ and $v_t$ for fixed values of the $\lambda_i$'s and $\lambda$, 
and in Figs.~\ref{fig:correlmu-1}.b,  
\ref{fig:correlmu-2}.b and \ref{fig:correlmu-2}.c,  
the correlations among $\mu$, $\sin \alpha$ and the ${\mathcal{CP}}_{even}$ 
Higgs masses (or equivalently $\lambda$), for fixed values of the $\lambda_i$'s and $v_t$, where we take into account 
the boundedness from below and unitarity constraints discussed in the previous sections.
Note that the chosen numbers in the figures are such that
 ${\cal V}_\lambda < 0$ in Figs.~\ref{fig:correlmu-1} and ${\cal V}_\lambda > 0$ in Fig.\ref{fig:correlmu-2}.a,
while Figs.\ref{fig:correlmu-2}.b and \ref{fig:correlmu-2}.c interpolate between these two regimes, 
see Eq.~(\ref{eq:mucregimes}).  For fixed $\mu$, increasing the magnitude of $v_t$ decreases $m_{h^0}$ and 
increases the mixing parameter $|s_\alpha|$ as can be seen from Figs.~\ref{fig:correlmu-1}.a and \ref{fig:correlmu-2}.a.
The upper-left white areas  in these plots correspond to $m_{h^0} \lsim  115$GeV where we took the latter value
as a fiducial lower bound for a standard-model like Higgs. Such a bound corresponds to $\lambda^{\rm SM} 
\simeq 0.44$, for $v_t < 1$GeV, while the upper bound for $m_{h^0}$ is around $120$GeV, corresponding to the value 
$\lambda =0.48$ chosen in the figures, cf. Eqs.~(\ref{eq:h0max}, \ref{eq:H0min}). 
It thus follows that the colored areas in the 
plots, indicating mainly very small $s_\alpha$ values, i.e. $h^0$ behaving like a SM Higgs, correspond to the small 
Higgs mass range $115$GeV $\leq m_{h^0} \lsim 120$GeV. Increasing the value of $\lambda$, keeping $\lambda^{\rm SM}$
fixed, would result in an increase of the Higgs mass range as well as of the regions with larger $|s_\alpha|$ 
(the red areas on the plots). In fact, there are two regions corresponding to $m_{h^0} \lsim  115$GeV, the white
area in the upper-left corner corresponding to small values of $\mu$  delimited by the red thin area, and 
another region at very large values of $\mu$  ($\gtrsim {\cal O}(1) - {\cal O}(10^3)$TeV), which are out of the scope 
of the $\mu$ range shown on the plots, that are delimited by green-blue areas.  
One should note that, in the former region, $|s_\alpha|$ reaches quickly $1$, so that $h^0$ carries essentially
a triplet component and is thus not excluded by a fiducial SM-like Higgs mass lower bound, even if it is lighter
than this bound. In contrast, in the latter region where $|s_\alpha|$ remains very small, a SM Higgs mass lower bound 
applies to $h^0$. It follows that such a bound does not put lower bounds on $\mu$, while it leads typically to 
very large upper bounds on $\mu$ as a function of $v_t$.  In the small  $\mu$ region, $H^0$ carries mainly the
SM-like component and should respect a SM Higgs mass lower bound. However, due to the very low sensitivity to $\mu$
in this regime (see Fig.~\ref{fig:fig2}), such a bound will translate merely into a lower bound on $\lambda$.
Therefore, exclusion of very small values of $\mu$ can only originate from exclusion limits on the lightest non-SM-like
${\mathcal{CP}}_{even}$ or ${\mathcal{CP}}_{odd}$ Higgses, which could be extracted for instance from existing limits 
for the  the minimal supersymmetric extension of SM in the non-decoupling regime \cite{Nakamura:2010zzi}.  

Complementary features, now with a fixed $v_t$ and varying $\lambda$, are illustrated in 
figures \ref{fig:correlmu-1}.b,  \ref{fig:correlmu-2}.b and \ref{fig:correlmu-2}.c. The gross features of
Figs.~\ref{fig:correlmu-1}.b and \ref{fig:correlmu-2}.b are in agreement with the previous discussion 
on the phenomenological bounds, related to Eqs.~(\ref{eq:solmuposexp}, \ref{eq:solmuposexpTOT}).
They illustrate how an information on $m_{h^0}$
constrains the allowed range for $\mu$ without any prior knowledge on $\lambda$. For a given $m_{h^0}$, the
allowed range of $\lambda$ is theoretically bounded from below by some $\lambda_{\rm min}$, in order
to satisfy  $m_{h^0} \leq m_{h^0}^{\rm max}$, see Eq.~(\ref{eq:solmuposexpTOT}). Then for each value of
$\lambda$ in the domain $\lambda_{\rm min} \leq \lambda \leq \lambda_{\rm max} \equiv \frac{16 \pi}{3}$ there
corresponds two values of $\mu$ consistent with a given $m_{h^0}$, according to Eq.({\ref{eq:solmuposexp}). 
Then it is easy to see,  from the shape of the $m_{h^0}(\mu)$ plots
shown in Fig.~\ref{fig:fig2},  that the largest spread between $\mu_{+}^{h^0}$ and $\mu_{-}^{h^0}$ is
reached for $\lambda = \lambda_{\rm max}$, since increasing $\lambda$ results in shifting upwards these
plots . The two branches of the envelop of the domains in Figs.~ \ref{fig:correlmu-1}.b, \ref{fig:correlmu-2}.b
correspond to  $\mu_{\pm}^{h^0}(\lambda_{\rm max})$. Furthermore, increasing $m_{h^0}$ with fixed 
$\lambda = \lambda_{\rm max}$ results in an increase of $\mu_{-}$ and decrease of $\mu_{+}$, as can be again seen
from the shape of $m_{h^0}(\mu)$ plots shown in Fig.~\ref{fig:fig2}, till the two branches join and terminate when
$m_{h^0}$ reaches its unitarity bound Eq.~(\ref{eq:unitarity8}). 
With the numbers chosen on the plots, $\mu$ is bounded to lie between $\mu_-\approx 0.3$ GeV and  
$\mu_+ \approx 10^5$ GeV. One can see that for small $\mu\leq 1$ GeV, 
$m_{h^0}$ must be less than about $200$ GeV. 
The latter bound on $m_{h^0}$ increases quickly to reach 
the unitarity bound Eq.~(\ref{eq:unitarity8}) when $\mu$ increases 
from 1 GeV to 10 GeV. Above, $\mu=10$ GeV, $m_{h^0}$ can be 
any number between the LEP limit (114 GeV) and this unitarity bound.
As noted previously, one should take into
account the actual doublet content of $h^0$ when reading out exclusion domains from these plots. 
In the plot, we have illustrated the size of $|s_\alpha|$.
In most of the cases the mixing angle is very small (blue to green areas),
which means that $h^0$ is dominated by doublet component.
In these regions where a SM Higgs exclusion limit can be readily applied, one might still need to combine this 
information with the search limits for the other charged, doubly-charged and ${\mathcal{CP}}_{odd}$ Higgs states, in order to reduce 
further the otherwise large allowed domain for $\mu$, see Eq.~(\ref{eq:mumin}). However, due to the $v_t$ suppression 
in Eq.~(\ref{eq:mumin}) of the lower bound $\mu_{\rm min}$, such a reduction is not expected to be significant
unless the experimental lower bounds, $(m_{H^{\pm \pm}})_{\rm exp}$ or $(m_{H^{\pm}})_{\rm exp}$ or
$(m_{H^{A^0}})_{\rm exp}$, become sufficiently higher than the electroweak scale. In contrast, 
bounds on $m_{h^0}$ alone would shrink significantly the spread of
the $\mu$ range whenever $|s_\alpha| > 10^{-2}$ (the green/red areas), 
reducing as well the order of magnitude of the size of $\mu$. In such a regime of small $\mu$  one starts being 
sensitive to values of the $\lambda_i$'s, as can be seen through the slight difference, in the green area, 
between Fig.~\ref{fig:correlmu-1}.b and  Fig.~\ref{fig:correlmu-2}.b. This effect will of course increase
for higher values of the $\lambda_i$'s consistent with unitarity and BFB constraints. 

Figure \ref{fig:correlmu-2}.c illustrates the behavior of $m_{H^0}$ as a function of 
$\mu$ and $\lambda$ which, as compared to Fig.~\ref{fig:correlmu-2}.b, shows a striking difference from the behavior 
of $m_{h^0}$. According to the previous discussion on neutral Higgses (see also Fig.~\ref{fig:mixingangle}),
 $|s_\alpha|$ is essentially either very small or very close to $1$. Thus, the red area corresponds to an 
$H^0$ behaving essentially like the SM Higgs. The dual sizes of the red areas in both plots can be understood
again from the mass shapes of Fig.~\ref{fig:fig2}: for small $\mu (< \bar{\mu})$, $m_{h^0}$ changes very quickly
with $\mu$ while $m_{H^0}$ is almost insensitive to $\mu$. It follows that a variation of $\lambda$, that amounts
to shifting upwards or downwards these mass shapes in  Fig.~\ref{fig:fig2}, results in a small change in $\mu$
for a fixed $m_{h^0}$ and a big change in $\mu$ for a fixed $m_{H^0}$, whence the narrow red strip in 
Fig.~\ref{fig:correlmu-2}.b and the large red area in Fig.~\ref{fig:correlmu-2}.c. (One can understand similarly
the dual sizes of the blue and green areas for large $\mu (> \bar{\mu})$.)  These features suggest a useful 
complementary strategy when using present or future exclusion limits, depending on whether one interprets these
limits in the small or large $|s_\alpha|$ regimes. We discuss this strategy only qualitatively here,
 summarizing its main points as follows:\\

\noindent
-I- in the {\sl small}  $|s_\alpha|$ regime, akin to moderate to large $\mu$ values, the typical Higgs spectrum
features a ${\mathcal{CP}}_{even}$ lightest state $h^0$ behaving like a SM-Higgs, the remaining Higgs states being
much heavier as illustrated in Fig.~\ref{fig:fig2} and Fig.~\ref{fig:higgsmass}.a. Interpreting the exclusion limits
within this regime amounts to applying a SM Higgs mass lower bound $m^{\rm (SM)}_{h}$ to 
$m_{h^0}$ that leads to a lower bound on $\lambda$, see Eq.~(\ref{eq:solmuposexpTOT}). To any $\lambda$ above this 
bound will correspond a lower, $\mu_{-}^{h^0}$, and an upper, $\mu_{+}^{h^0}$, bound on $\mu$. The lower bound 
 $\mu_{-}^{h^0}$ is, however, typically too small to be consistent with the small $|s_\alpha|$ regime and
should be superseded by a larger value ${\cal O}(\max \{\mu_c^{(1)}, \mu_c^{(2)}) \})$. Furthermore, one should keep in mind that $\mu_{+}^{h^0}$
is extremely sensitive to $m_{h^0}$ and decreases quickly with increasing $m_{h^0}$. This implies the important
feature that a slight improvement of the exclusion limit  $m^{\rm (SM)}_{h}$ results in a substantial {\sl decrease}
of the upper bound on $\mu$. The heavier ${\mathcal{CP}}_{even}$ state $H^0$ is not expected to bring significant
constraints. Indeed, in the considered regime, this state carries essentially the triplet component with suppressed
couplings to the SM sector. Its mass can thus be bounded only by $m^{\rm (non-SM)}_{h}$, the exclusion mass limit 
on non-SM-like Higgs particles.  
Since such an exclusion mass limit is expected to be weaker than the SM-like limit due to lower statistics, that is 
$m^{\rm (non-SM)}_{h} < m^{\rm (SM)}_{h}$,  then taking into account that one has theoretically  
$m_{H^0} > m_{h^0}$, one is trivially lead
to  $m^{\rm (non-SM)}_{h} < m_{H^0}^{\rm min}$  which implies no new constraints (cf. the discussion
following Eq.~(\ref{eq:solmuposexpTOT})). As stated previously, exclusion limits on the remaining Higgs states can 
also be used independently to improve the lower bound on $\mu$ based on Eq.~(\ref{eq:mumin}). 
One can, however, get further information within the present regime depending on whether these exclusion limits
are higher or lower than  $m^{\rm (SM)}_{h}$.  In particular if $(m_{H^{\pm \pm}})_{\rm exp} \gtrsim m^{\rm (SM)}_{h}$,
which excludes an $H^{\pm \pm}$ lighter than $h^0$, then one excludes all the $\lambda_4 > 0$ region, or else,
puts a stronger lower bound on $\mu$, namely $\mu > \mu^*$. (see Eq.~(\ref{eq:H++lightest}) and discussion thereof.)
In the case where $(m_{H^{\pm \pm}})_{\rm exp} \lsim m^{\rm (SM)}_{h}$, which is the present experimental situation,
there is a small window $\mu_{min} < \mu < \mu^*$ with $\lambda_4 > 0$, otherwise $\mu^* < \mu < \mu^{h^0}_{+}$
irrespective of the sign of  $\lambda_4$, and for all the allowed values of $\lambda$ discussed above. 
We have illustrated in Fig.~\ref{fig:higgsmass}.b a case where $H^{\pm \pm}$ can be the lightest Higgs state.  \\
 
\noindent
-II- in the {\sl large}  $|s_\alpha|$ regime, akin to small $\mu$ values, $H^0$ is the heaviest 
among all the Higgs states of the model and behaves like a SM-Higgs. This is a rather unusual configuration that
should help constrain more efficiently, or perhaps exclude, this regime. Also in this small $\mu$ regime, and
in contrast with the previous regime where only $\lambda$ was playing a role, there can be  now a somewhat increased 
sensitivity to the $\lambda_i$'s as well, in particular $\lambda_1 + \lambda_4$. The reason is that the size of the 
$\mu$ domain is of order $\hat{\mu}$, Eq.~(\ref{eq:muhat}), where in the latter $\lambda_1, \lambda_4$
do not suffer a $v_t$ suppression as compared to $\lambda$. However, as discussed in section \ref{sec:higgsbounds},
the parameter $k$ will characterize the sensitivity to the deviation of $H^0$ from a pure SM-Higgs state, which
can lead, for realistic experimental sensitivities, to a significant reduction of the sensitivity on
$\lambda_1 + \lambda_4$.   

One then has to consider two cases:

\begin{itemize}
\item[a)] $m^{\rm (SM)}_{h} < m_{H^0}^{\rm min}$: this case implies 
essentially a lower bound on $\lambda$ through Eq.~(\ref{eq:CPevenbounds2}), but no constraint on $\mu$ apart from the 
defining region in this regime, namely
$0 < \mu \leq \hat{\mu}$, whose size depends mainly on $\lambda$ and to a lesser extent on $\lambda_1 + \lambda_4$. 
The latter couplings are bounded by the combined unitarity and BFB constraints of section \ref{sec:comb}, so that 
there is an indirect sensitivity to $\lambda_2$ and $\lambda_3$ as well. The red area in Fig.~\ref{fig:correlmu-2}.c 
gives an illustration of
this least constrained case. The $\mu$ domain extends over all the red area, while the vertical boundary of this area
is determined by the maximal value of $\lambda = \frac{16 \pi}{3}$ given by unitarity. This boundary corresponds
to the unitarity upper bound on the SM-Higgs mass as well as the one on $m_{h^0}$, Eq.~(\ref{eq:unitarity8}). Of
course $H^0$ can escape this bound but at the expense of switching consistently to the small $|s_\alpha|$ regime as 
seen on Fig.~\ref{fig:correlmu-2}.c. 
\item[b)] $m^{\rm (SM)}_{h} \gtrsim m_{H^0}^{\rm min}$: in this case not only do we have an upper bound on 
$\lambda$ through  Eq.~(\ref{eq:CPevenbounds2}), but actually also a lower bound. Indeed, a too small $\lambda$, 
leading to a too low  $m_{H^0}^{\rm min}$ with respect to $m^{\rm (SM)}_{h}$, will eventually put $m^{\rm (SM)}_{h}$ 
just above all the values of $m_{H^0}$ corresponding to the  {\sl large}  $|s_\alpha|$ regime thus ruling out this 
regime altogether.
Furthermore, this configuration will immediately rule out the {\sl small} $|s_\alpha|$ regime as well, since
$m^{\rm (SM)}_{h}$ is by definition applicable only to SM-like states and we have 
$m^{\rm (SM)}_{h} \gtrsim m_{H^0}^{\rm min} > m_{h^0}$ for all $m_{h^0}$. Consequently, a too small $\lambda$ would
exclude the whole $\mu$ parameter space. One concludes that $\lambda$ should lie in a very narrow strip such that
 $m_{H^0}^{\rm min} \lsim m^{\rm (SM)}_{h} \lsim m_{H^0} (\mu=0)$. This strip is essentially giving the lower
bound on $\lambda$ of case a) and thus does not provide significantly new information. [Note, though, 
that for values of $\lambda_1 + \lambda_4$ close to its unitarity bound, and 
taking for instance $m^{\rm (SM)}_{h} \simeq 114$GeV and $v_t=1$GeV, case b) can still reduce the lower bound on 
$\lambda$,  from $\lambda\simeq 0.43$ to $\lambda \simeq 0.38$. But the effect will be smaller
for smaller values of $v_t$.]  One should, however, keep in 
mind that due to the high flatness of $m_{H^0}$ as a function of $\mu$ in the region  $0 < \mu \leq \hat{\mu}$, 
the slightest variation of $\lambda$ within the above noted strip would result in the exclusion of significant parts of
the  $0 < \mu \leq \hat{\mu}$ region. For instance, with  $\lambda_1 = 1.5$, 
$\lambda_2 = \lambda_3 = 0.1$, $\lambda_4 = -1$  and taking the SM-Higgs lower bound $m^{\rm (SM)}_{h} = 114$GeV,
if one reduces the lower bound of $\lambda$  (for which the whole range  $0 < \mu \leq \hat{\mu} \approx 0.3$   of 
 the {\sl large}  $|s_\alpha|$ regime is allowed) by just $1 \permil$, then the  SM-Higgs lower bound would
imply $\mu <  -5$GeV  or $\mu > 0.3$GeV thus ruling out the whole {\sl large}  $|s_\alpha|$ regime!
\end{itemize} 

\noindent
For all practical purposes and barring the fine-tuned effects just mentioned,
the above discussion of cases a) and b) shows that an experimental lower bound on the SM-like Higgs mass cannot
by itself cut into the {\sl large} $|s_\alpha|$/small $\mu$ regime; it either excludes it or allows all of it,
depending on whether $\lambda$ is respectively below or above the value $\bar{\lambda}$ that satisfies 
$m^{\rm (SM)}_{h} =  m_{H^0}^{\rm min}(\lambda = \bar{\lambda})$. Thus, the size of the $\mu$ domain 
$[0, \hat{\mu}]$ that is controlled mainly by $\lambda$ [but can also be sensitive to
$\lambda_1 + \lambda_4$] for each given value of $v_t$, see Eq.~(\ref{eq:muhat}), will not 
be reduced by the actual value of the experimental bound $m^{\rm (SM)}_{h}$. 
Moreover, an extra constraint from an experimental lower bound on the mass of a non-SM-like ${\mathcal{CP}}_{even}$ 
Higgs state would have a marginal effect since $m_{h^0}$ decreases very quickly in the region  
$\mu \lsim \hat{\mu}$.
An efficient reduction of the $\mu$ domain can come only from experimental 
lower bounds on the masses of the charged, doubly-charged and ${\mathcal{CP}}_{odd}$ Higgs states.
Indeed, these bounds translate into a lower bound on $\mu$ typically of the same size  
as $\hat{\mu}$ , Eq.~(\ref{eq:mumin}). 
As far as these experimental bounds are of the same order, 
$({m_{A^0}^2})_{\rm exp} \simeq ({m_{H^\pm}^2})_{\rm exp} \simeq ({m_{H^{\pm \pm}}^2})_{\rm exp}$, the relevant
bound will be given by $A^0$ (resp. $H^{\pm \pm}$)  when $\lambda_4 <0$ (resp.  $\lambda_4 > 0$).

Comparing $\mu_{\rm min}$ and $\hat{\mu}$, one determines easily the necessary and sufficient conditions for 
which the {\sl large} $|s_\alpha|$ regime would be excluded. They read:

\noindent
$\displaystyle ({m_{A^0}^2})_{\rm exp} \ge 
(k \lambda   - 2 (\lambda_1 + \lambda_4) - \sqrt{2 k} |\lambda - \lambda_1 - \lambda_4| ) \frac{v_d^2}{2 (k-2)}
\, + \, {\cal O}(v_t^2) $, 
  for $\lambda_4 < 0$,   and 

\noindent
$\displaystyle ({m_{H^{\pm \pm}}^2})_{\rm exp} \ge 
(  k (\lambda - \lambda_4) -2 \lambda_1 - \sqrt{2 k} |\lambda - \lambda_1 - \lambda_4|) \frac{v_d^2}{2 (k-2)}
\, + \, {\cal O}(v_t^2)$,  
 for $\lambda_4 > 0$, where we have taken into account the two-fold structure as discussed after Eq.~(\ref{eq:muhat}).


%

  

\begin{figure}[t] 
\begin{picture}(320,260)
\put(-10,0){\mbox{\psfig{file=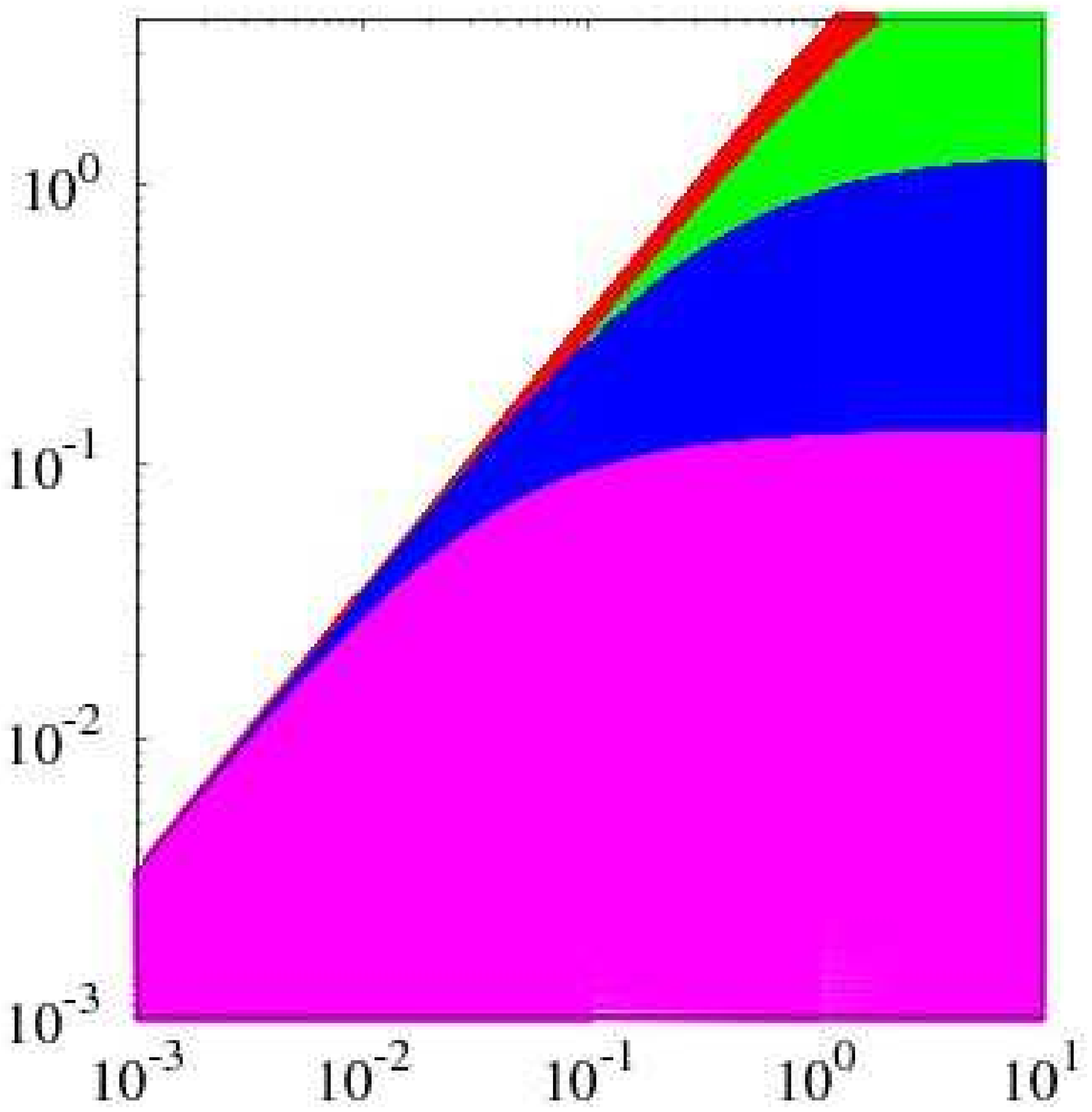,height=4.2in,width=3.3in}}}
\put(-5,95){\makebox(0,0)[bl]{\rotatebox{90}{\large{$v_t$(GeV) }}}}
\put(90,-10){\makebox(0,0)[bl]{\large{$\mu$ (GeV)}}}
\put(220,0){\mbox{\psfig{file=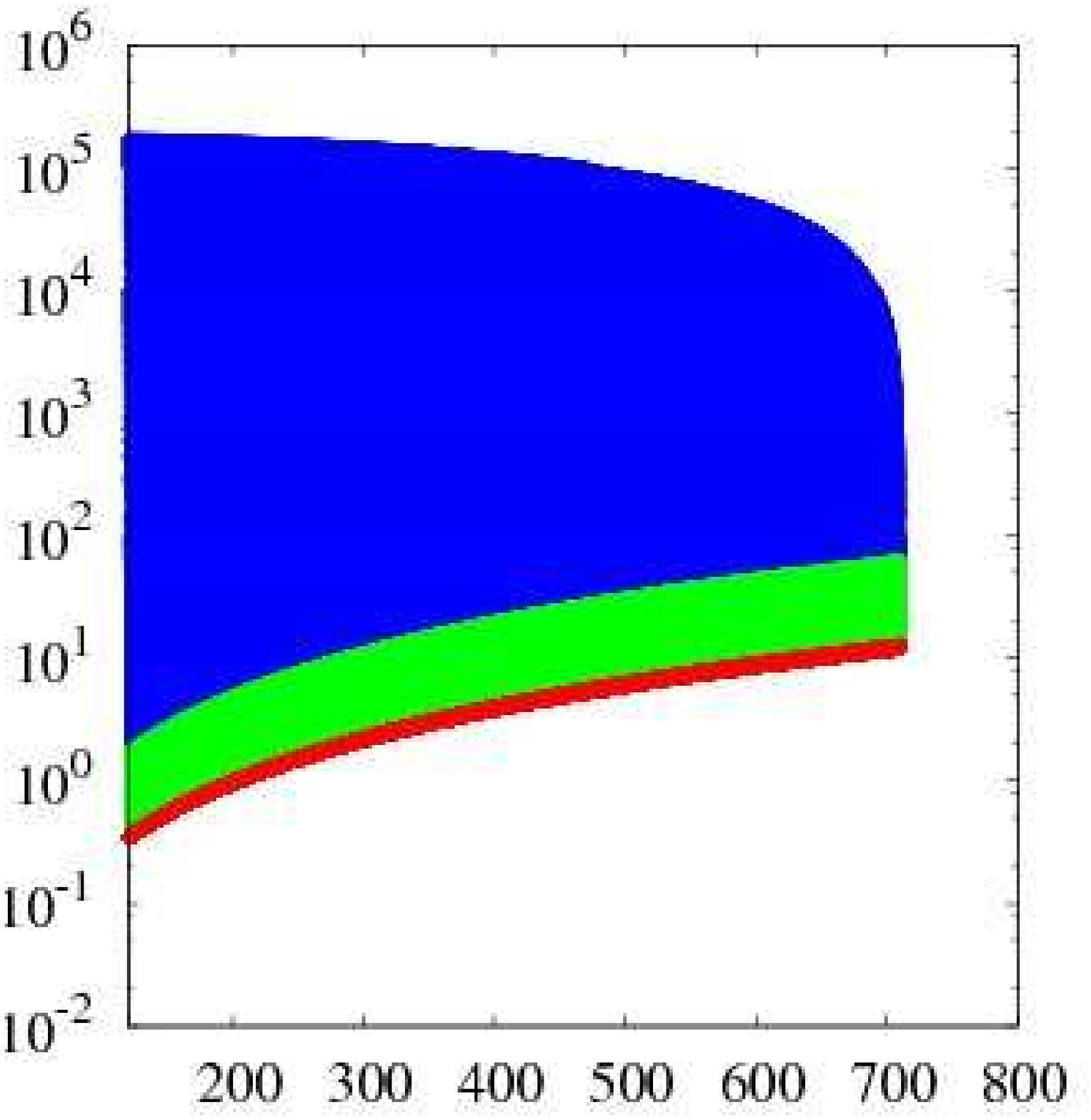,height=4.2in,width=3.3in}}}
\put(225,95){\makebox(0,0)[bl]{\rotatebox{90}{\large{$\mu$(GeV) }}}}
\put(310,-10){\makebox(0,0)[bl]{\large{$m_{h^0}$ (GeV)}}}
\put(310,40){\makebox(0,0)[bl]{\large{$v_t = 1$ GeV}}}
\put(40,180){\makebox(0,0)[bl]{\large{$m_{h^0}^{(\rm SM)} = 115$GeV}}}
\put(180,190){\makebox(0,0)[bl]{(a)}}
\put(410,190){\makebox(0,0)[bl]{(b)}}
\end{picture}
\caption{ 
(a) correlation between $\mu$ and $v_t$ with $m_{h^0} > m_{h^0}^{(\rm SM)}= 115$GeV and $\lambda = 0.48$, 
(b) correlation between $\mu$ and $m_{h^0}$, scanning over $\lambda$ in the range
$0.44 \le \lambda \le 16 \pi/3$, with $v_t = 1$GeV; color code:
$10^{-1}\le s_\alpha \le 1$ (red), 
$10^{-2}\le s_\alpha \le 10^{-1}$ (green),
$10^{-3}\le s_\alpha \le 10^{-2}$ (blue) and
 $s_\alpha \le 10^{-3}$ (magenta). 
The other parameters are $\lambda_1 = -\lambda_4 = 1$, $\lambda_2 = \lambda_3 = 0$ and $\kappa=8$.
${\cal V}_\lambda < 0$ for both figures.
}
\label{fig:correlmu-1}
\end{figure}

\begin{figure}[t!] 
\begin{picture}(320,460)
\put(-10,250){\mbox{\psfig{file=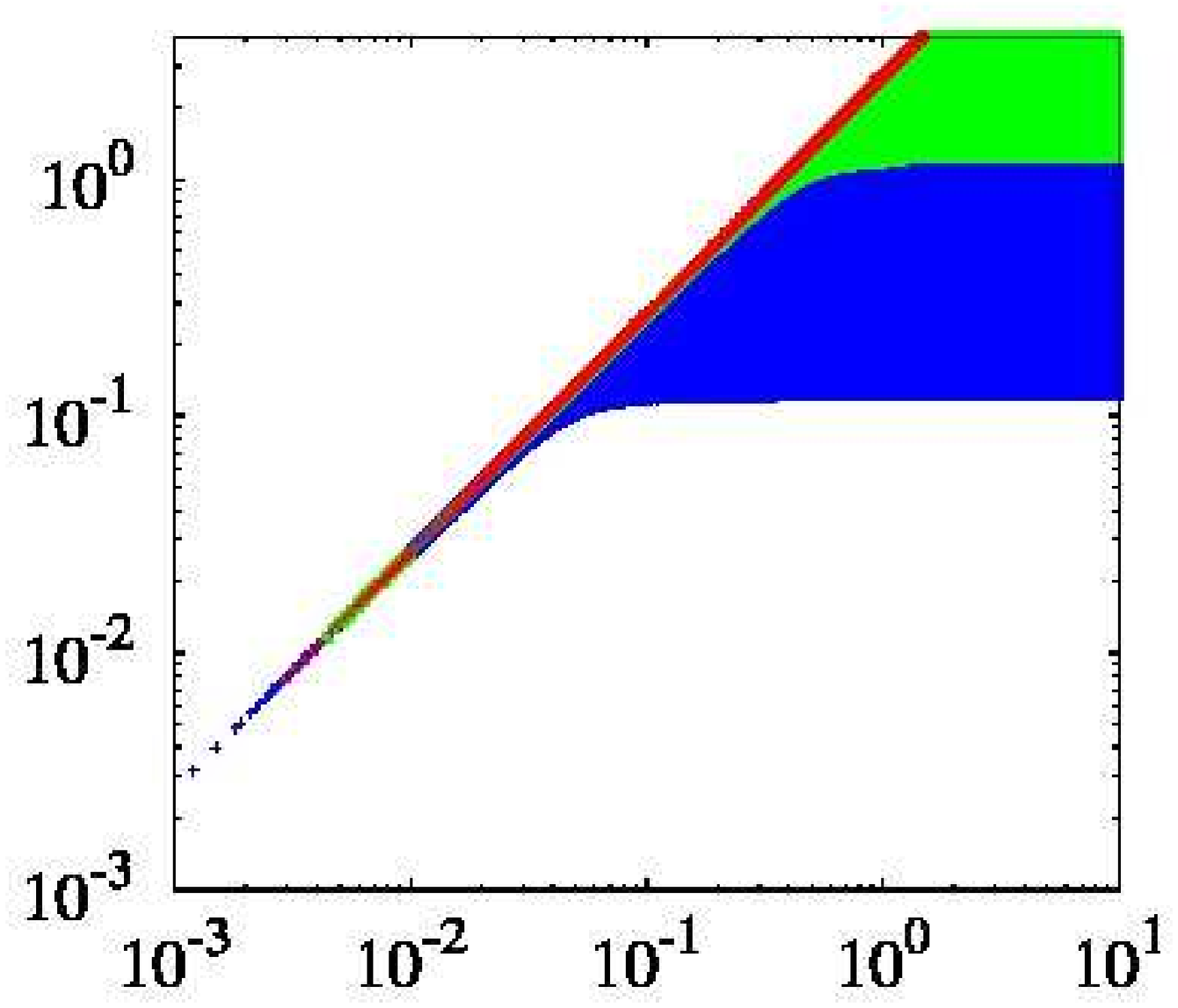,height=5.5in,width=3.3in}}}
\put(0,345){\makebox(0,0)[bl]{\rotatebox{90}{\large{$v_t$(GeV) }}}}
\put(220,250){\mbox{\psfig{file=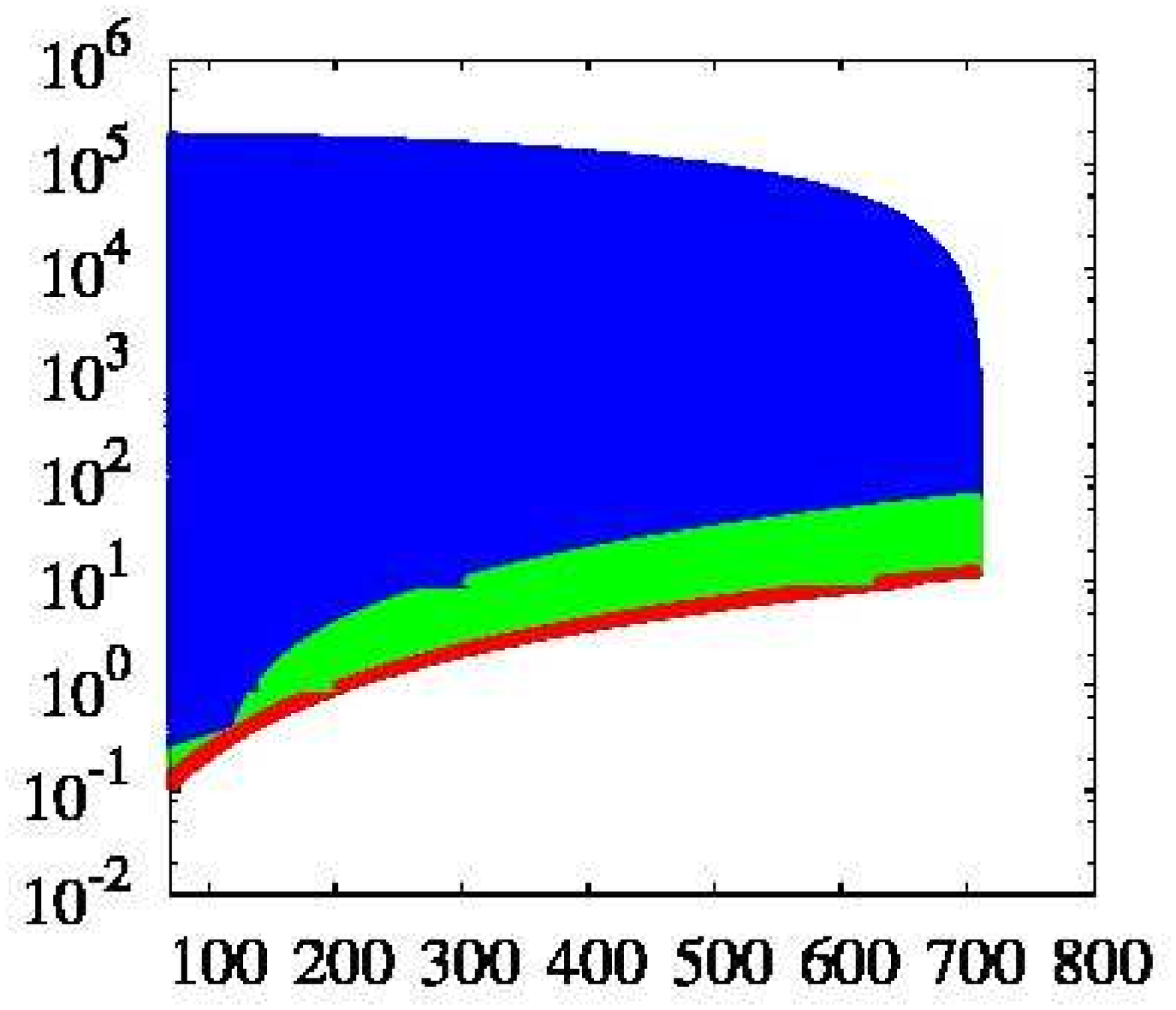,height=5.5in,width=3.3in}}}
\put(225,345){\makebox(0,0)[bl]{\rotatebox{90}{\large{$\mu$(GeV) }}}}
\put(310,290){\makebox(0,0)[bl]{\large{$v_t = 1$ GeV}}}
\put(45,430){\makebox(0,0)[bl]{\large{$m_{h^0}^{(\rm SM)} = 115$GeV}}}
\put(110,10){\mbox{\psfig{file=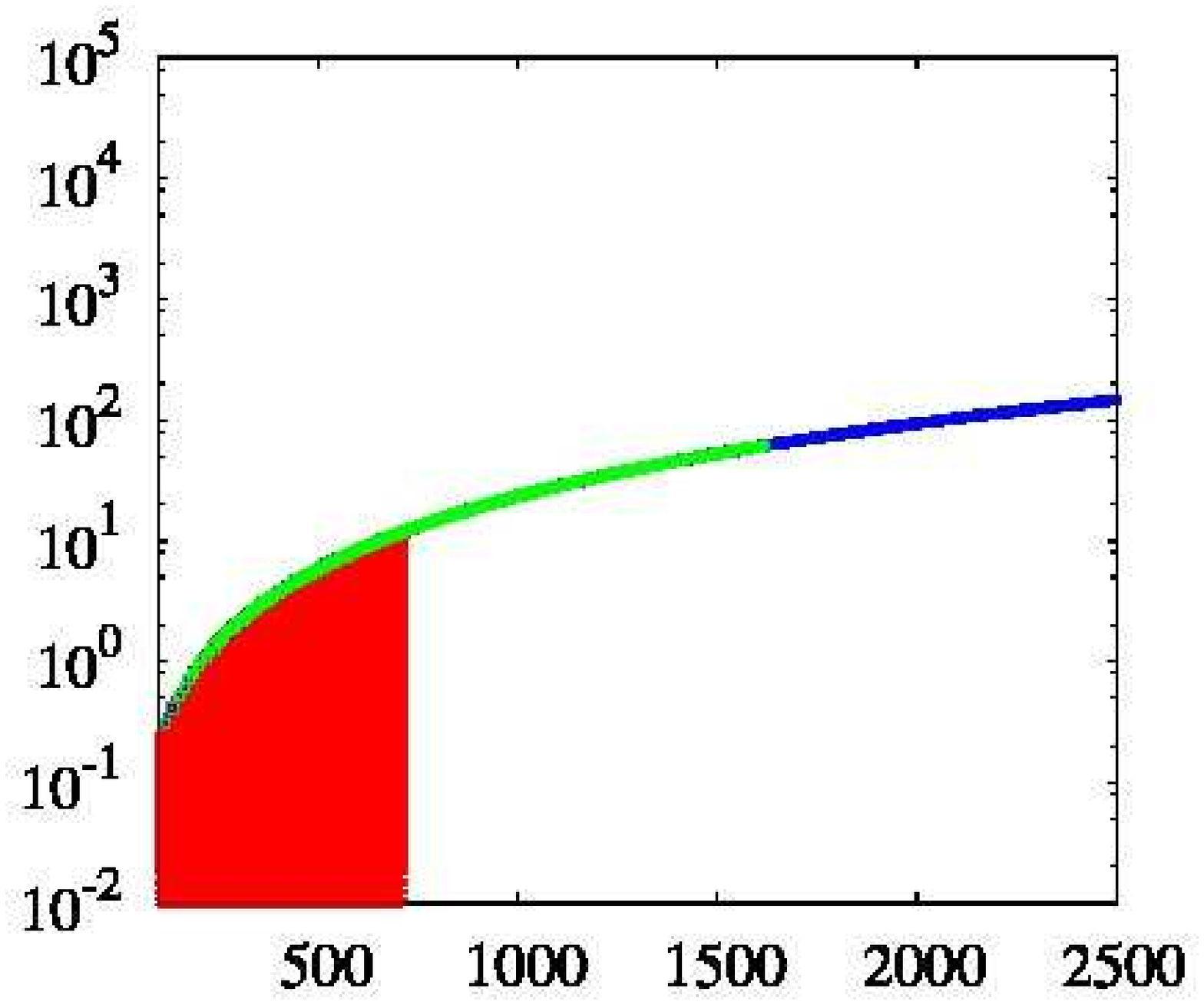,height=5.5in,width=3.3in}}}
\put(220,60){\makebox(0,0)[bl]{\large{$v_t = 1$ GeV}}}
\put(115,115){\makebox(0,0)[bl]{\rotatebox{90}{\large{$\mu$(GeV) }}}}
\put(200, 0){\makebox(0,0)[bl]{\large{$m_{H^0}$ (GeV)}}}
\put(310,240){\makebox(0,0)[bl]{\large{$m_{h^0}$ (GeV)}}}
\put(90,240){\makebox(0,0)[bl]{\large{$\mu$ (GeV)}}}
\put(180,440){\makebox(0,0)[bl]{(a)}}
\put(410,440){\makebox(0,0)[bl]{(b)}}
\put(300,210){\makebox(0,0)[bl]{(c)}}
\end{picture}
\caption{
(a) correlation between $\mu$ and $v_t$ with $m_{h^0} > m_{h^0}^{(\rm SM)}= 115$GeV and $\lambda = 0.48$, 
(b) correlation between 
$\mu$ and the light ${\mathcal{CP}}_{even}$ Higgs mass,
(c) between $\mu$ and the heavy ${\mathcal{CP}}_{even}$ Higgs mass, scanning over $\lambda$ in the range
$0.44 \le \lambda \le 16 \pi/3$, with $v_t = 1$GeV; color code: 
$10^{-1}\le |s_\alpha| \le 1$ (red),
$10^{-2}\le |s_\alpha| \le 10^{-1}$ (green),
$10^{-3}\le |s_\alpha| \le 10^{-2}$ (blue) and $|s_\alpha| \le 10^{-3}$ (white bottom area in (a)).
The other parameters are given by $\lambda_1 = 1.5$, 
$\lambda_2 = \lambda_3 = 0.1$, $\lambda_4 = -1$ and $\kappa = 8$. ${\cal V}_\lambda > 0$ in (a), while
in (b) and (c) ${\cal V}_\lambda $ changes sign with increasing Higgs masses.}
\label{fig:correlmu-2}
\end{figure}


\begin{figure}[t] 
\begin{picture}(320,260)
\put(-10,0){\mbox{\psfig{file=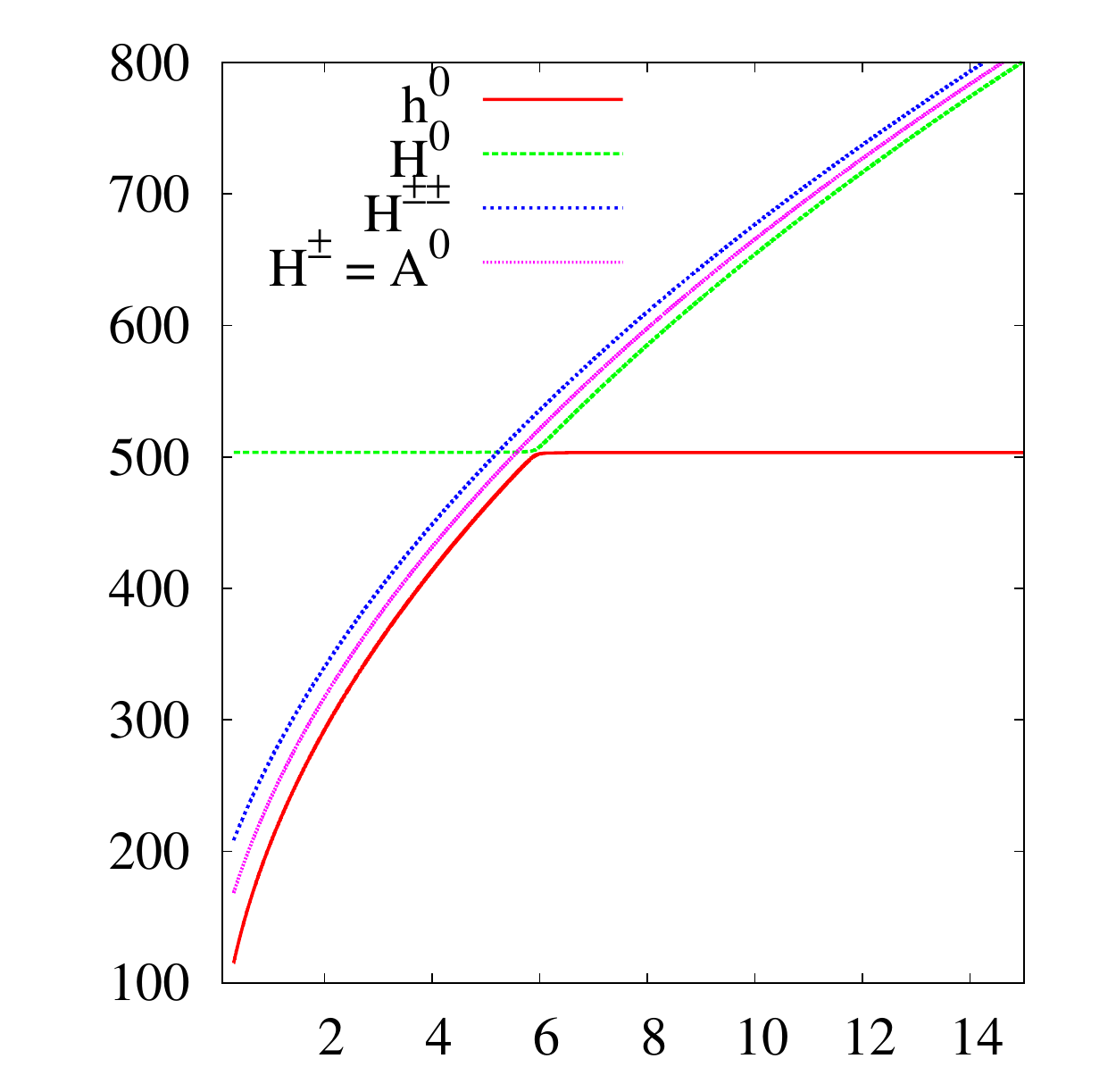,height=3.0in,width=3.0in}}}
\put(-10,75){\makebox(0,0)[bl]{\rotatebox{90}{\large{$m_{h^0, H^0, A^0, H^\pm, H^{\pm\pm}}$}}}}
\put(80,-10){\makebox(0,0)[bl]{\large{$\mu$ (GeV)}}}
\put(80,40){\makebox(0,0)[bl]{\large{${\cal V_{\lambda}} < 0, \, \lambda_4 = -1$ }}}
\put(220,0){\mbox{\psfig{file=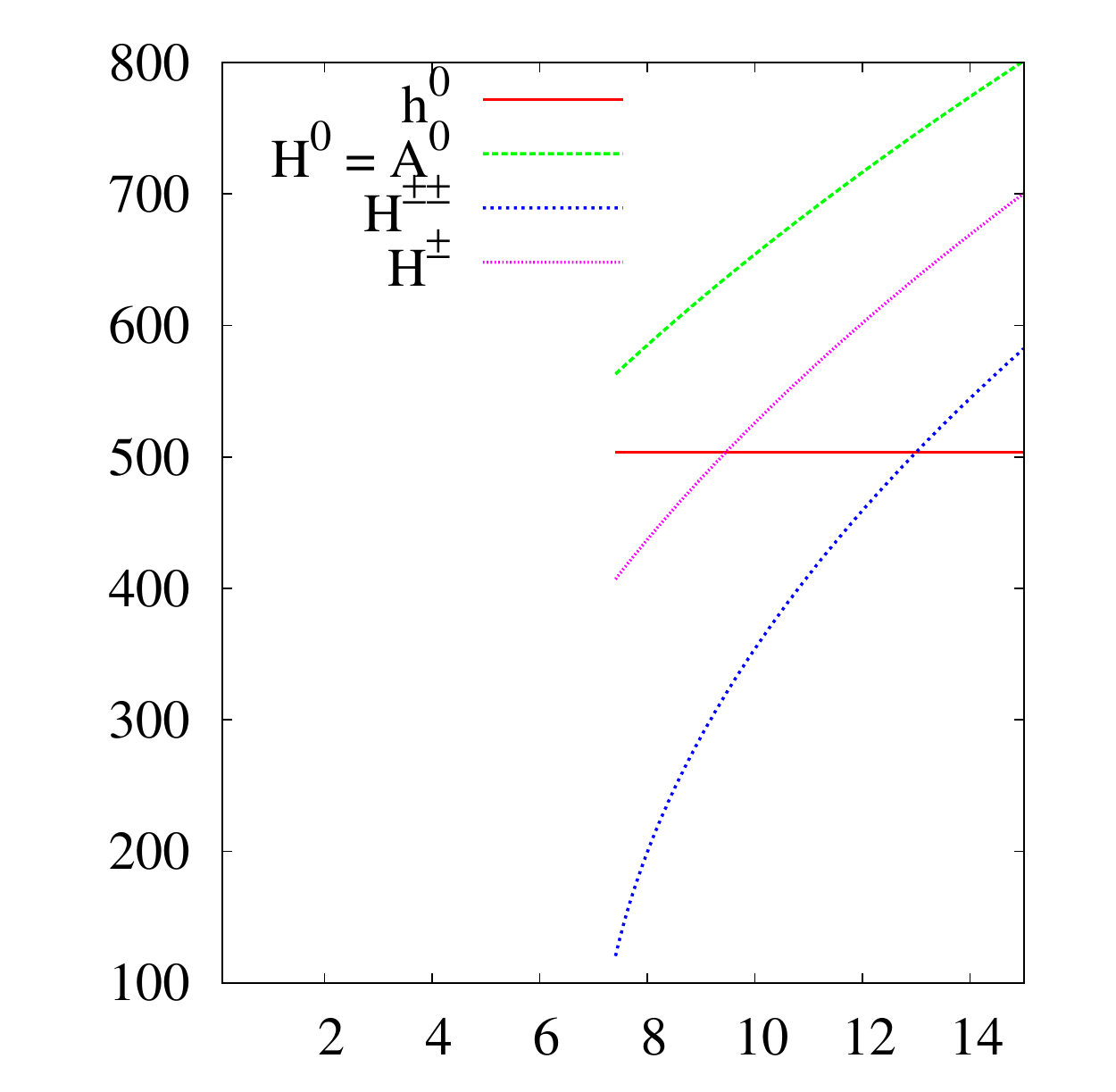,height=3.0in,width=3.0in}}}
\put(310,-10){\makebox(0,0)[bl]{\large{$\mu $ (GeV)}}}
\put(100,210){\makebox(0,0)[bl]{(a)}}
\put(330,210){\makebox(0,0)[bl]{(b)}}
\put(210,75){\makebox(0,0)[bl]{\rotatebox{90}{\large{$m_{h^0, H^0, A^0, H^\pm, H^{\pm\pm}}$}}}}
\put(310,40){\makebox(0,0)[bl]{\large{${\cal V_{\lambda}} > 0, \, \lambda_4 = 10$ }}}
\end{picture}
\caption{ 
Higgs boson masses as a function of $\mu$ with $v_t = 1$ GeV, 
$\lambda = 8\pi/3$, $\lambda_{1} = 0.5 $, $\lambda_{2} = \lambda_{3} = 0.1$, for ${\cal V_{\lambda}} < 0$, 
$\lambda_4 = -1$ (left) and ${\cal V_{\lambda}} > 0$, $\lambda_4 = 10$ (right), we note that in both panels 
$m_{H^0} = m_{A^0}$. }
\label{fig:higgsmass}
\end{figure}

\section{Conclusion}

We have carried out a detailed study of the renormalizable Higgs potential relevant to the
type II seesaw model, keeping the full set of the seven free parameters of the potential. We determined analytically 
the unitarity constraints on the various scalar couplings and fully solved the {\sl all directions} conditions for 
boundedness from below. These combined theoretical constraints delineate efficiently the physically allowed regions 
of the parameter space and should be taken into account in phenomenological studies.
We also examined the vacuum structure of the potential and determined general consistency constraints
on the $\mu$ parameter, as well as theoretical upper (resp. lower) bounds on the lighter (resp. heavier) 
${\mathcal{CP}}_{even}$ Higgs particle mass that can constrain further the phenomenological analyses. We also 
identified two distinct regimes respectively for large and small $\mu$. In the first regime the lightest Higgs 
particle is  the $h^0$, behaving as a SM-like Higgs, the remaining Higgses being typically too heavy to be of any 
phenomenlogical relevance. In the second regime, it is the heaviest Higgs $H^0$ that behaves as a SM-like Higgs,
the lighter charged, doubly charged and ${\mathcal{CP}}_{odd}$ states become accessible to the colliders,
with possibly the $H^{\pm \pm}$ being the lightest state, while the lighter ${\mathcal{CP}}_{even}$
decouples quickly from the SM sector.
We also initiated the study of possible consequences from existing experimental exclusion limits. 

Although we did not commit to any underlying GUT assumptions, thus allowing $\mu$ to vary between a few GeV and possibly
the GUT scale, we do retrieve, as a consequence of the (model-independent) dynamical constraints on $\mu$, 
a seesaw-like behavior that leads to tiny $v_t$ if $\mu$ is taken very large. 

Finally, the results of this study having been obtained at the tree-level, one should keep in mind that  
modifications due to quantum corrections to the effective potential can possibly be substantial in
some cases.
The inclusion of such corrections is, however, beyond the scope of the present paper given the non-trivial form of 
the constraints already at the tree-level.

\section*{Acknowledgments}
We would like to thank Borut Bajc, Ben Gripaios, Roberto Salerno and Goran Senjanovic for discussions. This work was 
supported  by {\sl Programme Hubert Curien, Volubilis, AI n${}^0:$ MA/08/186}. 
We also acknowledge the LIA ({\sl International Laboratory for Collider Physics - ILCP}) as well as 
the ICTP-IAEA  Training Educational Program for partial support. The work of R.B. was supported by CSIC.

\renewcommand{\theequation}{A.\arabic{equation}}
\setcounter{equation}{0}
\section*{Appendix A.}
\label{sec:appendixA}

As stated in section 3, the positivity of $m_{h^0}^2$ constrains $\mu$ to lie in the range $\mu_{-} \leq \mu \leq \mu_{+}$ 
 so as to satisfy Eq.~(\ref{eq:posh0H0-1}). We give here the full expression for $\mu_{\pm}$: 

\begin{equation}
\mu_{\pm} = \frac{\lambda \upsilon_d^2+8(\lambda_1+\lambda_4 )\upsilon_t^2 \pm \sqrt{\lambda 
(\lambda \upsilon_d^4+16\upsilon_t^2 ( (\lambda_1 +\lambda_4) \upsilon_d^2
+4 (\lambda_2+\lambda_3)\upsilon_t^2))}}{8\sqrt2 \upsilon_t} \label{eq:solmupos}
\end{equation}

\noindent

Note that due to the negative coefficient of $\mu^2$ in Eq.(\ref{eq:posh0H0-1}), $\mu_{\pm}$ should always 
be real-valued otherwise  Eq.(\ref{eq:posh0H0-1}) is not satisfied and $h^0$ is tachyonic for 
{\sl all} values of $\mu$. As can be seen from Eq.~(\ref{eq:solmupos}), 
this requirement leads in principle  to an extra constraint on top of Eqs.~(\ref{eq:posh0H0-0} - \ref{eq:posh0H0}), 
that is

\begin{equation}
\lambda 
(\lambda \upsilon_d^4+16\upsilon_t^2 ( (\lambda_1 +\lambda_4) \upsilon_d^2
+4 (\lambda_2+\lambda_3)\upsilon_t^2)) \geq 0  \label{eq:spurious}
\end{equation}  

\noindent
However, we show here that this extra constraint is automatically satisfied due to the BFB constraints: since $\lambda >0$, 
cf. Eq.~(\ref{eq:bound1}), it suffices to show that $(\lambda \upsilon_d^4+16\upsilon_t^2 ( (\lambda_1 +\lambda_4) \upsilon_d^2
+4 (\lambda_2+\lambda_3)\upsilon_t^2)) \geq 0$. Now using the first inequality of Eq.~(\ref{eq:bound3})
one obtains
\begin{eqnarray}
\lambda v_d^4 + 16 v_t^2 ((\lambda_1 + \lambda_4) v_d^2 + 4 (\lambda_2 + \lambda_3) v_t^2) &\geq&
(\lambda v_d^4 + 16 v_t^2 (- \sqrt{\lambda (\lambda_2 + \lambda_3)} v_d^2 + 4 (\lambda_2 + \lambda_3) v_t^2)) 
\nonumber\\
 & \geq &
(\sqrt{\lambda} v_d^2 - 8 \sqrt{(\lambda_2 + \lambda_3)} v_t^2)^2 
\end{eqnarray}

\noindent
that proves our statement.

\renewcommand{\theequation}{B.\arabic{equation}}
\setcounter{equation}{0}
\section*{Appendix B.}
\label{sec:appendixB}
In this appendix we give the form of the  Higgs potential in the field subspaces where only $2$ or only
$3$ fields are non-vanishing, dubbed respectively $2$-field and  $3$-field directions. We identify 
exhaustively 10 different directions for each of these two classes and give their corresponding 
BFB conditions.\\   

\noindent
The ten $2$-field directions:
\begin{eqnarray}
{}_2V^{(4)}_{{\rm dir.} 1}&=& \frac{\lambda}{4}\, (|\phi^0|^2 + |\phi^+|^2)^2 \label{eq:dir1} \\
{}_2V^{(4)}_{\rm dir. 2}&=&  (\lambda_2 + \lambda_3)\,|\delta^{++}|^4 + 
    (\lambda_1 + \lambda_4)\,|\delta^{++}|^2\,|\phi^+|^2 + \frac{\lambda}{4}\,|\phi^+|^4 \label{eq:dir2} \\ 
{}_2V^{(4)}_{\rm dir. 3}&=&  (\lambda_2 + \lambda_3)\,|\delta^{++}|^4 + \lambda_1 \,|\delta^{++}|^2\,|\phi^0|^2 + 
    \frac{\lambda}{4}\,|\phi^0|^4  \label{eq:dir3} \\
{}_2V^{(4)}_{\rm dir. 4}&=& (\lambda_2 + \frac{\lambda_3}{2})\,|\delta^+|^4 + 
      (\lambda_1 + \frac{\lambda_4}{2} )\,|\delta^+|^2\,|\phi^+|^2 + \frac{\lambda}{4}\,|\phi^+|^4 \label{eq:dir4} \\ 
{}_2V^{(4)}_{\rm dir. 5}&=& (\lambda_2 + \frac{\lambda_3}{2})\,|\delta^+|^4 + 
      (\lambda_1 + \frac{\lambda_4}{2} )\,|\delta^+|^2\,|\phi^0|^2 + \frac{\lambda}{4}\,|\phi^0|^4 \label{eq:dir5} \\ 
{}_2V^{(4)}_{\rm dir. 6}&=&   (\lambda_2 + \frac{\lambda_3}{2})\,|\delta^+|^4 + 
    2\,(\lambda_2 + \lambda_3)\,|\delta^+|^2\,|\delta^{++}|^2 + 
    (\lambda_2 + \lambda_3)\,|\delta^{++}|^4 \label{eq:dir6} \\ 
{}_2V^{(4)}_{\rm dir. 7}&=&  (\lambda_2 + \lambda_3)\,|\delta^0|^4 + \lambda_1 \,|\delta^0|^2\,|\phi^+|^2 + 
    \frac{\lambda}{4}\,|\phi^+|^4 \label{eq:dir7} \\ 
{}_2V^{(4)}_{\rm dir. 8}&=& (\lambda_2 + \lambda_3)\,|\delta^0|^4 + 
    (\lambda_1 + \lambda_4)\,|\delta^0|^2\,|\phi^0|^2 + \frac{\lambda}{4}\,|\phi^0|^4 \label{eq:dir8} \\
{}_2V^{(4)}_{\rm dir. 9}&=&  (\lambda_2 + \lambda_3)\,|\delta^0|^4 + 2\,\lambda_2\,|\delta^0|^2\,|\delta^{++}|^2 + 
    (\lambda_2 + \lambda_3)\,|\delta^{++}|^4 \label{eq:dir9} \\
{}_2V^{(4)}_{\rm dir. 10}&=& (\lambda_2 + \lambda_3)\,|\delta^0|^4 + 2\,(\lambda_2 + \lambda_3)\,|\delta^0|^2\,|\delta^+|^2 + 
    (\lambda_2 + \frac{\lambda_3}{2})\,|\delta^+|^4  \label{eq:dir10}
\end{eqnarray}

\begin{eqnarray}
&& {\rm direction} \;1: \lambda > 0 \label{eq:BFB2dir1}\\
&& {\rm directions} \; 2 \; {\rm and} \; 8: 
\lambda > 0, \lambda_2 + \lambda_3 >0, \lambda_1 + \lambda_4 + \sqrt{\lambda (\lambda_2 + \lambda_3)} >0 \\ 
&& {\rm direcitons} \; 3 \; {\rm and} \; 7:
 \lambda > 0, \lambda_2 + \lambda_3 >0, \lambda_1 + \sqrt{\lambda  (\lambda_2 + \lambda_3)} >0 \\ 
&& {\rm directions} \; 4 \; {\rm and} \; 5: \lambda > 0, \lambda_2 + \frac{\lambda_3}{2} >0,
\lambda_1 + \frac{\lambda_4}{2} + \sqrt{\lambda (\lambda_2 + \frac{\lambda_3}{2})} >0 \label{eq:BFB2dir4}\\ 
&& {\rm directions} \; 6, 9 \; {\rm and} \; 10:  \lambda_2 + \lambda_3 >0, \lambda_2 + \frac{\lambda_3}{2} >0 
\label{eq:BFB2dir10}
\end{eqnarray}

\newpage
\noindent
The ten $3$-field directions:

\begin{eqnarray}
{}_3V^{(4)}_{{\rm dir.} 1}&=&(\lambda_2 + \lambda_3)|\delta^0|^4\,+\,
2 ( \lambda_2 + \lambda_3)|\delta^0|^2|\delta^{+}|^2\,+\,
    (\lambda_2 + \frac{\lambda_3}{2})|\delta^{+}|^4\,+\,2\,\lambda_2\,|\delta^0|^2\,|\delta^{++}|^2 \nonumber \\
  &&  \,+\, 2 ( \lambda_2 + \lambda_3)\,|\delta^{+}|^2\,|\delta^{++}|^2\,+\,
    (\lambda_2 + \lambda_3)\,|\delta^{++}|^4 \nonumber \\
  && - \lambda_3\,\frac{\delta^{- -}}{\delta^0\,(\delta^{-})^2} |\delta^0|^2|\delta^{+}|^4 
     - \lambda_3\, \frac{\delta^0\,(\delta^{-})^2}{\delta^{- -}} \,|\delta^{++}|^2  \\ 
{}_3V^{(4)}_{{\rm dir.} 2}&=&
  (\lambda_2 + \lambda_3)\,|\delta^0|^4\,+\,2 ( \lambda_2 + \lambda_3)\,|\delta^0|^2\,|\delta^{+}|^2\,+\, 
    (\lambda_2 + \frac{\lambda_3}{2})\,|\delta^{+}|^4 \nonumber \\ 
  &&\,+\,(\lambda_1 + \lambda_4)\,|\delta^0|^2\,|\phi^0|^2\,+\,
    (\lambda_1\,+\,\frac{\lambda_4}{2})\,|\delta^{+}|^2\,|\phi^0|^2\,+\,\frac{\lambda}{4}\,|\phi^0|^4   
\label{eq:threedir2}\\
{}_3V^{(4)}_{{\rm dir.} 3}&=&
   (\lambda_2 + \lambda_3)\,|\delta^0|^4\,+\,2 ( \lambda_2 + \lambda_3)\,|\delta^0|^2\,|\delta^{+}|^2\,+\,
    (\lambda_2 + \frac{\lambda_3}{2})\,|\delta^{+}|^4  \nonumber \\
&&  \,+\,\lambda_1\,|\delta^0|^2\,|\phi^+|^2\,+\,
    (\lambda_1 + \frac{\lambda_4}{2})\,|\delta^{+}|^2\,|\phi^+|^2\,+\,\frac{\lambda}{4}\,|\phi^+|^4 \\
 {}_3V^{(4)}_{{\rm dir.} 4}&=&
   (\lambda_2 + \lambda_3)\,|\delta^0|^4\,+\,2\,\lambda_2\,|\delta^0|^2\,|\delta^{++}|^2\,+\,
    (\lambda_2 + \lambda_3)\,|\delta^{++}|^4  \nonumber \\
  &&\,+\,(\lambda_1 + \lambda_4)\,|\delta^0|^2\,|\phi^0|^2\,+\,
    \lambda_1\,|\delta^{++}|^2\,|\phi^0|^2\,+\,\frac{\lambda}{4}\,|\phi^0|^4 \\
{}_3V^{(4)}_{{\rm dir.} 5}&=&
   (\lambda_2 + \lambda_3)\,|\delta^0|^4\,+\,2\,\lambda_2\,|\delta^0|^2\,|\delta^{++}|^2\,+\,
    (\lambda_2 + \lambda_3)\,|\delta^{++}|^4 \nonumber \\
  &&\,+\,\lambda_1\,|\delta^0|^2\,|\phi^+|^2\,+\,
    (\lambda_1 + \lambda_4)\,|\delta^{++}|^2\,|\phi^+|^2\,+\,\frac{\lambda}{4}\,|\phi^+|^4 \\
{}_3V^{(4)}_{{\rm dir.} 6}&=&
   (\lambda_2 + \lambda_3)\,|\delta^0|^4\,+\,(\lambda_1 + \lambda_4)\,|\delta^0|^2\,|\phi^0|^2\,+\,
    \frac{\lambda}{4}\,|\phi^0|^4 \nonumber \\
  &&\,+\,\lambda_1\,|\delta^0|^2\,|\phi^+|^2\,+\,
    \frac{\lambda}{2}\,|\phi^0|^2\,|\phi^+|^2\,+\,\frac{\lambda}{4}\,|\phi^+|^4 \\
{}_3V^{(4)}_{{\rm dir.} 7}&=&
   (\lambda_2 + \frac{\lambda_3}{2})\,|\delta^{+}|^4\,+\,
    2 ( \lambda_2 + \lambda_3)\,|\delta^{+}|^2\,|\delta^{++}|^2\,+\,
    (\lambda_2 + \lambda_3)\,|\delta^{++}|^4 \nonumber \\
  &&\,+\,(\lambda_1 + \frac{\lambda_4}{2})\,|\delta^{+}|^2\,|\phi^0|^2\,+\,
    \lambda_1\,|\delta^{++}|^2\,|\phi^0|^2 + \frac{\lambda}{4}\,|\phi^0|^4 \\
{}_3V^{(4)}_{{\rm dir.} 8}&=&
   (\lambda_2 + \frac{\lambda_3}{2})\,|\delta^{+}|^4\,+\,
    2 ( \lambda_2 + \lambda_3)\,|\delta^{+}|^2\,|\delta^{++}|^2\,+\,
    (\lambda_2 + \lambda_3)\,|\delta^{++}|^4 \nonumber \\
  &&\,+\,(\lambda_1 + \frac{\lambda_4}{2})\,|\delta^{+}|^2\,|\phi^+|^2\,+\,
    (\lambda_1 + \lambda_4)\,|\delta^{++}|^2\,|\phi^+|^2\,+\,\frac{\lambda}{4}\,|\phi^+|^4 \\
{}_3V^{(4)}_{{\rm dir.} 9}&=&
    (\lambda_2 + \frac{\lambda_3}{2})\,|\delta^{+}|^4\,+\,(\lambda_1 + 
      \frac{\lambda_4}{2})\,|\delta^{+}|^2\,|\phi^0|^2\,+\,
    \frac{\lambda}{4}\,|\phi^0|^4 \nonumber \\
  &&\,+\,(\lambda_1  + \frac{\lambda_4}{2})\,|\delta^{+}|^2\,|\phi^+|^2\,+\,
    \frac{\lambda}{2}\,|\phi^0|^2\,|\phi^+|^2\,+\,\frac{\lambda}{4}\,|\phi^+|^4   \label{eq:threedir9}\\
{}_3V^{(4)}_{{\rm dir.} 10}&=&
   (\lambda_2\,+\,\lambda_3)\,|\delta^{++}|^4\,+\,\lambda_1\,|\delta^{++}|^2\,|\phi^0|^2\,+\,
    \frac{\lambda}{4}\,|\phi^0|^4 \nonumber \\
   &&\,+\,(\lambda_1 + \lambda_4)\,|\delta^{++}|^2\,|\phi^+|^2\,+\,
    \frac{\lambda}{2}\,|\phi^0|^2\,|\phi^+|^2\,+\,\frac{\lambda}{4}\,|\phi^+|^4 
\end{eqnarray}

\newpage

\noindent
The corresponding BFB conditions read:

\begin{eqnarray}
 {\bf direction \; 1:} \;\;  2 \lambda_2+\lambda_3>0\land \lambda_2+\lambda_3>0\land \Bigl(\lambda_3^2<4
    (\lambda_2+\lambda_3)^2\lor \lambda_3<0\Bigr) ~~~~~~~~~~~~~~~~~&& \label{eq:BFB3dir1}
\end{eqnarray}

\begin{eqnarray}
&& {\bf direction \; 2:}\nonumber \\
  &&  \lambda>0\land \lambda_2+\lambda_3>0\land 2 \lambda_2+\lambda_3>0\land
    \sqrt{\lambda (\lambda_2+\lambda_3)}+\lambda_1+\lambda_4>0\land
    \Bigl(\Bigl(2 \lambda (2 \lambda_2+\lambda_3)>(2
    \lambda_1+\lambda_4)^2 \nonumber  \\
&&\land \Bigl(\Bigl(\sqrt{2} \sqrt{\lambda_3
    (\lambda_2+\lambda_3) \Bigl((2 \lambda_1+\lambda_4)^2-2 \lambda (2
    \lambda_2+\lambda_3)\Bigr)}+2 \lambda_2 \lambda_4>2 \lambda_1
    \lambda_3\land \lambda_3<0\Bigr)\lor \nonumber \\
&& \Bigl(\frac{(2 \lambda_2+\lambda_3)
    ((2 \lambda_1+\lambda_4) (2 \lambda_1+3 \lambda_4)-4 \lambda
    (\lambda_2+\lambda_3))}{2 \lambda_1+\lambda_4}>0\land 2
    \lambda_1+\lambda_4<0\Bigr)\Bigr)\Bigr)\lor 2
    \lambda_1+\lambda_4>0\Bigr)
\end{eqnarray}

\begin{eqnarray}
&&{\bf direction \; 3:} \nonumber \\
    &&\lambda>0\land \lambda_2+\lambda_3>0\land 2 \lambda_2+\lambda_3>0\land
    \sqrt{\lambda (\lambda_2+\lambda_3)}+\lambda_1>0\land \Bigl(\Bigl(2
    \lambda (2 \lambda_2+\lambda_3)>(2 \lambda_1+\lambda_4)^2\land \nonumber \\
&&    \Bigl(\Bigl(2 \lambda_1+\lambda_4<0\land \frac{(2 \lambda_2+\lambda_3)
    \Bigl(4 \lambda (\lambda_2+\lambda_3)-4
    \lambda_1^2+\lambda_4^2\Bigr)}{2 \lambda_1+\lambda_4}<0\Bigr)\lor \nonumber \\
&&  \Bigl((\lambda_2+\lambda_3) (2 \lambda_2+\lambda_3-2)>0\land \sqrt{2}
    \sqrt{\lambda_3 (\lambda_2+\lambda_3) \Bigl((2 \lambda_1+\lambda_4)^2-2
    \lambda (2 \lambda_2+\lambda_3)\Bigr)}>2 \lambda_1 \lambda_3 \nonumber \\
&&+2  \lambda_4 (\lambda_2+\lambda_3)\Bigr)\Bigr)\Bigr)\lor 2
    \lambda_1+\lambda_4>0\Bigr)
\end{eqnarray}

\begin{eqnarray}
&& \hspace{-.6cm}{\bf direction \; 4:} \nonumber \\
&& \lambda>0\land \lambda_2+\lambda_3>0\land \sqrt{\lambda
    (\lambda_2+\lambda_3)}+\lambda_1+\lambda_4>0 \nonumber \\
&& \land
    \Bigl(\Bigl(\frac{(\lambda_2+\lambda_3) \Bigl(-\lambda
    \lambda_2^2+\lambda_1^2 (\lambda_2-\lambda_3)+2 \lambda_1 \lambda_2
    \lambda_4\Bigr)}{\lambda_1 \lambda_2}>0 \nonumber \\
&&\land \Bigl(\Bigl(\lambda_2>0\land
    \lambda (\lambda_2+\lambda_3)>\lambda_1^2\land \lambda_1<0\Bigr)\lor
    (\lambda_1>0\land \lambda_3 (2 \lambda_2+\lambda_3)>0\land
    \lambda_2<0)\Bigr)\Bigr) \nonumber \\
&&\lor (\lambda_1>0\land \lambda_2>0)\lor
    \Bigl(\lambda (\lambda_2+\lambda_3)>\lambda_1^2\land \lambda_3 (2
    \lambda_2+\lambda_3)>0 \nonumber \\
&&\land \sqrt{\lambda_3 (2 \lambda_2+\lambda_3)
    \Bigl(\lambda (\lambda_2+\lambda_3)-\lambda_1^2\Bigr)}+\lambda_1
    \lambda_3+\lambda_4 (\lambda_2+\lambda_3)>0\Bigr)\Bigr)
\end{eqnarray}

\begin{eqnarray}
&& {\bf direction \; 5:} \nonumber \\
  &&  \lambda>0\land \lambda_2+\lambda_3>0\land \sqrt{\lambda
    (\lambda_2+\lambda_3)}+\lambda_1>0\land \nonumber \\
&&   \Bigl(\Bigl(\frac{(\lambda_2+\lambda_3) \Bigl(\lambda
    \lambda_2^2+\lambda_1^2 (\lambda_3-\lambda_2)+2 \lambda_1 \lambda_3
    \lambda_4+\lambda_4^2 (\lambda_2+\lambda_3)\Bigr)}{\lambda_2
    (\lambda_1+\lambda_4)}<0\land \nonumber \\
&&\Bigl((\lambda_3 (2
    \lambda_2+\lambda_3)>0\land \lambda_1+\lambda_4>0\land \lambda_2<0)\lor
    \Bigl(\lambda_2>0\land \lambda
    (\lambda_2+\lambda_3)>(\lambda_1+\lambda_4)^2 \nonumber \\
&&\land \lambda_1+\lambda_4<0\Bigr)\Bigr)\Bigr)\lor (\lambda_2>0\land
    \lambda_1+\lambda_4>0)\lor \Bigl(\lambda
    (\lambda_2+\lambda_3)>(\lambda_1+\lambda_4)^2\land \lambda_3 (2
    \lambda_2+\lambda_3)>0  \nonumber \\
&&\land \sqrt{-\lambda_3 (2 \lambda_2+\lambda_3)
    \Bigl((\lambda_1+\lambda_4)^2-\lambda
    (\lambda_2+\lambda_3)\Bigr)}+\lambda_1 \lambda_3>\lambda_2
    \lambda_4\Bigr)\Bigr)
\end{eqnarray}

\begin{eqnarray}
&& \hspace{-.5cm} {\bf direction \; 6:}  \nonumber \\
&& \lambda>0\land \lambda_2+\lambda_3>0\land \sqrt{\lambda
    (\lambda_2+\lambda_3)}+\lambda_1>0 \nonumber \\ 
&&\land \Bigl(\lambda_1+\lambda_4>0\lor
    \Bigl(\lambda (\lambda_2+\lambda_3)>(\lambda_1+\lambda_4)^2\land
    \lambda_4<0\Bigr)\Bigr)~~~~~~~~~~~~~~~~~~~~~~~~~~~~~~~~~~~~~~~
\end{eqnarray}

\begin{eqnarray}
&& {\bf direction \; 7:} \nonumber \\
 &&   \lambda>0\land 2 \lambda_2+\lambda_3>0\land \lambda_2+\lambda_3>0\land
    \sqrt{\lambda (4 \lambda_2+2 \lambda_3)}+2 \lambda_1+\lambda_4>0\land
    \Bigl(\Bigl(\lambda (\lambda_2+\lambda_3)>\lambda_1^2\land \nonumber \\
&&    \Bigl(\Bigl(\lambda_1 (2 \lambda_2+3 \lambda_3)+2 \lambda_4
    (\lambda_2+\lambda_3)>\frac{2 \lambda
    (\lambda_2+\lambda_3)^2}{\lambda_1}\land \lambda_1<0\Bigr)\lor \sqrt{2}
    \sqrt{\lambda_3 (\lambda_2+\lambda_3) \Bigl(\lambda_1^2-\lambda
    (\lambda_2+\lambda_3)\Bigr)} \nonumber \\
&&+\lambda_4  (\lambda_2+\lambda_3)>0\Bigr)\Bigr)\lor \lambda_1>0\Bigr)
\end{eqnarray}

\begin{eqnarray}
&&{\bf direction \; 8:} \nonumber \\ 
  &&  \lambda>0\land 2 \lambda_2+\lambda_3>0\land \lambda_2+\lambda_3>0\land
    \sqrt{\lambda (4 \lambda_2+2 \lambda_3)}+2 \lambda_1+\lambda_4>0\land  \nonumber \\
&&
    \Bigl(\Bigl(\lambda (\lambda_2+\lambda_3)>(\lambda_1+\lambda_4)^2\land
    \Bigl(\Bigl(\sqrt{2} \sqrt{\lambda_3 (\lambda_2+\lambda_3)
    \Bigl((\lambda_1+\lambda_4)^2-\lambda
    (\lambda_2+\lambda_3)\Bigr)}>\lambda_4 (\lambda_2+\lambda_3)\land
    \lambda_3<0\Bigr) \nonumber \\
&& \lor \Bigl(2 \lambda_1 \lambda_2+3 \lambda_1
    \lambda_3+\lambda_3 \lambda_4>\frac{2 \lambda
    (\lambda_2+\lambda_3)^2}{\lambda_1+\lambda_4}\land
    \lambda_1+\lambda_4<0\Bigr)\Bigr)\Bigr)\lor
    \lambda_1+\lambda_4>0\Bigr)
\end{eqnarray}

\begin{eqnarray}
&& \hspace{-.2cm} {\bf direction \; 9:} \nonumber \\
&&   \lambda>0\land 2 \lambda_2+\lambda_3>0\land \sqrt{\lambda (4
    \lambda_2+2 \lambda_3)}+2 \lambda_1+\lambda_4>0 \nonumber \\ 
&&\land \Bigl(2 \lambda (2
    \lambda_2+\lambda_3)>(2 \lambda_1+\lambda_4)^2\lor 2
    \lambda_1+\lambda_4>0\Bigr)~~~~~~~~~~~~~~~~~~~~~~~~~~~~~~~~~~~~~~~~~~~~~~~~~~~~
\end{eqnarray}

\begin{eqnarray}
&&{\bf direction \; 10:} \nonumber \\
 &&   \lambda>0\land \lambda_2+\lambda_3>0\land \sqrt{\lambda
    (\lambda_2+\lambda_3)}+\lambda_1+\lambda_4>0\land \Bigl(\lambda_1>0\lor
    \lambda (\lambda_2+\lambda_3)>\lambda_1^2\lor \lambda_4>0\Bigr) ~~~~~~~~~~~~~   \label{eq:BFB3dir10}
\end{eqnarray}

\noindent
We emphasize that all the above BFB conditions are contained in the general solution given by 
Eqs.~(\ref{eq:bound1} - \ref{eq:bound3}).

\renewcommand{\theequation}{\arabic{section}.\arabic{equation}}

\renewcommand{\theequation}{C.\arabic{equation}}
\setcounter{equation}{0}
\section*{Appendix C.}

For completeness we give in subsections C.1 and C.2 a partial list of the couplings in the DTHM that are relevant 
respectively to the discussion in section \ref{sec:higgspheno}
 and to the derivation of the results of section \ref{sec:unitarity}.

\subsection*{C.1 Higgs-gauge boson couplings \& triple Higgs couplings}

Shifting the neutral fields according to Eq.~({\ref{eq:shiftedfields}), and using the relations between  
the physical and non-physical state bases,
 
\begin{eqnarray}
\left(
  \begin{array}{c}
    h \\
    \xi^0 \\
  \end{array}
\right)
&=& \mathcal{R}_{\alpha} \left(
               \begin{array}{c}
                 h^0 \\
                 H^0 \\
               \end{array}
             \right)
, \qquad
\left(
  \begin{array}{c}
    Z_1 \\
    Z_2 \\
  \end{array}
\right)
= \mathcal{R}_{\beta} \left(
               \begin{array}{c}
                 G^0 \\
                 A^0 \\
               \end{array}
             \right)
\end{eqnarray}
\begin{eqnarray}
\left(
  \begin{array}{c}
    \phi^\pm \\
    \delta^\pm \\
  \end{array}
\right)
&=& \mathcal{R}_{\beta^{'}} \left(
               \begin{array}{c}
                 G^\pm \\
                 H^\pm \\
               \end{array}
             \right)
\end{eqnarray}

\noindent
whith $\mathcal{R}_{\alpha}, \mathcal{R}_{\beta}$ and $\mathcal{R}_{\beta'}$ as defined in 
Eqs.(\ref{eq:rotamatbetaprime},
\ref{eq:rotamatalphabeta}), one extracts from the kinetic terms and the covariant 
derivatives, Eqs.~( \ref{eq:DTHM}, \ref{eq:covd1}, \ref{eq:covd2}), the couplings involving Higgs bosons and gauge 
bosons, and from the potential, Eq.~(\ref{eq:Vpot}), the triple Higgs couplings.

We list below some of the resulting Feynman rules and provide also
 approximate expressions 
in the limit  of very small mixing between the triplet and doublet Higgs multiplets, 
(i.e $s_\alpha= {\cal O} (v_t^2/v_d^2)$, $c_{\alpha,\beta,\beta'}=1 + {\cal O} (v_t^2/v_d^2)$, 
$s_{\beta} = 2 v_t/v_d + {\cal O} (v_t^2/v_d^2)$ and $s_{\beta'} =  \sqrt{2} v_t/v_d + {\cal O} (v_t^2/v_d^2)$).

\begin{eqnarray}
&&{h^0ZZ}=+i{g\over c_W} m_Z (c_{\alpha} c_{\beta} + 2 s_{\alpha} 
s_{\beta})g_{\mu\nu}
\approx i{g\over c_W} m_Z g_{\mu\nu} \label{eq:coup1}\\
&&{H^0ZZ}=
-i{g\over c_W} m_Z (s_{\alpha} c_{\beta} - 2 c_{\alpha} s_{\beta}) g_{\mu\nu}
\approx 4 i{g\over c_W} {v_t\over v_d} m_Z g_{\mu\nu}
\label{eq:coup2}\\
&&{h^0W^+W^-}=i g c_W m_Z (c_{\alpha} c_{\beta} + s_{\alpha} s_{\beta}) g_{\mu\nu}
\approx i g m_W g_{\mu\nu}
\label{eq:coup3}\\
&&{H^0W^+W^-}= -i g m_W (s_{\alpha} c_{\beta} - c_{\alpha} s_{\beta})g_{\mu\nu}
\approx 2 i g m_W   {v_t\over v_d}g_{\mu\nu}
\label{eq:coup4}\\
&&{h^0A^0Z}= -{g\over2c_W}(c_\alpha s_\beta-2c_\beta s_\alpha)(p_h-p_A)_\mu
\approx -{g\over c_W} {v_t\over v_d}(p_h-p_A)_\mu
\label{eq:coup5}\\
&&{H^0A^0Z}={g\over2c_W}(s_\alpha s_\beta+2 c_\alpha c_\beta)(p_H-p_A)_\mu
\approx {g\over c_W}  (p_H-p_A)_\mu
\label{eq:coup6} \\ 
&&h^0h^0H^0= i \,( \,(\frac32 \lambda c_\alpha^2  - \lambda_{14}^+) s_\alpha v_d - 
              6 \lambda_{23}^+ c_\alpha  s_\alpha^2 v_t + 
    (c_\alpha^2 - 2 s_\alpha^2) (\sqrt{2} c_\alpha \mu - \lambda_{14}^+ (s_\alpha v_d + c_\alpha v_t)\,)\,)   \nonumber \\
&&~~~~~~~~~~\approx   i \, ( \, \sqrt{2} \mu +  (3 \lambda - 5 (\lambda_1 + \lambda_4)) v_t \,) + {\cal O}(v_t^2) \label{eq:coup21} \\
&&{h^0W^+H^-}=i{g\over 2}(c_\alpha s_{\beta'}-\sqrt{2}s_\alpha c_{\beta'})(p_h-p_{H^-})_\mu
\approx +i{g\over\sqrt{2} } {v_t\over v_d}(p_h-p_{H^-})_\mu
\label{eq:coup7} \\
&&{H^0W^+H^-}=-i {g\over 2}(s_\alpha s_{\beta'}+\sqrt{2}c_\alpha c_{\beta'})(p_H-p_{H^-})_\mu
\approx -i {g\over \sqrt{2}} (p_H-p_{H^-})_\mu
\label{eq:coup8} 
\end{eqnarray}

\begin{eqnarray}
&&{A^0W^+H^-}= {g\over2}(\sqrt{2}c_{\beta'}c_{\beta}+s_{\beta'}s_{\beta}) (p_{A^0}-p_{H^-})_\mu
\approx  {g\over\sqrt{2}}(p_{A^0}-p_{H^-})_\mu ~~~~~~~~~~~~~~~~~~~~~~~~~~~\label{eq:coup9} \\
&&{Z_\mu W_\nu^+H^-} =g m_Z (c_{\beta} s_{\beta'} s_W^2 - \frac{s_{\beta}
c_{\beta'}}{\sqrt{2}}(1+s_W^2))g_{\mu\nu}
\approx -\sqrt{2} g \frac{m_Z  v_t}{v_d}g_{\mu\nu} \label{eq:coup10}\\
&&H^{++}H^-W^-_\mu = i g c_{\beta^{'}}(p_{H^{++}}-p_{H^-})_\mu \approx i g (p_{H^{++}}-p_{H^-})_\mu \label{eq:coup11}\\
&&H^{++}W^-_\mu W^-_\nu= -i\sqrt{2}g^2 v_{t} g_{\mu\nu} \label{eq:coup12}\\
&&H^{++}H^-H^- = -i 
( 2\mu s_{\beta'}^2 + c_{\beta'} \; ( \lambda_4 \, s_{\beta'} \, v_d - \sqrt{2} \, \lambda_3 \, c_{\beta'} \, v_t)  ) 
\label{eq:coup13}\\
&&H^{++}H^{--} V V' = 8 i e_V e_{V'} g_{\mu\nu} \label{eq:coup14}\\
&&H^{++}H^{--} V =-2 i e_V (p_{H^{++}}-p_{H^{--}})_\mu \label{eq:coup15}\\
&&H^{+}H^{-} V V' = 2 i e_V e_{V'} \ g_{\mu\nu} \label{eq:coup16}\\
&&H^{+}H^{-} V = -i e_V (p_{H^{+}}-p_{H^{-}})_\mu \label{eq:coup17}\\
&&G^{+}G^{-} V V' = 2 i e_V e_{V'} \ g_{\mu\nu} \label{eq:coup18}\\
&&G^{+}G^{-} V = -i e_V (p_{G^{+}}-p_{G^{-}})_\mu \label{eq:coup19}
\end{eqnarray}

\noindent
where in Eqs.~(\ref{eq:coup14} - \ref{eq:coup19}) we denote by $V$ and $V'$ the $\gamma$ or $Z$ gauge boson, 
with the couplings satisfying  $e_{\gamma} \equiv e$ and $e_{Z} \equiv e \cot 2 \theta_W$. 
We also adopted the convention that all momenta are incoming at each vertex.

\subsection*{C.2 Quartic scalar couplings in the doublet-triplet basis}
Here we give the complete list of Feynman rules for the quartic scalar couplings in the unrotated
basis which were used in section \ref{sec:unitarity} to determine the unitarity constraints:

\begin{eqnarray}
\begin{array}{ll}
\delta^+ \ \delta^+ \ \delta^- \ \delta^- = -2i(2\lambda_2+\lambda_3) \quad , & \quad
Z_2 \ Z_2 \ \phi^- \ \phi^+=-i (\lambda_1 )\nonumber  \\ [0.12cm]
\delta^+ \ \delta^- \ \delta^{--} \ \delta^{++}=-2i(\lambda_2+\lambda_3) \quad , &  \quad
Z_1 \ Z_1 \ \phi^- \ \phi^+ = -i\frac{1}{2}\lambda \nonumber  \\ [0.12cm]
\delta^{++} \ \delta^{++} \ \delta^{--} \ \delta^{--}=-4i(\lambda_2+\lambda_3) \quad , & \quad
\phi^- \ \phi^- \ \phi^+ \ \phi^+ = -i \lambda \nonumber  \\ [0.12cm]
\delta^{+} \ \delta^{+} \ \delta^{--} \ Z_2= \sqrt{2}\lambda_3 \quad    , & \quad
\delta^{+} \ \delta^{+} \ \delta^{--} \ \xi^0 =i\sqrt{2}\lambda_3 \nonumber  \\ [0.12cm]
\delta^{-} \ \delta^{-} \ \delta^{++} \ Z_2 =-\sqrt{2}\lambda_3 \quad , & \quad
\delta^- \ \delta^- \ \delta^{++} \ \xi^0 = i \sqrt{2}\lambda_3 \nonumber  \\ [0.12cm]
\delta^- \ \delta^+ \ Z_2 \ Z_2 = -2i(\lambda_2+\lambda_3) \quad , & \quad
\delta^+ \ Z_1 \ \phi^- \ \xi^0 = \frac{\lambda_4 }{2\sqrt{2}} \nonumber  \\ [0.12cm]
Z_2 \ Z_2 \ Z_2 \ Z_2 = -6i(\lambda_2+\lambda_3) \quad , & \quad
\delta^- \ Z_1 \ \phi^+ \ \xi^0 = \frac{-\lambda_4}{2\sqrt{2}} \nonumber  \\ [0.12cm]
\end{array}
\end{eqnarray}
\begin{eqnarray}
\begin{array}{ll}
\delta^{++} \ \delta^{--} \ Z_2 \ Z_2 = -2i\lambda_2 \quad , & \quad
\delta^- \ \delta^+ \ \xi^0 \ \xi^0 = -2i(\lambda_2+\lambda_3) \nonumber  \\ [0.12cm]
\delta^{+} \ \delta^{-} \ Z_1 \ Z_1 = -\frac{i}{2}(2\lambda_1+\lambda_4 ) \quad , & \quad
\delta^{--} \ \delta^{++} \ \xi^0 \ \xi^0 = -2i\lambda_2 \nonumber  \\ [0.12cm]
\delta^{++} \ \delta^{--} \ Z_1 \ Z_1 = -i(\lambda_1 ) \quad , & \quad
Z_2 \ Z_2 \ \xi^0 \ \xi^0 = -2i(\lambda_2+\lambda_3) \nonumber  \\ [0.12cm]
Z_2 \ Z_2 \ Z_1 \ Z_1 = -i(\lambda_1+\lambda_4) \quad , & \quad
Z_1 \ Z_1 \ \xi^0 \ \xi^0 = -i(\lambda_1+\lambda_4) \nonumber  \\ [0.12cm]
Z_1 \ Z_1 \ Z_1 \ Z_1 = -\frac{3}{2}i\lambda \quad , & \quad
\phi^- \ \phi^+ \ \xi^0 \ \xi^0 = -i(\lambda_1 ) \nonumber  \\ [0.12cm]
\delta^{++} \ \delta^{-} \ Z_1 \ \phi^- = -\frac{ \lambda_4 }{2} \quad , & \quad
\xi^0 \ \xi^0 \ \xi^0 \ \xi^0 = -6i(\lambda_2+\lambda_3) \nonumber  \\ [0.12cm]
\delta^{+} \ \phi^{-} \ Z_1 \ Z_2 = -\frac{i \lambda_4 }{2\sqrt{2}} \quad , & \quad
\delta^- \ \delta^{++} \ \phi^{-} \ h = \frac{i \lambda_4 }{2} \nonumber  \\ [0.12cm]
\delta^+ \ \delta^{--} \ \phi^{+} \ Z_1 = \frac{\lambda_4 }{2} \quad , & \quad
\delta^+ \ \phi^{-} \ Z_2 \ h = -\frac{ \lambda_4 }{2\sqrt{2}} \nonumber  \\ [0.12cm]
\delta^- \ \phi^{+} \ Z_2 \ Z_1 = -\frac{i \lambda_4 }{2\sqrt{2}} \quad , & \quad
\delta^+ \ \delta^{--} \ \phi^{+} \ h = \frac{i \lambda_4 }{2} \nonumber  \\ [0.12cm]
\delta^- \ \delta^{+} \ \phi^{+} \ \phi^{-} = -\frac{i}{2}(2\lambda_1+\lambda_4 ) \quad , & \quad
\delta^- \ Z_2 \ \phi^{+} \ h = \frac{\lambda_4 }{2\sqrt{2}} \nonumber  \\ [0.12cm]
\delta^{--} \ \delta^{++} \ \phi^{+} \ \phi^{-} = -i(\lambda_1+\lambda_4) \quad , & \quad
\delta^{+} \ \xi^0 \ h \ \phi^{-} = -\frac{i \lambda_4 }{2\sqrt{2}} \nonumber  \\ [0.12cm]
\delta^{-} \ \phi^{+} \ \xi^0 \ h = -\frac{i \lambda_4 }{2\sqrt{2}} \quad , & \quad
Z_2 \ Z_2 \ h \ h = -i(\lambda_4+\lambda_1) \nonumber  \\ [0.12cm]
\delta^{-} \ \delta^{+} \ h \ h = -\frac{i}{2}(2\lambda_1+\lambda_4 ) \quad , & \quad
Z_1 \ Z_1 \ h \ h = -i\frac{\lambda}{2} \nonumber  \\ [0.12cm]
\delta^{--} \ \delta^{++} \ h \ h = -i(\lambda_1 ) \quad , & \quad
\phi^{+} \ \phi^{-} \ h \ h = -i\frac{\lambda}{2} \nonumber  \\ [0.12cm]
h \ h \ \xi^0 \ \xi^0 = -i(\lambda_1+\lambda_4) \quad , & \quad
h \ h \ h \ h = -i\frac{3\lambda}{2} \nonumber  \\ [0.12cm]
\end{array}
\end{eqnarray}

\noindent
One can read off from this list the $\mu$-independent part of the triple scalar couplings, by
making the substitutions $Z_1 \to -i v_d, h \to v_d$ or
$Z_2 \to -i v_t, \xi^0 \to v_t$ (cf. Eq.~({\ref{eq:shiftedfields}) ) in the appropriate vertices
and modifying accordingly the symmetry factors for identical fields.

\newpage
\bibliographystyle{unsrt}
\bibliography{references-triplet-NEW}

\end{document}